\documentclass[12pt]{article}
\pdfoutput=1
\usepackage{makeidx}
\makeindex
\usepackage[a4paper]{geometry}
\usepackage{jheppub,amsmath,amssymb,amsfonts,amsxtra,mathrsfs,graphics,graphicx,amsthm,epsfig,ytableau,bm,longtable,float,color,tikz,mathtools,xfrac,footnote,rotating,lscape}
\usepackage{dsfont}
\usepackage{textgreek}
\usetikzlibrary{decorations.pathmorphing}
\usetikzlibrary{decorations.markings}
\usetikzlibrary{arrows, decorations.markings, calc, fadings, decorations.pathreplacing, patterns, decorations.pathmorphing, positioning}
\usepackage{tikz-cd}

\usepackage{fixmath} 
\usepackage{scalerel}
\newlength\bshft
\def\fakebold#1{\ThisStyle{\ooalign{$\SavedStyle#1$\cr%
  \kern-\bshft$\SavedStyle#1$\cr%
  \kern\bshft$\SavedStyle#1$}}}

\usetikzlibrary{positioning,shapes}
\usetikzlibrary{chains}
\usetikzlibrary{arrows,fit,decorations.pathreplacing}
\tikzstyle{every picture}+=[remember picture]
\tikzstyle{na} = [baseline=-.5ex]

\usepackage{empheq}
\usepackage{multirow}
\usepackage{booktabs}
\usepackage[american]{babel}

\usepackage[latin1]{inputenc}

\makeatletter
\newcommand{\vast}{\bBigg@{1}}
\newcommand{\Vast}{\bBigg@{5}}
\makeatother

\usepackage{hyperref}

\numberwithin{equation}{section}

\newcommand{\cf}{\textit{cf.}}
\newcommand{\eg}{\textit{e.g.}}

\newcommand{\ie}{\textit{i.e.}}

\numberwithin{equation}{section}

\newcommand{\nn}{\nonumber}

\newcommand{\be}{\begin{equation}} \newcommand{\ee}{\end{equation}}
\newcommand{\bea}{\begin{equation} \begin{aligned}} \newcommand{\eea}{\end{aligned} \end{equation}}

\def\U{\mathrm{U}}
\def\SO{\mathrm{SO}}

\def\SU{\mathrm{SU}}

\def\USp{\mathrm{USp}}

\def\E{\mathrm{E}}

\newcommand{\rd}{\mathrm{d}}

\newcommand{\wt}{\widetilde}

\DeclareMathOperator{\Tr}{Tr}

\DeclareMathOperator{\sign}{sign}

\DeclareMathOperator{\re}{\mathbb{R}e}
\DeclareMathOperator{\im}{\mathbb{I}m}

\DeclareMathOperator{\Li}{Li}

\newcommand{\cA}{\mathcal{A}}

\newcommand{\cC}{\mathcal{C}}

\newcommand{\cE}{\mathcal{E}}
\newcommand{\cF}{\mathcal{F}}
\newcommand{\cG}{\mathcal{G}}
\newcommand{\cH}{\mathcal{H}}
\newcommand{\cI}{\mathcal{I}}

\newcommand{\cM}{\mathcal{M}}
\newcommand{\cN}{\mathcal{N}}
\newcommand{\cO}{\mathcal{O}}

\newcommand{\cQ}{\mathcal{Q}}
\newcommand{\cR}{\mathcal{R}}

\newcommand{\cW}{\mathcal{W}}

\newcommand{\bC}{\mathbb{C}}

\newcommand{\bF}{\mathbb{F}}

\newcommand{\bP}{\mathbb{P}}

\newcommand{\bR}{\mathbb{R}}
\newcommand{\bZ}{\mathbb{Z}}

\newcommand{\fg}{\mathfrak{g}}

\newcommand{\fh}{\mathfrak{h}}

\newcommand{\fk}{\mathfrak{k}}
\newcommand{\fm}{\mathfrak{m}}

\newcommand{\fn}{\mathfrak{n}}
\newcommand{\fN}{\mathfrak{N}}
\newcommand{\fl}{\mathfrak{l}}

\newcommand{\fp}{\mathfrak{p}}
\newcommand{\fq}{\mathfrak{q}}
\newcommand{\fr}{\mathfrak{r}}
\newcommand{\fR}{\mathfrak{R}}
\newcommand{\fs}{\mathfrak{s}}
\newcommand{\ft}{\mathfrak{t}}

\usepackage[Symbol]{upgreek}
\usepackage{bm}

\usepackage{calligra}
\DeclareMathAlphabet{\mathcalligra}{T1}{calligra}{m}{n}

\newcommand{\mathbbm}[1]{\text{\usefont{U}{bbm}{m}{n}#1}}

\theoremstyle{plain}
\newtheorem{thm}{\protect\theoremname}[section]
  \theoremstyle{definition}
\newtheorem{example}[thm]{\protect\examplename}
\providecommand{\examplename}{Example}
\providecommand{\theoremname}{Theorem}

\makeatletter
\g@addto@macro\bfseries{\boldmath}
\makeatother

\title{Topologically twisted indices \\ in five dimensions and holography}

\author[a]{Seyed Morteza Hosseini,}
\author[a]{Itamar Yaakov}
\author[b,c]{and Alberto Zaffaroni}
\affiliation[a]{Kavli IPMU (WPI), UTIAS, The University of Tokyo, Kashiwa, Chiba 277-8583, Japan}
\affiliation[b]{Dipartimento di Fisica, Universit\`a di Milano-Bicocca, I-20126 Milano, Italy}
\affiliation[c]{INFN, sezione di Milano-Bicocca, I-20126 Milano, Italy}
\emailAdd{morteza.hosseini@ipmu.jp}
\emailAdd{itamar.yaakov@ipmu.jp}
\emailAdd{alberto.zaffaroni@mib.infn.it}

\preprint{IPMU18-0133}

\abstract{We provide a formula for the partition function of five-dimensional $\mathcal{N}=1$ gauge theories on $\mathcal{M}_4 \times S^1$, topologically twisted along $\mathcal{M}_4$ in the presence of general background magnetic fluxes, where $\mathcal{M}_4$ is a toric K\"ahler manifold. The result can be expressed as a contour integral of the product of copies of the K-theoretic Nekrasov's partition function, summed over gauge magnetic fluxes. The formula generalizes to five dimensions the topologically twisted index of three- and four-dimensional field theories. We analyze the large $N$ limit of the partition function and some related quantities for two theories: $\mathcal{N}=2$ SYM and the $\mathrm{USp}(2N)$ theory with $N_f$ flavors and an antisymmetric matter field. For $\mathbb{P}^1 \times \mathbb{P}^1 \times S^1$, which can be easily generalized to $\Sigma_{\mathfrak{g}_2} \times \Sigma_{\mathfrak{g}_1} \times S^1$, we conjecture the form of the relevant saddle point at large $N$. The resulting partition function for $\mathcal{N}=2$ SYM scales as $N^3$ and is in perfect agreement with the holographic results for domain walls in AdS$_7 \times S^4$. The large $N$ partition function for the $\mathrm{USp}(2N)$ theory scales as $N^{5/2}$ and gives a prediction for the entropy of a class of magnetically charged black holes in massive type IIA supergravity.}

\begin{document}

\setcounter{tocdepth}{2}
\maketitle

%
%

\date{Dated: \today}




\section{Introduction}
\label{sec:intro}

In this paper we will consider the partition functions of five-dimensional $\cN = 1$ gauge theories with an $\SU(2)$ R-symmetry on $\cM_4 \times S^1$,
partially topologically twisted on the toric K\"ahler manifold $\cM_4$ \cite{Witten:1988ze,Witten:1991zz}, with a view to holography. In particular, we are interested in  evaluating through localization \cite{Pestun:2007rz}
the \emph{topologically twisted index} of a five-dimensional theory on $\cM_4 \times S^1$, which is defined as the equivariant Witten index
\be
 Z_{\cM_4 \times S^1}(\fs_I, y_I) = \Tr_{\cM_4} (-1)^F e^{-\beta \{{\cal Q},\bar {\cal Q}\}} \prod_I y_I^{J_I} \, ,
\ee
of the topologically twisted theory on $\cM_4$, where $\fs_I$ are magnetic fluxes on $\cM_4$ that explicitly enter in the Hamiltonian and $y_I$ complexified fugacities for the flavor symmetries $J_I$ of the theory. 

In three and four dimensions, the topologically twisted index has proven useful in checking dualities and in the microscopic counting for the entropy of a class of asymptotically AdS$_{4/5}$
black holes/strings \cite{Benini:2015noa,Benini:2015eyy}. Since one of the goals of this work is to  extend the above analyses to higher dimensions, let us briefly review what is known in lower dimensions.

\subsection{The three- and four-dimensional indices}

The topologically twisted index of three-dimensional $\cN = 2$ and four-dimensional $\cN = 1$ gauge theories with an R-symmetry is the supersymmetric partition function on $\Sigma_{\fg_1} \times T^d$,
partially topologically $A$-twisted along the genus $\fg_1$ Riemann surface $\Sigma_{\fg_1}$, where  $T^d$ is a torus with $d= 1, 2$, respectively.
The index  can be computed in two different ways. It has been first derived  by  topological field theory arguments in \cite{Okuda:2012nx,Okuda:2013fea,Nekrasov:2014xaa,Gukov:2015sna}.
In this approach, further discussed and generalized in \cite{Okuda:2015yea,Gukov:2016gkn,Closset:2016arn,Closset:2017zgf,Dedushenko:2017tdw,Gukov:2017zao,Closset:2017bse,Closset:2018ghr},
the index is written as a sum of contributions coming from the Bethe vacua,
the critical points of the twisted superpotential of the two-dimensional theory obtained by compactifying on $T^d$.
The index has been also derived using localization in \cite{Benini:2015noa,Benini:2016hjo} and can be written as the contour integral 
\begin{equation}
 Z (\fs_I, y_I) = 
 \sum_{\fm \,\in\, \Gamma_\fh} \oint_\cC Z_{\text{int}} (\fm, x;  \fs_I , y_I) \, ,
\end{equation}
of a meromorphic differential form in variables $x$ parameterizing the Cartan subgroup and subalgebra of the gauge group, summed over
the lattice of gauge magnetic fluxes $\fm$ on $\Sigma_{\fg_1}$. Here $\fs_I $ and $y_I = e^{\mathbbm{i} \Delta_I}$ are, respectively, fluxes and fugacities for the global symmetries of the theory.

One remarkable application of the topologically twisted index is to  understand  the microscopic origin of the Bekenstein-Hawking entropy of asymptotically anti de Sitter (AdS) black holes.
In particular, the microscopic entropy of certain four-dimensional static, dyonic, BPS black holes \cite{Cacciatori:2009iz,DallAgata:2010ejj,Hristov:2010ri,Katmadas:2014faa,Halmagyi:2014qza},
which can be embedded in AdS$_4\times S^7$,  has been calculated in this manner \cite{Benini:2015eyy,Benini:2016rke},
by showing that the ABJM \cite{Aharony:2008ug} twisted index, in the large $N$ limit, agrees with the area law for the black hole entropy ---
$S_{\rm BH} = A / ( 4 G_{\rm N})$ with $A$ being the horizon area and $G_{\rm N}$ the Newton's constant.
Specifically, the statistical entropy $S_{\text{BH}}$ as a function of the magnetic and electric charges $(\fs , \fq)$ is given by the Legendre transform of the field theory twisted index $Z (\fs , \bar \Delta)$,
evaluated at its critical point $\bar \Delta_I$:
\be
 \label{intro:I-extremization}
 \cI ( \fs , \bar \Delta ) \equiv \log Z (\fs , \bar \Delta) - \mathbbm{i} \sum_{I} \fq_I \bar \Delta_I = S_{\text{BH}} (\fs , \fq) \, .
\ee
This procedure was dubbed $\cI$-extremization in \cite{Benini:2015eyy}.
These results have been generalized to black strings in five dimensions \cite{Hosseini:2016cyf,Hosseini:2018qsx,Hong:2018viz},
black holes with hyperbolic horizons \cite{Benini:2016hjo,Cabo-Bizet:2017jsl},
universal black holes \cite{Azzurli:2017kxo},%
\footnote{These can be embedded in all M-theory and massive type IIA compactifications, thus explaining the name universal.}
black holes in massive type IIA supergravity \cite{Hosseini:2017fjo,Benini:2017oxt},
M-theory black holes in the presence of hypermultiplets \cite{Bobev:2018uxk},
Taub-NUT-AdS/Taub-Bolt-AdS solutions \cite{Toldo:2017qsh},
and $N$ M5-branes wrapped on hyperbolic three-manifolds \cite{Gang:2018hjd}.%
\footnote{Other interesting progresses in this context include: computing the logarithmic correction to AdS$_4 \times S^7$ black holes \cite{Liu:2017vbl} (see also \cite{Liu:2017vll,Jeon:2017aif}),
evaluating the on-shell supergravity action for the latter black holes \cite{Halmagyi:2017hmw,Cabo-Bizet:2017xdr},
localization in gauged supergravity \cite{Hristov:2018lod} (see also \cite{Nian:2017hac}), relation between anomaly polynomial of $\cN = 4$ super Yang-Mills (6D $\cN = (2,0)$ theory)
and rotating, electrically charged, AdS black holes in five (seven) dimensions \cite{Hosseini:2017mds,Hosseini:2018dob}.} Another interesting general result is the Cardy behaviour
of the topologically twisted index of four-dimensional $\cN = 1$ gauge theories that flow to an infrared (IR) two-dimensional $\cN = (0,2)$ superconformal field theory (SCFT) upon twisted compactification on  $\Sigma_{\fg_1}$ \cite{Hosseini:2016cyf}:
\be
 \label{Cardy:4d:intro}
 \log Z_{\Sigma_{\fg_1} \times T^2} \approx \frac{\mathbbm{i} \pi}{12 \tau} c_r (\fs , \Delta)  \, ,
\ee
where $c_r (\fs , \Delta)$ is the trial right-moving central charge of the $\cN = (0,2)$ SCFT \cite{Benini:2012cz} and $\tau$ is the modular parameter of the torus $T^2$.%
\footnote{Here we use the chemical potentials $\Delta_I / \pi$ to parameterize a trial R-symmetry of the theory.} This result is valid at high temperature,
$\tau\rightarrow \mathbbm{i} 0^+$, and is a consequence of the fact that the index computes the elliptic genus of the two-dimensional CFT. 

Furthermore, it has been shown in \cite{Benini:2015eyy,Hosseini:2016cyf} that,  in the large $N$ limit, one particular Bethe vacuum dominates the partition function.
It has also been found in \cite{Hosseini:2016tor,Hosseini:2016cyf} (see \cite{Hosseini:2016ume} for examples) that the topologically twisted index of three- and four-dimensional field theories, at large $N$ or high temperature, can be compactly written as
\be
 \label{index:theorem:original:3d:4D}
 \begin{aligned}
  & \log Z_{\Sigma_{\fg_1} \times T^d} \approx\sum_{I} \fs_I \frac{\partial \wt\cW_d (\Delta_I) }{\partial \Delta_I}  \, , \qquad
  \wt\cW_d (\Delta_I) \propto \Bigg\{
   \begin{array}{lr}
    F_{S^3} (\Delta_I), & \text{for } d = 1\\
    a (\Delta_I), & \text{for } d = 2
   \end{array}
   \, ,
 \end{aligned}
\ee
where $\wt\cW_d (\Delta_I)$ is the effective twisted superpotential, evaluated on the dominant Bethe vacuum, 
$F_{S^3}(\Delta_I)$ is the $S^3$ free energy of 3D $\cN = 2$ theories, computed for example in \cite{Herzog:2010hf,Jafferis:2011zi,Fluder:2015eoa}, and $a(\Delta_I)$ is the conformal anomaly coefficient of 4D $\cN = 1$ theories. 
Here (and throughout the paper) $\approx$ denotes the equality at large $N$.%
\footnote{The  relation \eqref{index:theorem:original:3d:4D} is only valid when we use a set of chemical potentials such that $\wt\cW_d (\Delta_I)$ is a homogeneous function of the $\Delta_I$ (and a similar parameterization for the fluxes),
which is always possible \cite{Hosseini:2016tor,Hosseini:2016cyf}. Otherwise, \eqref{index:theorem:original:3d:4D} should be replaced by $\log Z_{\Sigma_{\fg_1} \times T^d} \approx( 1 - \fg_1 ) \mathfrak{D}^{(1)}_d \wt\cW_d (\Delta_I)$,
where $\mathfrak{D}^{(1)}_d \equiv \frac{(d + 1)}{\pi} + \sum_{I} \Big( \frac{\fs_I}{1 - \fg_1} - \frac{\Delta_I}{\pi} \Big) \frac{\partial}{\partial \Delta_I}$ for $\fg_1 \neq 1$.}
Based on \eqref{index:theorem:original:3d:4D} it has been conjectured in \cite{Hosseini:2016tor} that
\bea
 \label{intro:conjecture:index:attractor}
 \wt\cW_d (\Delta_I) \propto \cF_{\text{sugra}} ( X^\Lambda ) \, , \qquad
 \cI\text{-extremization} = \text{attractor mechanism} \, ,
\eea
where $\cF_{\text{sugra}} ( X^\Lambda )$ is the prepotential of the effective $\cN = 2$ gauged supergravity in four dimensions describing the horizon of the black hole or black string.
We refer the reader to section \ref{subsec:attractor} for details on the attractor mechanism in gauged supergravity.

\subsection{The five-dimensional index}

In this paper we take the first few steps in generalizing the above analysis to five dimensions.

We will consider the case of a generic $\cN = 1$ gauge theory on $\cM_4 \times S^1$, where $\cM_4$ is a toric K\"ahler manifold and $S^1$ a circle of length $\beta$.%
\footnote{See \cite{Kim:2012tr,Kim:2013nva} for another localization computation on $\bP^2 \times S^1$.}
One main complication compared to three and four dimensions is that, in the localization computation for five-dimensional gauge theories,  there are non-perturbative contributions due to the presence of instantons.
The topologically twisted index is still given by the contour integral 
\begin{equation}
 \label{5Dtwisted}
  Z _{\cM_4 \times S^1}(q,\fs_I, y_I) =
  \sum_{k=0}^\infty \sum_{\{\fm \} | \text{semi-stable}} \oint_\cC q^k Z_{\text{int}}^{(k\text{-instantons})} (\fm, a; q, \fs_I , y_I) \, ,
\end{equation}
of a meromorphic form in the complex variable $a$, which parameterizes the Coulomb branch of the four-dimensional theory obtained by compactifying on $S^1$.
Here $q=e^{-8\pi^2 \beta/g_{\text{YM}}^2}$ is the instanton counting parameter with $g_{\text{YM}}$ being the gauge coupling constant,
and $\fm$ are a set of gauge magnetic fluxes that depend on the toric data of $\cM_4$. 
 
We find it useful to work first on an $\Omega$-deformed background specified by equivariant parameters $\epsilon_1$ and $\epsilon_2$ for the toric $(\mathbb{C}^*)^2$
action on $\cM_4$, for which we write explicitly the supersymmetry background, the Lagrangian and the one-loop determinant around the classical saddle point configurations.
The non-perturbative contribution is given by contact instantons that, with the equivariant parameters turned on, are localized at the fixed point of the toric action on $\cM_4$ and wrap $S^1$.
There are $\chi(\cM_4)$  fixed points and each of these  contributes  a copy of the five-dimensional Nekrasov's partition function $Z_{\text{Nekrasov}}^{\mathbb{C}^2\times S^1}(a,\epsilon_1,\epsilon_2)$
on $\mathbb{C}^2\times S^1$, so that the partition function on $\cM_4\times S^1$ is given by the gluing formula
\be
 \label{intro:gluing}
 Z_{\cM_4 \times S^1}  = 
 \sum_{\{\fm \} | \text{semi-stable}} \oint_\cC  \rd a \prod_{l = 1}^{\chi (\cM_4)} Z_{\text{Nekrasov}}^{\bC^2 \times S^1} \big(a^{(l)}, \epsilon_1^{(l)}, \epsilon_2^{(l)} \big) \, ,
\ee
where $a^{(l)}$, $\epsilon_1^{(l)}$ and $\epsilon_2^{(l)}$ are determined by the toric data near each fixed point and encode the dependence on the fluxes $\fm$.
For brevity, we suppressed the dependence on the flavor fluxes and fugacities $\fs_I$ and $y_I$ that can be easily reinstated by considering them as components of a background vector multiplet.
This formula is the five-dimensional analogue of a similar four-dimensional expression that has been successfully used to evaluate equivariant Donaldson invariants \cite{Nekrasov:2003vi,Bawane:2014uka,Bershtein:2015xfa,Bershtein:2016mxz}.

In this paper we will be interested in the non-equivariant limit $\epsilon_1,\epsilon_2\rightarrow 0$. Despite the fact that $Z_{\text{Nekrasov}}^{\mathbb{C}^2\times S^1}(a,\epsilon_1,\epsilon_2)$ is singular in the non-equivariant limit, $Z_{\cM_4 \times S^1}$
is perfectly smooth. Moreover, with an eye to holography, we will be mostly interested in the large $N$ limit and therefore we will neglect instanton contributions, since they are exponentially suppressed in this limit.
As we will see, the classical and one-loop contributions to the non-equivariant partition function still yield a non-trivial and complicated matrix model.
As for the topologically twisted index in three dimensions, we can interpret the result as the Witten index of the quantum mechanics obtained by reducing the $\cN = 1$ gauge theory on $\cM_4$ in the presence of background magnetic fluxes $\fs_I$. This index receives contributions from infinitely many topological sectors specified by the gauge magnetic fluxes $\fm$. 

We will also initiate the study of the large $N$ limit of the topologically twisted index in five dimensions and of other related quantities, leaving a more complete analysis for the future.
We will focus on two five-dimensional  $\cN = 1$ field theories. The first is the $\USp(2N)$ theory with $N_f$ flavors and an antisymmetric matter field, which has a 5D ultraviolet (UV) fixed point with enhanced $E_{N_f+1}$ global symmetry \cite{Seiberg:1996bd}. The theory is 
dual to AdS$_6 \times_w S^4$ in massive type IIA supergravity \cite{Brandhuber:1999np}. The second theory is $\cN = 2$ super Yang-Mills (SYM), which we consider as the compactification of the $\cN = (2,0)$ theory in six dimensions on a circle of radius
$R_6 = g_{\text{YM}}^2/(8 \pi^2)$ \cite{Douglas:2010iu,Lambert:2010iw,Bolognesi:2011nh}.%
\footnote{An analogous argument has been used in \cite{Kim:2012tr,Kim:2013nva,Kim:2012ava} in order to study the superconformal index of the 6D $\cN = (2,0)$ theory.}
Using this interpretation, the index of $\cN = 2$ SYM can be considered as the partition function of the $\cN = (2,0)$ theory on $\cM_4 \times T^2$.
The topologically twisted index at large $N$ for the $\USp(2N)$ theory should then contain information about black holes with horizons AdS$_2 \times \cM_4$ in massive type IIA supergravity, while the index for $\cN = 2$ SYM should contain information
about AdS$_7 \times S^4$ black strings in M-theory.

An interesting object to study in the large $N$ limit is the Seiberg-Witten (SW) prepotential $\cF(a)$ of the four-dimensional theory obtained by compactifying on $S^1$,
which receives contributions from all the Kaluza-Klein (KK) modes on $S^1$ \cite{Nekrasov:1996cz}.
$\cF(a)$ is expected to play a role similar to the twisted superpotential $\wt \cW$ in three and four dimensions. We therefore study the distribution of its critical points in the large $N$ limit and we find that its critical value,
as a function of the chemical potentials $\Delta_\varsigma$ $(\varsigma = 1, 2)$,%
\footnote{The $\Delta_\varsigma$ parameterize the Cartan of the $\SU(2)$ R-symmetry and the $\SU(2)$ flavor symmetry of
the $\USp(2N)$ theory and the Cartan of the $\SO(5)$ R-symmetry of the $\cN = (2,0)$ theory, respectively.
They satisfy the constraint  $\sum_{\varsigma = 1}^2 \Delta_\varsigma = 2 \pi$. 
Similarly, in the case of $\Sigma_{\fg_1}\times \Sigma_{\fg_2}\times S^1$ discussed below, the fluxes fulfill  the constraints $\sum_{\varsigma = 1}^2 \fs_\varsigma = 2 (1 - \fg_1)$, $\sum_{\varsigma = 1}^2 \ft_\varsigma = 2 (1 - \fg_2)$.
With such a choice, all expressions in \eqref{index:theorem:F:5D}, \eqref{index:theorem:W:5D} and \eqref{index:theorem:logZ:5D} are homogeneous functions of
$\Delta_\varsigma$, $\fs_\varsigma$ and $\ft_\varsigma$. Details are given in section \ref{sec:largeN}.} is given by

\be
 \label{index:theorem:F:5D}
  \cF(\Delta_\varsigma) \propto F_{S^5} (\Delta_\varsigma) \, ,
\ee
where $F_{S^5} (\Delta_\varsigma)$ is the free energy on $S^5$ of the corresponding $\cN=1$ theory, in perfect analogy with \eqref{index:theorem:original:3d:4D}.
One of the reasons for analysing the critical points of $\cF(a)$ is the expectation that they play a role similar to the Bethe vacua for the five-dimensional partition function, and, in particular, one such critical point dominates the large $N$ limit of the index.
We have no real evidence that this is the case but we will see that working under this assumption leads to interesting results.
Some motivations for this conjecture are discussed in section \ref{sec:largeN}.

We will consider the particular example of an $\cN=1$ field theory on $\mathbb{P}^1 \times \mathbb{P}^1 \times S^1$.
With no effort, we can  generalize all results to the non-toric manifold $\Sigma_{\fg_1}\times \Sigma_{\fg_2}\times S^1$,
where $\Sigma_{\fg_1}$ and $\Sigma_{\fg_2}$ are two Riemann surfaces of genus $\fg_1$ and $\fg_2$, respectively.
We denote by $\fs_\varsigma$ and $\ft_\varsigma$ $(\varsigma = 1, 2)$ the background magnetic fluxes on $\Sigma_{\fg_1}$ and $\Sigma_{\fg_2}$.
We will be able to define an effective twisted superpotential $\wt \cW$ for the three-dimensional theory that we obtain by compactifying the five-dimensional $\cN=1$ theory on $\Sigma_{\fg_2}$.
We refer for details to section \ref{sec:largeN}. We will find that the value of $\wt \cW$, evaluated at the combined critical points of $\cF$ and $\wt \cW$, as a function of the chemical potentials $\Delta_\varsigma$ and fluxes $\ft_\varsigma$, satisfies
\be
 \label{index:theorem:W:5D}
 \begin{aligned}
  \wt\cW ( \ft_\varsigma , \Delta_\varsigma)\propto \sum_{\varsigma = 1}^{2}  \ft_\varsigma \frac{\partial F_{S^5} (\Delta_\varsigma) }{\partial \Delta_\varsigma }  \propto \Bigg\{
   \begin{array}{lr}
    F_{\Sigma_{\fg_2} \times S^3} ( \ft_\varsigma , \Delta_\varsigma) \, , & \text{ for } ~~ \USp(2N) \\
    a ( \ft_\varsigma , \Delta_\varsigma) \, , & \text{ for } \, \cN = (2,0)
   \end{array}
   \, .
 \end{aligned}
\ee
Here $F_{\Sigma_{\fg_2} \times S^3}( \ft_\varsigma , \Delta_\varsigma)$ is the $S^3$ free energy of the three-dimensional $\cN = 2$ theory obtained by compactifying
the $\USp(2N)$ theory on $\Sigma_{\fg_2}$, recently computed holographically in \cite{Bah:2018lyv},
and $a( \ft_\varsigma , \Delta_\varsigma)$ is the conformal anomaly coefficient of the four-dimensional $\cN = 1$ theory obtained by compactifying the $\cN = (2,0)$ theory on $\Sigma_{\fg_2}$, computed in \cite{Bah:2011vv,Bah:2012dg}.
We verified the statement for the $\USp(2N)$ theory only upon extremization with respect to $\Delta_\varsigma$, but we expect it to be true for all values of the chemical potentials.\footnote{This was confirmed in \cite{Crichigno:2018adf} that appeared after the completion of this work.}

We shall also consider the large $N$ limit of the topologically twisted index itself. The matrix model is too hard to compute directly even in the large $N$ limit.
The main difficulty compared to the three- and four-dimensional cases is the quadratic dependence on the gauge and background fluxes that do not allow for a simple resummation in \eqref{intro:gluing}.
The case of $\mathbb{P}^1\times \mathbb{P}^1\times S^1$ is technically simpler, since there are two sets of gauge fluxes, one for each $\mathbb{P}^1$, but still too hard to attack directly. 
By resumming one set of gauge magnetic fluxes (call them $\fm$), we obtain a set of Bethe equations for the eigenvalues $a_i$ (these are just the Bethe vacua of the effective twisted superpotential $\wt \cW$ of the compactification on $\Sigma_{\fg_2}$).
The result still depends on the second set of gauge magnetic fluxes (call them $\fn$). We expect that, in the large $N$ limit, one single distribution of eigenvalues $a_i$ and one single set of fluxes $\fn_i$ dominate the partition function.
At this point we shall pose the conjecture that the extremization of $\cF(a)$ provides the missing condition for determining both $a_i$ and $\fn_i$. Under this conjecture we obtain
\be 
 \label{index:theorem:logZ:5D}
 \log Z_{\Sigma_{\fg_1} \times \Sigma_{\fg_2}\times S^1}\propto \sum_{\varsigma = 1}^{2}  \fs_\varsigma \frac{\partial \wt \cW (\ft_\varsigma , \Delta_\varsigma) }{\partial \Delta_\varsigma } \propto\sum_{\varsigma , \varrho = 1}^{2} \fs_\varsigma \ft_\varrho \frac{\partial^2 F_{S^5} (\Delta) }{\partial \Delta_\varsigma \partial \Delta_\varrho} \, ,
\ee
where we generalized the result to $\Sigma_{\fg_1} \times \Sigma_{\fg_2} \times S^1$.

It is remarkable that \eqref{index:theorem:logZ:5D}, which is based on a conjecture, is completely analogous to the three- and four-dimensional result \eqref{index:theorem:original:3d:4D}.
Even more remarkably, we can compare \eqref{index:theorem:logZ:5D} with the existing results for the twisted compactification of the 6D $\cN = (2,0)$ theory on $\Sigma_{\fg_1} \times \Sigma_{\fg_2}$ \cite{Benini:2013cda}.
\eqref{index:theorem:logZ:5D} is expected to compute the leading behaviour, in the limit $\tau \to \mathbbm{i} 0^+$, of the elliptic genus of the two-dimensional CFT  obtained by the twisted compactification.
We find that \eqref{index:theorem:logZ:5D} indeed leads to the correct Cardy behaviour
\be
 \label{N=1*:SYM:Cardy:intro}
 \log Z_{\Sigma_{\fg_1} \times \Sigma_{\fg_2}\times S^1} \approx \frac{\mathbbm{i} \pi}{12 \tau} c_r (\fs_\varsigma , \ft_\varsigma , \Delta_\varsigma) \, ,
\ee
where  $c_r (\fs_\varsigma , \ft_\varsigma , \Delta_\varsigma)$ precisely coincides with the trial central charge of the two-dimensional CFT computed in \cite{Benini:2013cda}.
Moreover, we will show in section \ref{sec:4D domain-walls} that \eqref{index:theorem:logZ:5D} is equivalent to the attractor mechanism for the corresponding black strings in AdS$_7$.
All this is in complete analogy with the four-dimensional results \eqref{Cardy:4d:intro} and \eqref{intro:conjecture:index:attractor}.
It would be very interesting to see if the conjectured result for the $\USp(2N)$ theory matches the entropy of magnetically charged AdS$_6 \times_w S^4$ black holes in massive type IIA supergravity, which are still to be found.
Work in this direction is in progress \cite{Hosseini:2018mIIA}.

\subsection{Overview}

The structure of this paper is as follows. In section \ref{sec:localization} we analyse the conditions of supersymmetry and the Lagrangian for a five-dimensional $\cN=1$ gauge theory on $\cM_4\times S^1$,
where $\cM_4$ is a toric manifold, in a $\Omega$-background for the torus action $(\mathbb{C}^*)^2$.
We determine the classical saddle points and compute explicitly the one-loop determinants.
We finally write an expression for the (equivariant) topologically twisted index as a gluing of various copies of the K-theoretic Nekrasov's partition function,
one for each fixed point of the toric action. We then study in detail the non-equivariant limit in the sector with no instantons.
We also write explicitly the SW prepotential $\cF(a)$ that will play an important role in the rest of the paper.

In section \ref{sec:largeN} we discuss the large $N$ limit of the topologically twisted index and of related quantities.
We first motivate the importance of finding the critical points of $\cF(a)$.
Then we consider the partition function on $\Sigma_{\fg_1} \times \Sigma_{\fg_2} \times S^1$ of two theories, $\cN = 2$ SYM, which decompactify to the $\cN = (2,0)$ theory in six dimensions
and the $\USp(2N)$ theory with $N_f$ flavors and an antisymmetric matter field, corresponding to a 5D UV fixed point.
Under some assumptions, we will derive \eqref{index:theorem:F:5D}, \eqref{index:theorem:W:5D} and \eqref{index:theorem:logZ:5D}.

Special attention will be devoted to the 6D $\cN = (2,0)$ theory where we can compare the results with the existing holographic literature about domain walls and black strings.
For this theory, in section \ref{sec:4D domain-walls} we will be able to interpret our results as the counterpart of the attractor mechanism in four-dimensional $\cN = 2$ gauged supergravity.

We conclude in section \ref{sec:discussion and outlook} with discussions and future problems to explore.
Some details regarding toric varieties, conventions, computation of anomaly coefficients in twisted compactifications, and polylogarithms are collected in four appendices.

\paragraph*{Note added:} while we were writing this work, we became aware of \cite{Crichigno:2018adf} which has some overlaps with the results presented here.

\section[Localization on \texorpdfstring{$\cM_4 \times S^1$}{M(4) x S**1}]{Localization on $\cM_4 \times S^1$}
\label{sec:localization}

In this section we evaluate the twisted indices of five-dimensional $\mathcal{N}\ge1$
theories, \ie\;the partition function on $\mathcal{M}_{4}\times S^{1}$,
using localization. We begin in section \ref{subsec:Geometry-of-M4} by describing
the geometry of the toric K\"ahler manifold $\mathcal{M}_{4}$. Although
the twisted theory is semi-topological, \ie\;does not depend on the
metric on $\mathcal{M}_{4}$, we find it useful to have a canonical
set of coordinates and a canonical metric. In section \ref{subsec:Supergravity-Background},
we describe the rigid supergravity background to which we couple the
theory in order to produce the twist and the $\Omega$-deformation.
We describe the relevant supersymmetry algebra and supersymmetric
actions in section \ref{subsec:Supersymmetry-algebra}.
The localization procedure is carried out in section \ref{subsec:Localization}. In section \ref{subsec:Nekrasov partition function} we present the relevant expression for the K-theoretic Nekrasov's partition function.
Finally, in section \ref{completePF} we present the complete partition function on $\mathcal{M}_{4}\times S^{1}$.

\subsection[Geometry of \texorpdfstring{$\cM_{4}$}{M(4)}]{Geometry of $\cM_{4}$}
\label{subsec:Geometry-of-M4}

We review the construction of a canonical invariant metric for a toric
K\"ahler manifold $\mathcal{M}_{4}$ in symplectic coordinates \cite{2000math......4122A,guillemin1994}.

A K\"ahler manifold $M^{2n}$ is a complex manifold of real dimension
$2n$ with an integrable almost complex structure 
\be
J^{2}=-\mathbbm{1}_{2n} \, .
\ee
It is also a symplectic manifold with symplectic form $\omega$,
satisfying a compatibility condition on the metric defined
by the bilinear form 
\be
g\equiv\omega\left(\cdot,J\cdot\right) \, ,
\ee
which states that $g$ is symmetric and positive definite. 

A toric K\"ahler manifold is a K\"ahler manifold with an effective (faithful),
Hamiltonian, and holomorphic action of a real $n$-torus $\mathbb{T}^{n}$.
Given a Hamiltonian action, there exists a vector field $\tilde{v}$
for each element of the Lie algebra of $\mathbb{T}^{n}$ and a smooth
function $\mu$, the moment map, such that 
\begin{equation}
\tilde{v}=\omega^{-1} \rd \mu \, .
\label{eq:moment-map-equation}
\end{equation}
The moment map should be thought of as an equivariant map from the
Lie algebra of $\mathbb{T}^{n}$ to the space of smooth functions
on $M^{2n}$. It is defined only up to the addition of a constant. 

The image of the moment map, the orbit space
\be
\Delta\equiv M^{2n}/\mathbb{T}^{n} \, ,
\ee
is a convex $n$-dimensional polytope called the moment polytope. It can be written as
\be
\Delta=\left\{ x\in\mathbb{R}^{n}|\left\langle x,u_{i}\right\rangle -\lambda_{i}\ge0,\; i\in\left\{ 1, \ldots, d\right\} \right\} ,
\ee
for an appropriate set of data 
\be
u_{i}\in\mathbb{Z}^{n} \, ,\qquad \lambda_{i}\ge0 \, .
\ee
Its vertices are located at the fixed points of the torus action and
$\Delta$ is the convex hull.  The moment polytope is related to the
combinatorial description of $M^{2n}$ as a toric variety with an
associated toric fan, dual to $\Delta$, which constructed out of the vectors $n_{j}$ (see appendix \ref{sec:toric geometry}).
It will be important in the following that the number of vertices $d$ of the polytope, or equivalently the number of vectors $n_{j}$ of the fan,  is equal to the 
number of fixed points of the toric action. It is also equal to the 
Euler characteristic of $M^{2n}$, $d=\chi(M^{2n})$.

One may describe all three structures appearing in the definition
of $M^{2n}$ explicitly using symplectic coordinates: $x^{i}$ for
$\Delta$ and $y^{i}$ for $\mathbb{T}^{n}$. Define the functions
\be
l_{r}\left(x\right)\equiv\left\langle x,u_{r}\right\rangle -\lambda \, ,
\ee
 and an auxiliary potential function 
\bea
p\left(x\right)\equiv g_{p}\left(x\right)+h\left(x\right) \, ,\qquad
g_{p}\equiv\frac{1}{2}\sum_{r=1}^{d}l_{r}\left(x\right)\log l_{r}\left(x\right) \, ,\qquad
G_{ij}\left(x\right)\equiv\partial_{x^{i}}\partial_{x^{j}}p\left(x\right) \, .
\eea
The function $h\left(x\right)$ must be such that there exists a smooth,
strictly positive function $\delta\left(x\right)$ satisfying 
\be
\frac{1}{\det G(x)} = \delta(x)\prod_{r=1}^{d}l_{r}(x) \, .
\ee
The complex structure, symplectic (K\"ahler) form, and $\mathbb{T}^{n}$
invariant K\"ahler metric are then given by
\bea
J=\begin{pmatrix}0 & -G^{-1} \\
G & 0
\end{pmatrix} \, ,\qquad
\omega=dx^{i}\wedge dy_{i} \, ,\qquad
g=G_{ij}dx^{i}dx^{j}+\left(G^{-1}\right)^{ij}dy_{i}dy_{j} \, .
\eea
Note that $\det g=1$.

All smooth symplectic toric manifolds are simply connected \cite{fulton1993introduction}.
Compact simply connected topological four-manifolds are mostly classified
by their intersection form. Note, in particular, that 
\be
b_{2}^{+}=1 \, ,
\ee
for any symplectic toric four-manifold. One can check with the metric
above that%
\footnote{The orientation for which this is true is such that $\varepsilon^{x_{1}y^{1}x_{2}y^{2}}=1$.}
\be
\star\omega=\omega \, .
\ee

\subsection{Nekrasov's conjecture}
\label{sec:Nekrasov's conjecture}

There is a standard way, reviewed in the next section, of putting
any four-dimensional $\mathcal{N}=2$ Lagrangian field theory on a smooth four-manifold
while preserving supersymmetry. This is done using the Witten twist
\cite{Witten:1988ze}. The resulting computations are insensitive
to, or at least piece-wise constant under, variations of the metric.
This is an example of a cohomological topological quantum field theory (TQFT),
usually called the Donaldson-Witten TQFT.
The relevant observables reside in the cohomology of the preserved supercharge. 

Nekrasov has introduced a generalization of this TQFT which is valid
when the four-manifold admits a metric with an isometry \cite{Nekrasov:2003vi}.%
\footnote{In this section we restrict ourselves to describing the $\U(N)$ theory.}
The toric manifolds described in the previous section are prime examples of this construction.
The construction can be seen as a generalization of the computation
of the equivariant partition function for theories on $\bR^4$,
that can be used to recover the exact effective prepotential \cite{Nekrasov:2002qd,Nekrasov:2003rj}.
The latter can be defined as \cite{Nekrasov:2002qd}
\be
 \mathcal{F}_{0}(\Lambda , a) \equiv \lim_{\epsilon_{1,2} \to 0} \epsilon_{1} \epsilon_{2} \log Z_{\text{Nekrasov}}^{\bC^2} ( q, a, \epsilon_{1} , \epsilon_{2} ) \, , \qquad q \to \Lambda^{2h^{\lor}(G) - k (\fR)} \, ,
\ee
where $Z_{\text{Nekrasov}}^{\bC^2}$ is the so-called Nekrasov's partition function,
coinciding with the partition function on $\mathbb{R}^{4}$ in
the presence of the $\Omega$ deformation with parameter $\vec\epsilon = ( \epsilon_{1},\epsilon_{2})$.
$\Lambda$ is the dynamically generated scale, and $a$ represents
the vacuum expectation value for the scalar field in the vector multiplet at a specific point on the Coulomb branch.
Moreover, $h^{\lor}(G)$ is the dual Coxeter number of the gauge group $G$ and $k (\fR)$ is the quadratic Casimir normalized such that it is $2h^{\lor}(G)$ for the adjoint representation.

It has been argued in \cite{Nekrasov:2003vi} that the analogous partition function on a compact toric manifold $\mathcal{M}_{4}$ takes the form
\be
 \label{Nekrasov}
 Z_{\mathcal{M}_{4}}=\sum_{\mathfrak{p}_{l} \in \mathbb{Z}^{N}} \oint_\cC \rd a \prod_{l = 1}^{\chi (\cM_4)} Z_{\text{Nekrasov}}^{\bC^2} \big(a + \epsilon_{1}^{(l)} \mathfrak{p}_{l} + \epsilon_{2}^{(l)} \mathfrak{p}_{l+1}, \epsilon^{(l)}_{1} , \epsilon^{(l)}_{2} ; q \big) .
\ee
In the equation above we have chosen to disregard insertions of operators
and the dependence on characteristic classes for non-simply connected
gauge groups, both of which are not relevant for our purposes. The
main new ingredient in this formula, in comparison to the formula
on $\mathbb{R}^{4}$, is the appearance of a sum over a set of fluxes
$\mathfrak{p}_l$. These are associated with equivariant divisors
on $\mathcal{M}_{4}$, and thus with vectors in the toric fan. The
deformation parameters $\epsilon^{(l)}_{1}$, $\epsilon^{(l)}_{2}$ are also given by the data
in the fan. Note that the modulus $a$ is now integrated over,
as should be the case on a compact space. The result presented in
\cite{Nekrasov:2003vi} is a conjecture. Specifically, the exact form
of the sum over the integers $\mathfrak{p}_{l}$ and the contour
for the integral over the modulus $a$ are not known. 

It is expected that the results for the Donaldson-Witten theory are recovered in the non-equivariant limit, $\epsilon_{1,2} \to 0$.
Nekrasov conjectured that this limit is given by%
\footnote{We correct a misprint in \cite{Nekrasov:2003vi} here.}
\bea
Z_{\mathcal{M}_{4}} = \sum_{\fk^{(i)} \in \mathbb{Z}^{N} } \oint_\cC \rd a \exp & \Bigg[\int_{\mathcal{M}_{4}}\mathcal{F}_{0} \bigg(a+\sum_{i} \fk^{(i)} c_{1} (L_{i}) \bigg)
 +c_{1}(\mathcal{M}_{4}) \cH_{\frac{1}{2}} \bigg (a+\sum_{i} \fk^{(i)} c_{1}(L_{i}) \bigg)\\
 & + \chi(\mathcal{M}_{4}) \mathcal{F}_{1}(a) + \left( 3 \sigma(\mathcal{M}_{4}) + 2 \chi(\mathcal{M}_{4}) \right) \mathcal{G}_{1}(a)\Bigg].
\eea
The $ L_{i}$ are line bundles supported on two-cycles of $\mathcal{M}_{4}$,
which do not have a flat space analogue, and $ c_{1} (L_{i})$ their first Chern class. $\chi(\mathcal{M}_{4})$ is the Euler characteristic of $\mathcal{M}_{4}$ and $\sigma(\mathcal{M}_{4})$ its signature.
The additional terms in the exponential, relative to the usual effective action on $\mathbb{R}^{4}$,
come from subleading terms in $Z_{\text{Nekrasov}}^{\bC^2}$:
\be
\log Z_{\text{Nekrasov}}^{\bC^2} \left(a,\epsilon_{1},\epsilon_{2};q\right)=\frac{1}{\epsilon_{1}\epsilon_{2}}\mathcal{F}_{0}+\frac{\epsilon_{1}+\epsilon_{2}}{\epsilon_{1}\epsilon_{2}}\mathcal{H}_{\frac{1}{2}}+\mathcal{F}_{1}+\frac{\left(\epsilon_{1}+\epsilon_{2}\right)^{2}}{\epsilon_{1}\epsilon_{2}}\mathcal{G}_{1}+\ldots \, .
\ee

The authors of \cite{Bawane:2014uka,Bershtein:2015xfa,Bershtein:2016mxz}
have began to verify \eqref{Nekrasov} using localization. Our twisted
indices are a generalization of the partition functions on $\mathcal{M}_{4}$,
and we will follow closely the arguments used in these papers.
We will not have anything to add regarding the part of the calculation
involving the sum over fluxes. However, we will comment on the similarities 
between the present setup and the calculation using localization of the 
twisted indices in three and four dimensions, in which a similar contour integral arises and is given 
by an explicit prescription.

\subsection[Supersymmetry on \texorpdfstring{$\cM_4 \times S^1$}{M(4) x S**1}]{Supersymmetry on $\cM_4 \times S^1$}
\label{subsec:Supergravity-Background}

Supersymmetric theories can sometimes be coupled to a curved background
while preserving some supersymmetry. This was originally achieved
by twisting the theory --- identifying a new euclidean rotation group
with a diagonal subgroup of rotations and R-symmetry transformations.
A subset of the supercharges become scalars under the new rotation
group, and are conserved on an arbitrary curved manifold, as long
as the coupling to the metric is implemented using this new group.
As a bonus, the energy momentum tensor turns out to be the supersymmetry
variation of a scalar supercharge $Q$. The twisted theory, where
observables are restricted to be $Q$-closed operators, then becomes
a TQFT of cohomological type \cite{Witten:1988ze}. 

A more general procedure for preserving supersymmetry, initiated in
\cite{Festuccia:2011ws} and continued for four dimensions in \cite{Dumitrescu:2012ha,Klare:2012gn}, is to couple the theory to rigid supergravity
and to search for backgrounds which are fixed points of the supersymmetry
transformations. Technically, this means choosing a configuration
for the bosonic fields in the supergravity multiplet such that the
supersymmetry variation of the gravitino vanishes for some spinor.
The vanishing of the gravitino variation yields a linear differential
equation known as a generalized Killing spinor equation whose solutions
are known as generalized Killing spinors. A variation using these
spinors constitutes a rigid supersymmetry.

One can expand the scope of this construction by considering superconformal
tensor calculus instead of a specific Poincare supergravity \cite{Klare:2013dka,Klare:2012gn}. In superconformal
tensor calculus, the gravitino is part of the Weyl multiplet which
includes another fermion, the dilatino, whose supersymmetry variation
must also vanish. The resulting solutions are generalized Killing
spinors which generate an action of a subalgebra of the superconformal
algebra on the dynamical fields. In order to use this algebra to localize,
one should avoid including transformations which are not true symmetries
of the theory such as dilatations. To the best of our knowledge, this
is the most general context in which this program of preserving rigid
supersymmetry on curved backgrounds has been pursued. 

A five-dimensional $\mathcal{N}=1$ theory with R-symmetry group $\SU(2)$
can be formulated while preserving supersymmetry on any five-manifold
as long as the holonomy group is contained in $\SO(4)$.
The necessary supergravity background is simply a twist. We derive
the rigid 5D $\mathcal{N}=1$ supergravity background corresponding
to the $\Omega$-background on a manifold with topology $\mathcal{M}_{4}\times S^{1}$,
where $\mathcal{M}_{4}$ is a toric K\"ahler four-manifold. Supersymmetry
is preserved using a five-dimensional uplift of the Witten twist on
$\mathcal{M}_{4}$ augmented to include the $\Omega$-deformation.
The rigid supergravity background for a twisted four-dimensional $\mathcal{N}=2$
theory, with the background corresponding to the $\Omega$-deformation,
was explicitly constructed for any toric K\"ahler manifold in \cite{Rodriguez-Gomez:2014eza}.
These backgrounds can be lifted to the 5D $\mathcal{N}=1$ theory
on $\mathcal{M}_{4}\times S^{1}$ in a straightforward manner, implicitly
described in \cite{Alday:2015lta,Pini:2015xha}. We review this below. Our spinor
and metric conventions are spelled out in appendix \ref{sec: spinor conventions}.

We consider $X\equiv\mathcal{M}_{4}\times S^{1}$ and choose coordinates
such that the $S^{1}$ is parameterized by $x_{5}\in\left[0,\beta\right)$.
The construction of a $T^{2}$ invariant K\"ahler metric $g$ for $\mathcal{M}_{4}$ was reviewed in section \textcolor{red}{\ref{subsec:Geometry-of-M4}}.
Let us define
\be
\tilde{v}=\epsilon_{i}\partial_{y^{i}} \, , \qquad
x_{2}\equiv y_{1} \, ,\qquad x_{4}=y_{2} \, ,
\ee
and let 
\be
e_{a}^{\phantom{a}m} \, , \qquad a\in\{ 1,2,3,4\} \, , \qquad m\in\{ 1,2,3,4\} \, ,
\ee
be a vielbein for $g$. We define the metric on $X$ by augmenting $e_{a}^{\phantom{a}m}$ with 
\be
e_{5}^{\phantom{5}m}=\tilde{v}^{m} \, , \qquad
e_{5}^{\phantom{5}5}=1 \, .
\ee
The associated spin connection still has $\U(2)$ holonomy.

The Weyl multiplet of five-dimensional superconformal tensor calculus
is described, for instance, in \cite{deWit:2017cle}. Along with the
vielbein, it contains the following independent bosonic fields: an
$\SU(2)$ R-symmetry gauge field which we denote $A_{m}^{(\text{R})}$,
and an anti-symmetric tensor $T_{mn}$, a vector $b_{m}$, and
a scalar $D$. The remaining bosonic fields are determined in terms
of these, and of the fermions, by constraints. We will turn off $T_{mn}$
and $b_{m}$. After some renaming, the variation of the gravitino
in the remaining background can be written as 
\be
\delta\psi_{I m}=D_{m}\xi_{I}-\Gamma_{m}\eta_{I} \, ,
\ee
where%
\footnote{Throughout this paper, $D_m$ will denote a generic covariant derivative. The covariance is with respect to the spin, R-symmetry, gauge, and background flavor symmetry connections.
We will specify the concrete form of the derivative when appropriate.}
\be
D_{m}\xi_{I}\equiv\partial_{m}\xi_{I}+\frac{1}{4} \omega_{m}^{\phantom{m}ab}\Gamma_{ab}\xi_{I} + \left(A_{m}^{(\text{R})}\right)_{I}^{\phantom{I}J}\xi_{J} \, .
\ee

We perform the twist by setting 
\be
A_{m}^{(\text{R})}=\frac{1}{4} \omega_{m}^{\phantom{m}ab} \sigma_{ab} \, .
\ee
One can easily check that the spinor 
\be
\xi= - \frac{1}{\sqrt{2}}\begin{pmatrix}\tau^2\\
0
\end{pmatrix} \, , \qquad\eta=0 \, ,
\ee
is a solution. Note that the components of both $\omega$ and $A^{(\text{R})}$
in the $x_{5}$ direction vanish in the non-equivariant limit, $\epsilon_{1,2} \to 0$.
One may verify explicitly that $\xi$ satisfies the dilatino equation with an appropriate value of $D$.  

\subsection{Supersymmetry transformations and Lagrangian}
\label{subsec:Supersymmetry-algebra}

We record the supersymmetry transformations for the vector and hypermultiplets,
following the conventions of \cite{Hosomichi:2012ek,Kallen:2012va}.
For the purposes of localization it is simpler to use twisted fields
defined using the Killing spinor $\xi$. Note that $\xi$ satisfies
\bea
& \xi_I\xi^I = 1 \, , \qquad 
v^{m}v_{m}=1 \, , \qquad
v^{m}\partial_{m}\equiv\xi_I\Gamma^{m}\xi^I\partial_{m}=\epsilon_{1}\partial_{3}+\epsilon_{2}\partial_{4}+\partial_{5} \, .
\eea
For consistency of notation with \cite{Qiu:2016dyj}, we define 
\be
 \label{not:a:contact:form}
 \kappa_{m}\equiv g_{mn}v^{n} \, .
\ee
Note that $\kappa=\rd x^5$ is \emph{not} a contact form for $\mathcal{M}_{4}\times S^{1}$.

\subsubsection{Vector multiplet}

The five-dimensional $\mathcal{N}=1$ vector multiplet has $8+8$ off-shell components.
It comprises a connection $A_{m}$, a real scalar $\sigma$, an $\SU(2)$
Majorana spinor $\lambda^{\alpha}_{\phantom{\alpha}I}$, and a triplet of auxiliary
fields $D_{IJ}$ satisfying the reality condition 
\be
\left(D^{*}\right)^{IJ}=\varepsilon^{IK}\varepsilon^{JL}D_{KL} \, ,
\ee
all in the adjoint representation of the Lie algebra $\mathfrak{g}$.
We use the physics convention where all gauge fields are hermitian. We define the gauge covariant derivative acting on fields in the adjoint representation and the field strength as
\be
\label{gauge:covariant:derivative}
D_{m}\equiv\partial_{m}-\mathbbm{i}[A_{m},\cdot] \, , \qquad F_{mn}\equiv\partial_{m}A_{n}-\partial_{n}A_{m}-\mathbbm{i}[A_{m},A_{n}] \, .
\ee
A gauge transformation with parameter $\alpha$ reads
\be
G_{\alpha} = \mathbbm{i}[\alpha, \cdot ] \, , \qquad G_{\alpha}A_{m}=D_{m}\alpha \, .
\ee
In order to ensure convergence of the actions in section \ref{subsec:Supersymmetric-Lagrangians},
we will preemptively rotate both $D_{IJ}$ and $\sigma$ into the imaginary plane 
\be
D_{IJ}\rightarrow\mathbbm{i}D_{IJ} \, ,\qquad
\sigma\rightarrow-\mathbbm{i}\sigma \, .
\ee
The new reality conditions are such that 
\be
\left(D^{*}\right)^{IJ}=-\varepsilon^{IK}\varepsilon^{JL}D_{KL} \, .
\ee

The supersymmetry transformations read \cite{Hosomichi:2012ek}
\bea
& \delta A_{m}=\mathbbm{i}\xi_{I}\Gamma_{m}\lambda^{I} \, , \qquad
\delta\sigma=-\xi_{I}\lambda^{I} \, , \\
& \delta\lambda_{I}=-\frac{1}{2}\Gamma^{mn}F_{mn}\xi_{I}-\mathbbm{i}D_{m}\sigma\Gamma^{m}\xi_{I}-\mathbbm{i}D_{IJ}\xi^{J} - 2 \mathbbm{i} \tilde{\xi}_{I}\sigma \, , \\
& \delta D_{IJ}=-\xi_{I}\Gamma^{m}D_{m}\lambda_{J}-\xi_{J}\Gamma^{m}D_{m}\lambda_{I}-\left[\sigma,\xi_{I}\lambda_{J}+\xi_{J}\lambda_{I}\right]+\tilde{\xi}_{I}\lambda_{J}+\tilde{\xi}_{J}\lambda_{I} \, .
\eea
The spinor $\tilde{\xi}$ is defined as 
\be
\tilde{\xi}\equiv\frac{1}{5}\Gamma^{m}D_{m}\xi \, ,
\ee
and therefore vanishes in the present context.

Following \cite{Kallen:2012cs}, we define the twisted fields%
\footnote{The orientation here is the opposite of that used in \cite{Kallen:2012cs}, and corresponds with the one used in \cite{Qiu:2016dyj}.
Due to this choice, some forms which were anti-self-dual in \cite{Kallen:2012cs} are now self-dual.}
\bea
& \Psi_{m}\equiv\xi_{I}\Gamma_{m}\lambda^{I} \, , \qquad
H_{mn}\equiv2F_{mn}^{+}+\mathbbm{i}\xi^{I}\Gamma_{mn}\xi^{J}D_{IJ} \, , \\
& \chi_{mn}\equiv\xi_{I}\Gamma_{mn}\lambda^{I}+v_{n}\xi_{I}\Gamma_{m}\lambda^{I}-v_{m}\xi_{I}\Gamma_{n}\lambda^{I} \, ,
\eea
where
\be
F^{+}\equiv\frac{1}{2}\left(1+i_{v}\star\right)\left(1-\kappa\wedge i_{v}\right)F \, .
\ee
The two projection operators appearing in the definition of $F^{+}$
split the two-forms on $\mathcal{M}_{4}\times S^{1}$ into vertical
and horizontal forms. The latter are further split into self-dual
and anti-self-dual forms on $\mathcal{M}_{4}$:
\bea
& F=F_{H}+F_{V}=\left(1-\kappa\wedge i_{v}\right)F+\left(\kappa\wedge i_{v}\right)F \, , \\
& F_{H}=F_{H}^{+}+F_{H}^{-}=\frac{1}{2}\left(1+i_{v}\star\right)F_{H}+\frac{1}{2}\left(1-i_{v}\star\right)F_{H} \, .
\eea
The supersymmetry algebra now takes the standard cohomological form,
up to the addition of the equivariant deformation
\bea
\label{eq:vector_multiplet_susy_transformations}
& \delta A_{m}=\mathbbm{i}\Psi_{m} \, , \qquad
\delta\sigma=-v^{m}\Psi_{m} \, , \qquad
\delta\Psi_{m}=i_{v}F-\mathbbm{i}D_{m}\sigma \, , \\
& \delta\chi_{mn}=H_{mn} \, , \qquad
\delta H_{mn}=\mathbbm{i}\mathcal{L}_{v}^{A}\chi_{mn}+\mathbbm{i}\left[\sigma,\chi_{mn}\right] \, .
\eea
The square of the transformation $\delta$ contains a translation
and a gauge transformation 
\be
\delta^{2}=\mathbbm{i}\mathcal{L}_{v}+G_{\Phi} \, ,
\ee
where $\mathcal{L}$ is the Lie derivative on forms and 
\be
\Phi\equiv\sigma-\mathbbm{i}v^{m}A_{m} \, , \qquad
\mathcal{L}_{v}^{A}\equiv\mathcal{L}_{v}-\mathbbm{i}\left[v^{m}A_{m},\cdot\right].
\ee
Note that $\delta\Phi=0$.

\subsubsection{Hypermultiplet}

A hypermultiplet comprises a pair of complex scalars $q_{I}^{A}$
and a fermion $\psi^{A}$ satisfying 
\begin{equation}
\left(q_{I}^{A}\right)^{*}=\Omega_{AB}\varepsilon^{IJ}q_{I}^{A} \, ,
\qquad \left(\psi^{A}\right)^{*}=\Omega_{AB}C\psi^{B} \, ,
\label{eq:hypermultiplet_reality_condition}
\end{equation}
where 
\be
\Omega_{AB}=\begin{pmatrix}0 & \mathbbm{1}_{N}\\
-\mathbbm{1}_{N} & 0
\end{pmatrix},
\ee
is the invariant tensor of $\USp(2N)$, which is a symmetry group of $N$ free hypermultiplets.
$A,B,C,\ldots$ indices are raised and lowered using $\Omega$.
The gauge group is a subgroup of $\USp(2N)$ whose indices we sometimes suppress. 

The supersymmetry transformations read
\be
\delta q_{I}=-2\mathbbm{i}\xi_{I}\psi \, , \qquad
\delta\psi=\Gamma^{m}\xi_{I}D_{m}q^{I}+\mathbbm{i}\xi_{I}\sigma q^{I} \, ,
\ee
where $D_{m}$ is covariant with respect to $A_{m}$ and $\SU(2)_{R}$,
and both $A_{m}$ and $\sigma$ act in the appropriate representation
\be
(\sigma q)^{A}\equiv\sigma^{A}_{\phantom{A}B}q^{B} \, .
\ee
After twisting, the field $q_{I}^{A}$ becomes a spinor
\be
q\equiv\xi_{I}q^{I} \, .
\ee
This spinor is actually pseudo-real, and contains only $4$ degrees
of freedom
\be
\left(q^{A}\right)^{*}=\Omega_{AB}Cq^{B} \, .
\ee
Its variation includes only the part of $\psi$ given by the projection%
\be
\psi_{+} \equiv\frac{1}{2}\left(\mathbbm{1}_{4}+v_{m}\Gamma^{m}\right) \psi
=\frac{1}{2}\left(\mathbbm{1}_{4}+\Gamma^{5}\right)\psi \, .
\ee
The supersymmetry transformations can be closed off-shell by introducing
a superpartner $F$ for the component\footnote{See \cite[sect.\,4.2]{Qiu:2013pta} for a more complete explanation.} 
\be
\psi_{-}=\frac{1}{2}\left(\mathbbm{1}_{4}-\Gamma^{5}\right)\psi \, .
\ee
The twisted supersymmetry transformations are then given by 
\bea
\label{eq:hypermultiplet_susy_transformations}
& \delta q=\mathbbm{i}\psi_{+} \, , \qquad
\delta\psi_{+}=\left(\mathcal{L}_{v}-\mathbbm{i}G_{\Phi}\right)q \, , \\
& \delta\psi_{-}=F \, , \qquad
\delta F= \left( \mathbbm{i} \mathcal{L}_{v}+G_{\Phi} \right) \psi_{-} \, .
\eea

\subsubsection[Supersymmetric actions on \texorpdfstring{$\cM_4 \times S^1$}{M(4) x S**1}]{Supersymmetric actions on $\cM_4 \times S^1$}
\label{subsec:Supersymmetric-Lagrangians}

The action for twisted theories is a covariantized version of the
flat space action. This is in contrast to the additional terms which
appear, for instance, in the superconformal index, on the five sphere,
and on a general contact manifold. Such actions are still supersymmetric
because the Killing spinor is covariantly constant. 

The flat space Yang-Mills term is given by \cite{Hosomichi:2012ek}
\begin{equation}
S_{\text{YM}}^{\mathbb{R}^5}=\frac{1}{g_{\text{YM}}^{2}}\int \text{Tr}\left(\frac{1}{2}F^{mn}F_{mn} - D^{m}\sigma D_{m}\sigma
-\frac{1}{2}D^{IJ}D_{IJ}+\mathbbm{i}\lambda_{I}\Gamma^{m}D_{m}\lambda^{I}-\lambda_{I}\left[\sigma,\lambda^{I}\right] \right) ,
\label{eq:Yang_Mills_action}
\end{equation}
where $D_{m}$ is the covariant derivative with respect to the connection $A_{m}$, see \eqref{gauge:covariant:derivative}.
The action on $\mathcal{M}_{4}\times S^{1}$ can be written as
\be
S_{\text{YM}}^{\mathcal{M}_{4}\times S^{1}}=\frac{1}{g_{\text{YM}}^{2}}\int\sqrt{g}\text{Tr}\left(\frac{1}{2}F^{mn}F_{mn}-D^{m}\sigma D_{m}\sigma
-\frac{1}{2}D^{IJ}D_{IJ}+\mathbbm{i}\lambda_{I}\Gamma^{m}D_{m}\lambda^{I}-\lambda_{I}\left[\sigma,\lambda^{I}\right]\right) ,
\ee
where $D_{m}$ is covariant with respect to $A_{m}$, the
spin connection and the $\SU(2)$ R-symmetry. 

In order to evaluate the Euclidean path integral with action 
\be
\exp\left(-S_{\text{YM}}^{\mathcal{M}_{4}\times S^{1}}\right),
\ee
we must choose a contour for the bosonic fields. An appropriate contour
which ensures convergence of the integral reads
\be
A_{m}^{\dagger}=A_{m} \, , \qquad
\sigma^{\dagger}=-\sigma \, , \qquad
\left(D^{*}\right)^{IJ}=-\varepsilon^{IK}\varepsilon^{JL}D_{KL} \, .
\ee
Integration of fermionic fields is an algebraic procedure and does
not require such a choice of contour. Note the change in reality conditions
for the auxiliary field $D_{IJ}$. In the rotated variables, appearing
in the supersymmetry transformations and in the rest of the paper,
\be
S_{\text{YM}}^{\mathcal{M}_{4}\times S^{1}}=\frac{1}{g_{\text{YM}}^{2}}\int\sqrt{g}\text{Tr}\left(\frac{1}{2}F^{mn}F_{mn}+D^{m}\sigma D_{m}\sigma+\frac{1}{2}D^{IJ}D_{IJ}+\mathbbm{i}\lambda_{I}\Gamma^{m}D_{m}\lambda^{I}+\mathbbm{i}\lambda_{I}\left[\sigma,\lambda^{I}\right]\right) .
\label{eq:twisted_Yang_Mills_action}
\ee

The action for a hypermultiplet is similarly given by
\bea
S_{\text{\ensuremath{\fR}-hyper}}^{\mathcal{M}_{4}\times S^{1}} = \int \sqrt{g} \Big( & D^{m}q_{I}^{A}D_{m}q_{A}^{I}+q_{I}^{A}\sigma_{AB}\sigma^{BC}q_{C}^{I}
-2\mathbbm{i}\psi^{A}\Gamma^{m}D_{m}\psi_{A} \\
& +2\mathbbm{i}\psi^{A}\sigma_{AB}\psi^{B}-4\psi^{A}\lambda_{ABI}q^{BI}+q_{I}^{A}D_{AB}^{IJ}q^{BI} \Big) ,
\label{eq:twisted_hypermultiplet_action}
\eea
where the matrices $\sigma^{A}_{\phantom{A}B},\lambda_{I\phantom{A}B}^{\phantom{I}A}$, and
${D^{A}}_{B}$ act in the representation $\fR$.
This action is convergent with the contour implied by the reality condition \eqref{eq:hypermultiplet_reality_condition}.

\subsection{Localization onto the fixed points}
\label{subsec:Localization}

The actions in the previous section are invariant under the fermionic
transformation $\delta$. By a standard argument, expectation values
of $\delta$-closed observables, and in particular the partition function,
are invariant under $\delta$-exact deformations of the action
\be
S_{\text{total}}=S+t \delta\mathcal{V} \, ,
\ee
provided we choose the fermionic functional $\mathcal{V}$ in such
a way that $\delta^{2}\mathcal{V}|_{\text{bosonic}}=0$, $\delta\mathcal{V}|_{\text{bosonic}}\ge0$, and
all configurations which yield a finite result when evaluated using
$S_{\text{total}}$ also yield a finite result when evaluated using $S$.
In order to localize the theory with Euclidean measure 
\be
\exp\left(-S_{\text{total}}\right),
\ee
we take the limit $t \to \infty$.
All configurations with $\delta\mathcal{V}|_{\text{bosonic}}\ne0$ have infinite action
in this limit and the theory localizes onto the moduli space $\delta\mathcal{V}|_{\text{bosonic}}=0$.
The semi-classical approximation around this moduli space yields the
\emph{exact} result for the functional integral.

In order to localize the five-dimensional $\cN = 1$ twisted theories, we can add the following localizing terms
\bea
\label{eq:localizing_terms}
& \delta\mathcal{V}_{\text{gauge}}\equiv\delta\int\text{Tr}\left(2\mathbbm{i}\chi\wedge\star F^{+}+\frac{1}{2}\Psi\wedge\star\left(\delta\Psi\right)^{*}\right) \, , \\
& \delta\mathcal{V}_{\text{matter}}\equiv\delta\int\sqrt{g}\left(\psi_{+}^{A}\left(\delta\psi_{+}\right)_{A}^{*}+\psi_{-}^{A}\left(\delta\psi_{-}\right)_{A}^{*}+\psi_{-}^{A}\Gamma^{m}D_{m}q_{A}\right) . 
\eea
The bosonic parts of which are 
\be
\begin{aligned}
& \delta\mathcal{V}_{\text{gauge}}|_{\text{bosonic}}=\int\text{Tr}\left(2\mathbbm{i}H\wedge\star F^{+}+\frac{1}{2}\left(i_{v}F-\mathbbm{i}D\sigma\right)\wedge\star\left(i_{v}F-\mathbbm{i}D\sigma\right)^{*}\right) , \\
& \delta\mathcal{V}_{\text{matter}}|_{\text{bosonic}}=\int\sqrt{g}\left(\left[\left(\mathcal{L}_{v}-\mathbbm{i}G_{\Phi}\right)q\right]^{A}\left(\mathcal{L}_{v}+\mathbbm{i}G_{\Phi^{*}}\right)q_{A}+F^{A}F_{A}+F^{A}\Gamma^{m}D_{m}q_{A}\right) .
\end{aligned}
\ee
The field $H$ acts as a Lagrange multiplier, setting
\be
F^{+}=0 \, .
\ee
The rest of the condition $\delta\mathcal{V}_{\text{gauge}}|_{\text{bosonic}}=0$, then requires 
\begin{equation}
F^{+}=0 \, ,\qquad i_{v}F=0 \, ,\qquad D\sigma=0 \, .
\label{eq:vector_multiplet_localization_equations}
\end{equation}

A similar procedure for the hypermultiplet localizing term yields
\begin{equation}
\mathcal{L}_{v}q=0 \, ,\qquad G_{\Phi}q=0 \, ,\qquad\Gamma^{i}D_{i}q=0 \, ,
\label{eq:hypermultiplet_localization_equations}
\end{equation}
where in the last term we have made explicit use of the fact that
\be
F\Gamma^{5}\nabla_{5}q=0 \, ,
\ee
by summing $i \in \{1, \ldots , 4\}$.
These equations may admit solutions for certain representations of
the gauge group, which would indicate that there are moduli coming
from the hypermultiplets. However, we will consider the situation
in which the hypermultiplets are also coupled to background vector
multiplets frozen to supersymmetric configurations which effectively
give all hypermultiplets a generic mass. In this situation, there
are no solutions to \eqref{eq:hypermultiplet_localization_equations}.
In what follow, we consider only solutions of the vector multiplet
equations.

\subsubsection{Bulk solutions}

An obvious set of solutions to \eqref{eq:vector_multiplet_localization_equations}
is given by flat connections and covariantly constant $\sigma$.
The topology of our spacetime satisfies 
\be
\pi_{1}(\mathcal{M}_{4}\times S^{1})\simeq\mathbb{Z} \, .
\ee
Flat connections are therefore parameterized by the holonomies around
the $S^{1}$ factor, restricted only by large gauge transformations.
Using an appropriate gauge transformation, these can be brought to
the form of a constant Cartan subalgebra valued connection $A^{(0)}$:
\be
A_{i}^{(0)}=0 \, ,\qquad i\in\{ 1,2,3,4\} \, , \qquad A_{5}^{(0)} \in \text{Cartan} (\fg) \, .
\ee
Large gauge transformations identify 
\be
\big(A_{5}^{(0)}\big)_{j}\sim\big(A_{5}^{(0)}\big)_{j}+\frac{2\pi}{\beta}n \, , \qquad n\in\mathbb{Z} \, ,
\ee
where $j$ is an index in the Cartan subalgebra. At generic values
of the holonomy, a covariantly constant scalar is then also constant
and Cartan valued, \ie\;
\be
\partial_{m}\sigma^{(0)}=0 \, , \qquad \left[\sigma^{(0)},A_{5}\right]=0 \, .
\ee
We denote 
\be
a\equiv-\mathbbm{i}\Phi^{(0)}=-A_{5}^{(0)}-\mathbbm{i}\sigma^{(0)} \, , \qquad
a\sim a+\frac{2\pi}{\beta}\mathbb{Z} \, .
\ee
Note that the equivariant action acts as 
\be
\delta^{2}=\mathbbm{i} (\mathcal{L}_{v} + G_{a}) \, .
\ee

In principle, the localization calculation includes an integral over
the $\text{rank }\mathfrak{g}$ cylinders parameterized by $a$.
Later on we will find it more convenient to move to exponentiated
coordinates on this space, whereby the integration region becomes $\left(\mathbb{C}^{*}\right)^{\text{rk}(\mathfrak{g})}$.

\subsubsection{Fermionic zero modes}

The quadratic approximation of $\delta\mathcal{V}_{\text{gauge}}$
around a bulk configuration specified by $a$ allows fermionic zero
modes for both $\chi$ and $\Psi$. In the presence of such zero modes
the functional integral naively vanishes. However, following \cite{Bershtein:2016mxz,Bershtein:2015xfa,Bawane:2014uka},
we will take this as an indication that the localizing term needs
to be improved to include a fermion mass term which will soak up the
zero modes. Since the additional term is by definition $\delta$-exact,
the value of the coefficient with which it is added, as long as it
is nonzero, does not change the final result. 

The $\Psi$ zero mode can be read off from the $\Psi$ kinetic term
which is proportional to 
\be
\int\Tr \left(\Psi\wedge\star\left(\mathcal{L}_{v}-G_{a^{*}}\right)\Psi\right) \, .
\ee
The zero mode is a constant profile for the Cartan part of $\Psi$
given by
\be
\Psi_{m}^{(0)}\propto v_{m}\propto\Psi_{5}^{(0)} \, .
\ee
It is the superpartner of the holonomy.

The zero mode for $\chi$ is also Cartan valued and can be identified
using the projection operator
\be
\pi: \mathcal{M}_{4}\times S^{1}\rightarrow\mathcal{M}_{4} \, ,
\ee
with a multiple of the pullback of the K\"ahler form on $\mathcal{M}_{4}$:
\be
\chi^{(0)}\propto\pi^{*}\omega \, .
\ee
Indeed, one can check that 
\be
\mathcal{L}_{v}\chi^{(0)}=G_{a}\chi^{(0)}=0 \, .
\ee

We can construct a nowhere vanishing off-diagonal mass term by pairing
the two sets of zero modes using 
\be
\mathcal{V}^{(0)} \equiv \int \Tr \left( \sigma^{(0)}\wedge\star\chi^{(0)}\wedge\pi^{*}\omega \right) \, ,
\ee
such that 
\begin{equation}
\delta\mathcal{V}^{(0)} = \int \Tr \left(  i_{v}\Psi^{(0)}\wedge\star\chi^{(0)}\wedge\pi^{*}\omega+\sigma^{(0)}\wedge\star H^{(0)}\wedge\pi^{*}\omega \right) \, ,
\label{eq:fermion_zero_modes_localizing_term}
\end{equation}
where we have defined 
\be
H^{(0)}\equiv\delta\chi^{(0)} \, .
\ee
Note that the mass is nowhere vanishing due to the property
\be
\pi^{*}\omega\wedge\star\pi^{*}\omega\ne0 \, .
\ee
We add to the localizing $\mathcal{V}$ the term $- \mathbbm{i}s\mathcal{V}^{(0)}$.
Note that the reality conditions on $H$ and $\sigma$ require $s$ to be real. 

\subsubsection{Fluxes}\label{subsubsec:fluxes}

As shown in \cite{Bershtein:2015xfa}, the four-dimensional equations
defined on $\mathcal{M}_{4}$ 
\be\label{BTequiv}
i_{\tilde{v}}F=\rd \phi \, ,
\ee
admit Abelian solutions corresponding to equivariant line bundles
supported on 
$
H^{2}\left(\mathcal{M}_{4}\right)
$.
Specifically, the flux is viewed now as a symplectic form for the
torus action represented by $\tilde{v}$, and $\phi$ represents the
moment map for the symplectic action.

The addition of $\delta\mathcal{V}^{(0)}$ to the localizing
action relaxes the constraint imposed by the Lagrange multiplier in
\eqref{eq:vector_multiplet_localization_equations} from
$
F^{+}=0
$
to
\be
F^{+} = \mathfrak{j} \pi^{*}\omega \, ,
\ee
where $\mathfrak{j}$ is some element of the Cartan subalgebra.
The combined equations 
\be
i_{v}F=0 \, ,\qquad F^{+}= \mathfrak{j} \pi^{*}\omega \, ,
\ee
are five-dimensional versions of those analyzed in \cite{Bershtein:2015xfa},
with $A_{5}$ playing the role of $\phi$. To make the connection,
use indices $i,j,\ldots$ for $\mathcal{M}_{4}$ and write the equation
\be
i_{v}F=0 \, ,
\ee
for an Abelian field strength in 4+1 notation  as 
\be\label{equivariantcohomology}
\tilde{v}^{i}F_{ij}+\partial_{5}A_{j}-\partial_{j}A_{5}=0 \, , \qquad
\tilde{v}^{i}\partial_{i}A_{5}-\tilde{v}^{i}\partial_{5}A_{i}=0 \, .
\ee
If we set $\partial_{5}A_{j}=0$,
then the first equation is the symplectic moment map condition. The
second equation follows from the first after applying $i_{\tilde{v}}$. 

The solutions above correspond to solutions of the moment map equation
\eqref{BTequiv} and define equivariant cohomology classes. The resulting equivariant line bundles
are associated to the equivariant divisors on $\mathcal{M}_{4}$.
The relationship between these divisors and the description of $\mathcal{M}_{4}$
using the toric fan is explained in appendix \ref{sec:toric geometry}. In particular, there
is an equivariant divisor $D_l$ for each vector  in the fan, and therefore for each fixed point of the torus action.
The total flux is then associated with a linear combination of divisors $\sum_{l=1}^d \fp_l D_l$,
where $\fp_l$ lives in the Cartan subalgebra.
We denote the resulting field strengths $F^{(0)}$.
Note that
\be
F^{(0)}=\pi^{*}F_{4}^{(0)} \, ,
\ee
for some two-form $F_{4}^{(0)}$on $\mathcal{M}_{4}$.

Notice that, due to \eqref{equivariantcohomology}, the field $a$  acquires a nonzero profile on the manifold $\cM_4$.
Near the fixed points of the torus action, the field becomes
 \be\label{replacement}
a^{(l)}=a+\epsilon_{1}^{(l)}\mathfrak{p}_{l}+\epsilon_{2}^{(l)}\mathfrak{p}_{l+1} \, ,
\ee
where the identification of the parameters $\epsilon_{1}^{(l)},\epsilon_{2}^{(l)}$  is given in appendix \ref{sec:toric geometry}.

\subsubsection{Instantons}

Near the fixed circles of $v$, the complex determined in section \ref{subsec:Supersymmetry-algebra}
coincides with the one considered by Nekrasov \cite{Nekrasov:2002qd}.
We therefore conjecture, in the spirit of \cite{Pestun:2007rz,Kallen:2012cs,Hosomichi:2012ek},
that these points support point-like instantons which are accounted
for by the five-dimensional or K-theoretic version of the Nekrasov's partition function
\be
Z^{\mathbb{C}^{2}\times S^{1}}_{\text{inst}} ( g_{\text{YM}} , k , a , \Delta, \epsilon_{1}, \epsilon_{2} , \beta ) \, ,
\ee
described in section \ref{subsec:Nekrasov partition function}.
The parameters appearing in this partition
function can be read off from the classical action, the toric geometry,
the metric, the fluxes, and the mass parameters. Specifically, the
identification of the parameters $\epsilon_{1},\epsilon_{2}$ and
$a$ is given in appendix \ref{sec:toric geometry} and corresponds to the values appearing in \eqref{replacement}. 

The authors of \cite{Kallen:2012cs} identified a class of solutions
to the equations 
\be
F^{+}=0 \, ,\qquad i_{v}F=0 \, , \qquad
i_{v}\kappa=1 \, ,
\ee
on any contact five-manifolds with contact structure determined by
a one-form $\kappa$. These solutions were dubbed contact instantons.
Although the one-form $\kappa$ defined in \eqref{not:a:contact:form}
is \emph{not} a contact form, the instantons appearing in our
partition function are the same solutions.

\subsubsection{Gauge fixing}

As discussed in \cite{Pestun:2007rz}, one can add a BRST-closed term
to the action in order to gauge fix without disturbing the localization
procedure. A convenient gauge for our calculation is the background
gauge 
\be
\rd_{A^{(0)}}^{\dagger}A=0 \, ,
\ee
where $A^{(0)}$ represents the value of $A$ at a point
in moduli space. This gauge is part of the definition of the Atiyah-Hitchin-Singer
complex for the instanton moduli space \cite{Atiyah:1978wi}. We will,
in addition, gauge fix the scalar moduli such that their non-Cartan
elements vanish. The modulus $a$ will be Cartan valued 
\be
a^{i} \, , \qquad i \in 1, \ldots , \text{rk}(G) \, .
\ee
Doing so incurs a determinant in the matrix model which is, however,
already taken into account in the one-loop determinant described below.
An additional factor of the inverse volume of the Weyl group, $\left|\mathfrak{W}\right|^{-1}$, is also present.

\subsubsection{Integration}

Localization takes effect when the coefficient $t$ of the localizing
action is taken to be very large. The value of $s$ is up to us. Following
\cite{Bershtein:2015xfa}, we choose to take a limit 
\be
s \to \infty \, .
\ee
In order to keep the moduli finite, we rescale
\be
\sigma^{(0)} \to \frac{1}{s}\sigma^{(0)} \, .
\ee
In taking the limit, the fermion zero modes acquire a large mass can
be trivially integrated out. In addition, $\sigma^{(0)}$
drops out of all terms except \eqref{eq:fermion_zero_modes_localizing_term}.
Following \cite{Bershtein:2015xfa}, we write the remaining integral
over the scalar moduli as
\be\label{JK}
\int \rd a \, \rd \bar{a} \frac{\partial}{\partial\bar{a}}\int\frac{\rd H_{0}}{H_{0}} \, e^{i\bar{a}H_{0}}\times\left(\bar{a}\text{ independent terms}\right) ,
\ee
where we have used $H_{0}$ to mean 
\be
\int\star H^{(0)}\wedge\pi^{*}\omega \, .
\ee
The fact that the integrand is a total derivative in $\bar{a}$ should also follow from the algebra of supersymmetry
of the zero mode supermultiplet \cite{,Benini:2015noa,Closset:2015rna}.

The  integral, being a total derivative in $\bar{a}$,
reduces to a contour integral. After the integral over $H_{0}$, which takes the residue at $H_{0}=0$,  we will be left with the contour integral of a meromorphic function of $a$.  
Following \cite{Benini:2013nda,Hori:2014tda,Benini:2015noa,Closset:2015rna} we expect  that the interplay between a  proper regularization of the integrand 
and the use of the zero mode $H^0$ as a regulator will lead to the determination of the correct contour of integration.
We also expect that the appropriate contour is given by some Jeffrey-Kirwan prescription  \cite{JeffreyKirwan}.
In related contexts,  this prescription appears in the calculation of the
instanton partition function \cite{Hori:2014tda,Hwang:2014uwa}. It
also makes an appearance in the calculation of the partition functions
of the two-dimensional $A$-model \cite{,Benini:2015noa,Closset:2015rna} and the three-
and four-dimensional topologically twisted indices \cite{Benini:2015noa}, which are lower-dimensional
analogues of the partition function on $\mathcal{M}_{4}$ and the
five-dimensional twisted index.
We postpone to future work the determination of the correct contour of integration.

\subsubsection{Classical contribution}
\label{sebsec:classical}

Classical contributions to the localization calculation come from
evaluating \eqref{eq:twisted_Yang_Mills_action} and \eqref{eq:twisted_hypermultiplet_action}
on the moduli space identified above. Since all hypermultiplet fields
vanish on the moduli space, there is no contribution from \eqref{eq:twisted_hypermultiplet_action}.
Moreover, the bulk moduli do not contribute even to \eqref{eq:twisted_Yang_Mills_action}.
This is in contrast to the contact manifold case \cite{Qiu:2016dyj}. The contribution
of instantons will be discussed when we discuss the Nekrasov's partition
function. All that is left is the contribution of the fluxes and the
auxiliary field to \eqref{eq:twisted_Yang_Mills_action}. Recall that
the evaluation of the classical action is on the configurations such that the right hand side of the fermion transformations vanish.
Specifically, this implies that $H=0$, regardless of the reality conditions of $D_{IJ}$.
In fact, $H$ appears alone on the right hand side of the transformation of $\chi$, see \eqref{eq:vector_multiplet_susy_transformations},
meaning that we are free to add an arbitrarily large quadratic term for it in the localizing action. The equation $H=0$ imposes the relation
\begin{equation}
F_{mn}^{(0)+}=-\frac{\mathbbm{i}}{2}\xi^{I}\Gamma_{mn}\xi^{J}D_{IJ}^{(0)} \, .
\label{eq:F_vs_D_relation}
\end{equation}
Similar relations involving fluxes appear in the three-dimensional computations of
the twisted indices \cite{Benini:2015noa,Benini:2016hjo}. The relevant
part of the classical action is
\be
\begin{aligned}
& \exp\left[-\frac{1}{g_{\text{YM}}^{2}}\int\left(F^{(0)}\wedge\star F^{(0)}+\frac{1}{2}D^{(0)IJ} \wedge \star D_{IJ}^{(0)}\right)\right] \\
& = \exp \left[ -\frac{1}{g_{\text{YM}}^{2}}\int\left(F_{V}^{(0)}\wedge\star F_{V}^{(0)}+F_{H}^{(0)+}\wedge\star F_{H}^{(0)+}
+F_{H}^{(0)-}\wedge\star F_{H}^{(0)-}+\frac{1}{2}D^{(0)IJ} \wedge \star D_{IJ}^{(0)}\right) \right] .
\end{aligned}
\ee
Due to the properties of $F^{(0)}$, we can rewrite this as
\be
\begin{aligned}
 & \exp \left[ -\frac{\beta}{g_{\text{YM}}^{2}}\int_{\mathcal{M}_{4}} \left( F_{4}^{(0)}\wedge\star_{4}F_{4}^{(0)}+\frac{1}{2}D^{(0)IJ}D_{IJ}^{(0)}\right) \right] \\
 & =\exp\left[-\frac{\beta}{g_{\text{YM}}^{2}}\int_{\mathcal{M}_{4}}\left(F_{4}^{(0)+}\wedge F_{4}^{(0)+}-F_{4}^{(0)-}\wedge F_{4}^{(0)-}+\frac{1}{2}D^{(0)IJ}D_{IJ}^{(0)}\right)\right] \\
 & =\exp\left[-\frac{\beta}{g_{\text{YM}}^{2}}\int_{\mathcal{M}_{4}}\left(2F_{4}^{(0)+}\wedge F_{4}^{(0)+}-F_{4}^{(0)}\wedge F_{4}^{(0)}+\frac{1}{2}D^{(0)IJ}D_{IJ}^{(0)}\right)\right] \\
 & =\exp\left(\frac{\beta}{g_{\text{YM}}^{2}}\int_{\mathcal{M}_{4}}F_{4}^{(0)}\wedge F_{4}^{(0)}\right) ,
\end{aligned}
\ee
where $F^{(0)}=\pi^{*}F_{4}^{(0)}$
and we have used \eqref{eq:F_vs_D_relation}. 

Given the relationship between the field strength and the differential
representative of the first Chern class
\be
c_{1}(A)=-\frac{1}{2\pi}F \, ,
\ee
the classical contribution can be written as 
\be
\label{Zcl:localization:computation}
\begin{aligned}
& Z_{\text{cl}}^{\cM_4 \times S^1} (g_{\text{YM}} , \fp , \beta) = \exp \left[ \frac{4\pi^{2}\beta}{g_{\text{YM}}^{2}}\left(\int_{\mathcal{M}_{4}}c_{1}( \{\fp\} )\wedge c_{1}(\{\fp\})\right)\right]
= \exp\left(\frac{4\pi^{2}\beta}{g_{\text{YM}}^{2}}c (\fp)\right) , \\
& c(\fp )\equiv\bigg(\sum_{l=1}^{d}\mathfrak{p}_{l}D_{l}\bigg)\cdot\bigg(\sum_{l=1}^{d}\mathfrak{p}_{l}D_{l}\bigg) .
\end{aligned}
\ee
The factor $c\left(\fp \right)$ can be evaluated
for any given fan and choice of $\mathfrak{p}_{l}$ using the techniques
in appendix \ref{sec:toric geometry}. We will give explicit examples in section \ref{completePF}.

\subsubsection{One-loop determinants via index theorem}
\label{subsubsec:1-loop via index theorem}

We can  compute the one-loop determinant from the equivariant index theorem.
We follow the derivation in \cite{Pestun:2007rz,Gomis:2011pf}. The
fields appearing in the supersymmetry algebra \eqref{eq:vector_multiplet_susy_transformations}
and \eqref{eq:hypermultiplet_susy_transformations} can be put into
the canonical form 
\be
\delta\varphi_{e,o}=\hat{\varphi}_{o,e} \, , \qquad
\delta\hat{\varphi}_{o,e}=\mathcal{R}\varphi_{e,o} \, .
\ee
where $\varphi_{e},\hat{\varphi}_{e}$ are bosonic and $\varphi_{o},\hat{\varphi}_{o}$
are fermionic. The expression above is meant to represent the $\delta$-complex
linearized around a point in the moduli space. We can identify 
\be
\mathcal{R}=\mathbbm{i} ( \mathcal{L}_{v} + G_{a}) \, .
\ee

The localizing functional contains a term of the form
\be
\mathcal{V}=\varphi_{o}D_{oe}\varphi_{e} \, .
\ee
According to \cite{Gomis:2011pf}, the result of the Gaussian integral
around a point in moduli space is given by 
\begin{equation}
Z_{1\text{-loop}}= \frac{\text{det}_{\text{coker}\,D_{oe}}\mathcal{R}|_o}{\text{det}_{\text{ker}\,D_{oe}}\mathcal{R}|_e} \,  .
\label{eq:one_loop_expression}
\end{equation}
Using \eqref{eq:localizing_terms} and the supersymmetry algebra, we
can identify 
\bea
D^{\text{vector}}= (1+i_{v}\star ) (1- \kappa \wedge i_{v} ) \rd_A \, , \qquad
D^{\text{hyper}}=\Gamma^{i}D_{i}=\pi^{*}D_{\text{Dirac}} \, .
\eea
The operator $D^{\text{vector}}$ is the projection
of the covariantized exterior derivative operator on $\mathcal{M}_{4}$, acting
on one-forms, to the self-dual two-forms.  One should take into account the gauge
fixing. Together with $D^{\text{vector}}$, this forms the self-dual complex
\be
D_{\text{SD}}:\Omega^{0}\xrightarrow{\rd}\Omega^{1}\xrightarrow{\rd_{+}}\Omega^{2+} \, ,
\ee
tensored with the adjoint representation. $D^{\text{hyper}}$ is simply the Dirac operator on $\mathcal{M}_{4}$.
The relevant complex is the Dirac complex
\be
D_{\text{Dirac}}: S^{+} \to S^{-} \, ,
\ee
tensored with the representation of the gauge and flavor groups.
The bundles $S^\pm$ are the positive and negative chirality spin bundles on $\mathcal{M}_4$.
If $\mathcal{M}_4$ is not spin, these should be replaced by an appropriate bundle associated with a $\text{spin}^c$ structure.
Neither of these complexes are
elliptic. However, both are transversely elliptic with respect to
the action of $\mathcal{R}$.

The data entering \eqref{eq:one_loop_expression} can be extracted from
the computation of the $\cR$-equivariant index for the operator $D_{oe}$:
\be
\text{ind} \, D_{oe}=\text{Tr}_{\text{ker}\,D_{oe}}e^{\mathcal{R}} - \text{Tr}_{\text{coker}\,D_{oe}}e^{\mathcal{R}} \, .
\ee
The computation of this index is described in \cite{atiyah2006elliptic}.
We will follow the exposition in \cite{Pestun:2016qko}. Let $\mathcal{E}$
be a $G$-equivariant complex of linear differential operators acting
on sections of vector bundles $E_{i}$ over $X$:
\be
\mathcal{E}:\Gamma (E_{0} )\xrightarrow{D_{0}}\Gamma (E_{1} )\xrightarrow{D_{1}}\Gamma (E_{2} )\to \ldots \, , \qquad
D_{i}D_{i+1}=0 \, .
\ee
The equivariant index of the complex $\cE$ is the virtual character of the $G$ action
on the cohomology classes $H^{k}(\mathcal{E})$:
\be
\text{ind}_{G} D (g)=\sum_{k} (-1)^{k} \Tr_{H^{k}(\mathcal{E})} g \, , \qquad g \in G \, .
\ee
If the set of fixed points of $G$ is discrete, the index can be determined
by examining the action of $G$ at those points, denoted $X^{G}$:
\be
\text{ind}_{G} D=\sum_{x\in X^{G}}\frac{\sum_{k}(-1)^{k}\text{ch}_{G}(E_{k})|_{x}}{\text{det}_{T_{x}X}(1-g^{-1})} \, .
\ee
The numerator in this expression encodes the action of $G$ on the
bundles, while the denominator is the action on the tangent space
of $X$. We refer the reader to \cite{Gomis:2011pf} for more information,
and to \cite{atiyah2006elliptic} for a complete treatment.

The index and the one-loop determinant can be computed using the equivariant
index theorem on $\mathcal{M}_{4}$. There is one fixed point at the
origin of every cone in the fan which determines $\mathcal{M}_{4}$.
The copy of $\mathbb{C}^{2}$ associated to the cone is acted upon
by the equivariant parameters, and feels the flux from equivariant
divisors associated to the two neighboring vectors, as determined
in appendix \ref{sec:toric geometry}. All that is needed to construct the character for
$\mathcal{M}_{4}$ is to add up the contributions. The subtleties
in the calculation involve the use of the right complex for the index
theorem, and the necessity of regularizing the infinite products that
it yields. 

The complex identified above for the vector multiplet is the self-dual complex.
Following the discussion in \cite{Pestun:2016jze}, we use the Dolbeault complex
(the ``holomorphic projection of the vector multiplet'') and find a match to the gluing calculation in section \ref{completePF}.
The two complexes are related on a K\"ahler manifold (see \eg\;of \cite[sect.\,2.3.1]{Marino:1996sd}).
The relevant index for the twisted Dolbeault operator $\bar{\partial}$ on $\mathbb{C}^{2}\times S^{1}$ is given by
\be
\text{ind}_{\mathcal{R}}\pi^{*}(\bar{\partial})|_{\mathbb{C}^{2}\times S^{1}}=\sum_{\alpha\in G}\sum_{n=-\infty}^{\infty}\frac{e^{\mathbbm{i}\frac{2\pi n}{\beta}}e^{\mathbbm{i}\alpha\left(\mathfrak{p}_{1}\right)\epsilon_{1}}e^{\mathbbm{i}\alpha\left(\mathfrak{p}_{2}\right)\epsilon_{2}}e^{\mathbbm{i}\alpha\left(a\right)}}{\left(1-e^{-\mathbbm{i}\epsilon_{1}}\right)\left(1-e^{-\mathbbm{i}\epsilon_{2}}\right)}\,,
\ee
where we have incorporated the free action of the rotation on $S^{1}$
and denoted by $\mathfrak{p}_{1,2}$ the coefficients of the two divisors.
Here, $\alpha$ are the roots of the gauge group $G$ and the $\epsilon_{1,2}$ are arbitrary complex deformation parameters,
which we will take to zero at the end. The complete index reads 
\begin{equation}
\text{ind}_{\mathcal{R}}\pi^{*}(\bar{\partial})|_{\mathcal{M}_{4}\times S^{1}}=
\sum_{l=1}^{d}\text{ind}_{\mathcal{R}^{(l)}}\pi^{*}(\bar{\partial})|_{\mathbb{C}^{2}\times S^{1}} \, ,
\label{eq:equivariant_index_on_M}
\end{equation}
where $d$ is the number of cones in the fan determining $\mathcal{M}_{4}$
and $\mathcal{R}^{(l)}$ signifies the use of the coefficients
and equivariant parameters relevant to that cone.

We will explicitly evaluate the one-loop determinant resulting from \eqref{eq:equivariant_index_on_M}
only in the non-equivariant limit. Define the degeneracy 
\be\label{indexfixedpoint}
d (\mathfrak{p} )\equiv\lim_{\epsilon_{1,2} \to 0}\sum_{l=1}^{d}\frac{e^{\mathbbm{i}\alpha(\mathfrak{p}_{l})\epsilon_{1}^{(l)}}e^{\mathbbm{i}\alpha (\mathfrak{p}_{l+1} )\epsilon_{2}^{(l)}}}
{\left(1-e^{-\mathbbm{i}\epsilon_{1}^{(l)}}\right)\left(1-e^{-\mathbbm{i}\epsilon_{2}^{(l)}}\right)} \, .
\ee
Our flux conventions are those in \cite{Pestun:2016qko}. 
The limit reduces the equivariant index to the Hirzebruch-Riemann-Roch
theorem
\bea\label{HRR}
d (\mathfrak{p})= \int_{\cM_4} \text{ch} (E) \text{td} (\cM_4) \, ,
\eea
where $E$ corresponds to $\sum_{l=1}^d \fp_l D_l$. We will give explicit examples of degeneracies in section \ref{completePF}.

The one-loop determinant for a vector multiplet is then given by
\be
Z_{1\text{-loop}}^{\text{vector}}\left(a,\mathfrak{p},\beta\right)=\prod_{\alpha\in G}\left[\prod_{n=-\infty}^{\infty}\left(\mathbbm{i}\frac{2\pi}{\beta}n+\mathbbm{i}\alpha(a)\right)\right]^{d\left(\alpha(\mathfrak{p})\right)}\,.
\ee
The infinite product above requires regularization. A physically acceptable
regularization is given by
\bea\label{choicereg}
& \prod_{n=-\infty}^{\infty}\left(\mathbbm{i}\frac{2\pi}{\beta}n+\mathbbm{i}\alpha(a)\right)=\left(\frac{1-x^{\alpha}}{x^{\alpha/2}}\right)\,,\\
& x^{\alpha}\equiv e^{\mathbbm{i}\beta\alpha\left(a\right)}\,,\qquad\alpha(a)=\alpha_{i}a^{i}\,.
\eea
This choice has simple transformation properties under parity and
correctly accounts for the induced Chern-Simons term when used in
the three-dimensional calculations \cite{Benini:2015noa}. See also \cite{Hori:2014tda}
for an example of its use. We thus obtain
\be
\label{1-loop:index:theorem:vector}
Z_{1\text{-loop}}^{\text{vector}}\left(a, \mathfrak{p} ,\beta\right)=\prod_{\alpha\in G}\left(\frac{1-x^{\alpha}}{x^{\alpha/2}}\right)^{d\left( \alpha(\mathfrak{p}) \right)}.
\ee

The index for a hypermultiplet is based on the twisted Dirac complex
instead of the Dolbeault complex. It also incorporates background
scalar moduli and fluxes, which we denote $\Delta$ and $\mathfrak{t}_{l}$
respectively. Given the relation on $\mathcal{M}_{4}$ between these
two complexes, one needs to change the index for the vector multiplet
by an overall flux corresponding to the square root of the canonical
bundle and take into account the opposite grading between the two
complexes \cite{Pestun:2016qko}. The canonical bundle is minus the
sum of all the equivariant divisors. In addition, there is an $\epsilon$
dependent choice of the origin of the flavor mass parameter. The choice
corresponding to the superconformal fixed point in five dimensions was discussed in \cite{Kim:2011mv,Bullimore:2014upa,Nishioka:2016guu}, following the correction to the four sphere partition function found in \cite{Okuda:2010ke}.
The authors of \cite{Bershtein:2015xfa} found a match with the results derived by Vafa and Witten in \cite{Vafa:1994tf} for the $\mathcal{N}=4$ theory with yet another choice.
If the manifold $\mathcal{M}_4$ is not spin, there may be other complications related to the choice of $\text{spin}^c$ structure.
Specifically, we have made no attempt to identify the canonical $\text{spin}^c$ structure associated with the almost complex structure of $\mathcal{M}_4$ since this choice can be shifted by background fluxes.
For notational convenience we choose a common shift of the mass parameter for all manifolds 
\be
 \label{shift:Delta}
 \Delta \to \Delta-\frac{1}{2}(\epsilon_{1}+\epsilon_{2}) \, ,
\ee
although this may require an appropriate redefinition of the origin of the background fluxes in particular examples.

Incorporating both of these leads to
\be
d^{\text{hyper}} (\mathfrak{p},\mathfrak{t} )\equiv-\lim_{\epsilon_{1,2}\to0}\sum_{l=1}^{d}\frac{e^{\mathbbm{i}\left(\rho (\mathfrak{p}_{l} )+\nu (\mathfrak{t}_{l} )-1\right)\epsilon_{1}^{(l)}}
e^{\mathbbm{i}\left(\rho(\mathfrak{p}_{l+1})+\nu(\mathfrak{t}_{l+1})-1\right)\epsilon_{2}^{(l)}}}{\left(1-e^{-\mathbbm{i}\epsilon_{1}^{(l)}}\right)\left(1-e^{-\mathbbm{i}\epsilon_{2}^{(l)}}\right)}\,,
\ee
where $\fR$ is the representation under the gauge group $G$, $\rho$ the corresponding weights, and
$\nu$ is the weight of the hypermultiplet under the flavor symmetry group.
The complete result can then be written as
\bea
\label{1-loop:index:theorem:hyper}
& Z_{1\text{-loop}}^{\text{hyper}}\left(a,\Delta, \mathfrak{p} , \mathfrak{t} ,\beta\right)=
\prod_{\rho\in \mathfrak{R}}\left(\frac{1-x^{\rho}y^{\nu}}{x^{\rho/2}y^{\nu/2}}\right)^{d_{\text{hyper}}\left( \rho(\mathfrak{p}) , \nu(\mathfrak{t}) \right)} \, , \\
& y^{\nu}\equiv e^{\mathbbm{i}\beta\nu(\Delta)} .
\eea

\subsection{The Nekrasov's partition function}
\label{subsec:Nekrasov partition function}

In this section we collect the expressions for the K-theoretic Nekrasov's partition function:
\be
 \label{K-theoretic:Nekrasov}
 Z_{\text{Nekrasov}}^{\mathbb{C}^{2}\times S^{1}} (g_{\text{YM}} , k , a , \Delta , \epsilon_{1},\epsilon_{2} , \beta ) \equiv Z^{\bC^2 \times S^1}_{\text{pert}} Z_{\text{inst}}^{\bC^2 \times S^1} \, .
\ee
As we will discuss in the next section, the topologically twisted index on $\cM_4 \times S^1$ can be obtained by gluing copies of the Nekrasov's partition functions in the spirit of \eqref{Nekrasov}. 

\subsubsection{Perturbative contribution}

 The perturbative part of the partition function on $\mathbb{C}^2\times S^1$  consists of a classical and a one-loop contribution. 

\paragraph*{Classical contribution.}

The classical contribution to \eqref{K-theoretic:Nekrasov} is given by \cite{Gottsche:2006bm,Bershtein:2018srt}
\be
 \label{Z:flat:space:cl}
 Z_{\text{cl}}^{\mathbb{C}^{2}\times S^{1}} ( g_{\text{YM}} , k, a,\epsilon_{1},\epsilon_{2} ) =
 \exp \left( \frac{4 \pi ^2 \beta}{g_{\text{YM}}^2 \epsilon_1 \epsilon_2} \Tr_{\text{F}} (a)^2 + \frac{\mathbbm{i} k \beta}{6 \epsilon_1 \epsilon_2} \Tr_{\text{F}}(a^3) \right) \, ,
\ee
where $k$ is the Chern-Simons level of $G$ and $\Tr_{\text{F}}$ is the trace in fundamental representation.%
\footnote{The generators $T_a$ are normalized as $\Tr_{\fR} ( T_ a T_b ) = k (\fR) \delta_{ab}$ with $k(\text{F}) = 1/2$ for the fundamental representation of $\SU(N)$.}

\paragraph*{One-loop contribution.}

We will  use the perturbative part of the partition function on $\mathbb{C}^2\times S^1$ as defined in \cite{Nakajima:2005fg}. For a gauge group $G$ the contribution of a vector multiplet to the perturbative part is given by
\be
 \label{Nakajima:pert}
 Z_{\text{pert-vector}}^{\mathbb{C}^{2}\times S^{1}}\left(\epsilon_{1},\epsilon_{2} , a ; \Lambda , \beta \right)=\exp\bigg(-\sum_{\alpha\in G}\tilde{\gamma}_{\epsilon_{1},\epsilon_{2}}\left(\alpha(a) | \beta ; \Lambda\right)\bigg),
\ee
where 
\bea
 \tilde{\gamma}_{\epsilon_{1},\epsilon_{2}}\left(a | \beta; \Lambda \right) & =
 \frac{1}{\epsilon_{1}\epsilon_{2}}\left(\frac{\pi^{2}a}{6\beta}-\frac{\zeta(3)}{\beta^{2}}\right)+\frac{\epsilon_{1}+\epsilon_{2}}{2\epsilon_{1}\epsilon_{2}}\left(a\log\left(\beta\Lambda\right) +\frac{\pi^{2}}{6\beta}\right) \\
 & + \frac{1}{2\epsilon_{1}\epsilon_{2}}\left[-\frac{\beta}{6}\left(a+\frac{1}{2}\left(\epsilon_{1}+\epsilon_{2}\right)\right)^{3}+a^{2}\log\left(\beta\Lambda\right)\right] \\
 & + \frac{\epsilon_{1}^{2}+\epsilon_{2}^{2}+3\epsilon_{1}\epsilon_{2}}{12\epsilon_{1}\epsilon_{2}}\log\left(\beta\Lambda\right)
 + \sum_{n=1}^{\infty}\frac{1}{n}\frac{e^{-\beta na}}{\left(e^{\beta n\epsilon_{1}}-1\right)\left(e^{\beta n\epsilon_{2}}-1\right)} \, .
\eea
With this definition of the perturbative part of the partition function, the authors of \cite{Nakajima:2005fg} derived a formula for the partition function for the blowup
of $\mathbb{C}^2$ at a point, which is just an example of the gluing procedure we will discuss in section \ref{completePF}.
This expression is written in the conventions of \cite{Nakajima:2005fg} and also includes what we already defined as the classical contribution.
One can swap between the conventions by
\be
 \label{swap:conventions}
 a^{\text{there}} = -\mathbbm{i} a^{\text{here}} \, , \qquad \epsilon_{i}^{\text{there}} = - \mathbbm{i} \epsilon_{i}^{\text{here}} \, .
\ee

The one-loop contribution from a vector multiplet in our conventions is given by
\be
 \label{vec1loop}
 \begin{aligned}
  & Z_{1\text{-loop, vector}}^{\mathbb{C}^{2}\times S^{1}} (a, \epsilon_1 , \epsilon_2) = Z_{\text{parity, vector}}^{\mathbb{C}^{2}\times S^{1}} (a, \epsilon_1 , \epsilon_2) \prod_{\alpha \in G} ( x^{\alpha}; p, t)_\infty \, , \\
  & Z_{\text{parity, vector}}^{\mathbb{C}^{2}\times S^{1}} (a, \epsilon_1 , \epsilon_2) = \prod_{\alpha \in G} \exp \bigg[ \frac{1}{\epsilon_1 \epsilon_2} \bigg( \frac{\mathbbm{i}}{2 \beta ^2} g_3 \left( - \alpha( \beta a ) \right)
  - \frac{\mathbbm{i} (\epsilon_1+\epsilon_2)}{4 \beta} g_2 \left( - \alpha( \beta a) \right) \\
  & \hspace{3.cm} + \frac{\mathbbm{i} ( \epsilon_1 + \epsilon_2 )^2}{16} g_1 \left( - \alpha( \beta a ) \right)
  - \frac{\mathbbm{i} \beta}{96} (\epsilon_1+\epsilon_2)^3 + \frac{\mathbbm{i} \pi}{48} \left( \epsilon_1^2 + \epsilon_2^2 \right) - \frac{\zeta (3)}{\beta^2} \bigg) \bigg] \, .
 \end{aligned}
\ee
Here, we defined the double $(p,t)$-factorial as
\be
 ( x ; p , t)_\infty = \prod_{i , j = 0}^{\infty} (1 - x p^i t^j ) \, ,
\ee
where $p = e^{- \mathbbm{i} \beta \epsilon_1}$ and $t = e^{- \mathbbm{i} \beta \epsilon_2}$.
The polynomial functions $g_s(a)$ are given in \eqref{PolyLog:inversion formulae}.
The parity contribution in \eqref{vec1loop} is related to the choice of regularization we made in \eqref{choicereg}. 
It can be partially understood as an effective one-half Chern-Simons contribution \eqref{Z:flat:space:cl}.
The contribution of a hypermultiplet to the one-loop determinant instead reads
\be
 \begin{aligned}
  & Z_{1\text{-loop, hyper}}^{\mathbb{C}^{2}\times S^{1}} (a, \Delta, \epsilon_1 , \epsilon_2) = Z_{\text{parity, hyper}}^{\mathbb{C}^{2}\times S^{1}} (a, \Delta, \epsilon_1 , \epsilon_2) \prod_{\rho \in \fR} (x^{\rho} y^{\nu}; p, t)_{\infty}^{-1} \, , \\
  & Z_{\text{parity, hyper}}^{\mathbb{C}^{2}\times S^{1}} (a, \Delta, \epsilon_1 , \epsilon_2) = \prod_{\rho \in \fR} \exp \bigg[ \frac{1}{\epsilon_1 \epsilon_2}
  \bigg( \frac{\mathbbm{i}}{2 \beta ^2} g_3 \big(  \rho( \beta a ) + \nu ( \beta \Delta ) \big) \\
  & \hspace{3.cm} + \frac{\mathbbm{i} (\epsilon_1+\epsilon_2)}{4 \beta} g_2 \big(  \rho( \beta a ) + \nu ( \beta \Delta ) \big)
  + \frac{\mathbbm{i} ( \epsilon_1 + \epsilon_2 )^2}{16} g_1 \big(  \rho( \beta a ) + \nu ( \beta \Delta ) \big) \\
  & \hspace{3.cm} + \frac{\mathbbm{i} \beta}{96} (\epsilon_1+\epsilon_2)^3
  + \frac{\mathbbm{i} \pi}{48} \left( \epsilon_1^2 + \epsilon_2^2 \right) \bigg) \bigg] \, .
 \end{aligned}
\ee

Putting everything together, the perturbative part of the Nekrasov's partition function can be written as
\be
Z_{\text{pert}}^{\mathbb{C}^{2}\times S^{1}} ( g_{\text{YM}} , k , a , \Delta , \epsilon_{1} , \epsilon_{2} , \beta ) =
Z_{\text{cl}}^{\mathbb{C}^{2}\times S^{1}} Z_{1\text{-loop, vector}}^{\mathbb{C}^{2}\times S^{1}} \, Z_{1\text{-loop, hyper}}^{\mathbb{C}^{2}\times S^{1}} \, .
\ee

\subsubsection{Instantons contribution}

The localization calculation for a five-dimensional gauge theory conjecturally
includes non-perturbative contributions from contact instantons \cite{Kallen:2012cs,Hosomichi:2012ek}.
With the equivariant deformation turned on, these configurations are
localized to the fixed points of the action on $\mathcal{M}_{4}$ and
wrap the $S^{1}$. Their contribution to the matrix model is given
by the five-dimensional version of the Nekrasov's instanton partition
function, which we describe below.

Nekrasov's instanton partition function, \cite{Nekrasov:2002qd,Nekrasov:2003rj},
is the equivariant volume of the instanton moduli space on $\mathbb{R}^{4}$
with respect to the action of 
\begin{equation}
\U(1)_{a}\times \U(1)_{\epsilon_{1}}\times \U(1)_{\epsilon_{2}} \, .
\label{eq:equivariant_action}
\end{equation}
The three factors correspond to constant gauge transformations and
to rotations in two orthogonal two-planes inside $\mathbb{R}^{4}$,
respectively. The five-dimensional version of the partition function
counts instantons extended along an additional $S^{1}$ factor in
the geometry, of circumference $\beta$. The four-dimensional partition
function can be recovered by letting the size of this $S^{1}$ shrink to zero.
As with the perturbative contribution, there is an ambiguity related to the regularization 
of the KK modes on the extra circle. Different looking expressions are found in \cite{Bullimore:2014upa}.

The K-theoretic instanton partition function for gauge group $\U(N)$ in our conventions, as derived from \cite{Tachikawa:2004ur, Nakajima:2005fg,Tachikawa:2014dja}, is given by

\bea
\label{eq:Nekrasov_partition_function}
& Z_{\text{\text{inst}}}^{\mathbb{C}^{2}\times S^{1}}(q,k,a,\Delta,\epsilon_{1},\epsilon_{2},\beta)=
\sum_{\vec{{\bf Y}}}q^{|\vec{{\bf Y}}|}Z_{\vec{{\bf Y}},k}^{\text{CS}}\left(a,\epsilon_{1},\epsilon_{2},\beta\right)Z_{\vec{{\bf Y}}}\left(a,\Delta,\epsilon_{1},\epsilon_{2},\beta\right) , \\
& Z_{\vec{{\bf Y}}}\left(a,\Delta,\epsilon_{1},\epsilon_{2},\beta\right)=Z_{\vec{{\bf Y}}}^{\text{vector}}\left(a,\epsilon_{1},\epsilon_{2},\beta\right)\prod_{f=1}^{N_{f}}Z_{\vec{{\bf Y}}}^{\text{hyper}}\left(a,\Delta_{f},\epsilon_{1},\epsilon_{2},\beta\right) . 
\eea
We have
\begin{align}
 & Z_{\vec{{\bf Y}},k}^{\text{CS}} (a,\epsilon_{1},\epsilon_{2},\beta )= \prod_{i=1}^{N} \prod_{s \in\mathbf{Y}_{i}} e^{- \mathbbm{i} \beta k \phi ( a_i , s ) } \, , \nn \\
 & Z_{\vec{{\bf Y}}}^{\text{vector}} (a,\epsilon_{1},\epsilon_{2},\beta )=\prod_{i,j=1}^{N}\left(N_{i,j}^{\vec{{\bf Y}}}(0)\right)^{-1} \, , \nn \\  \displaybreak[4]
 & Z_{\vec{{\bf Y}}}^{\text{adj-hyper}} (a,\Delta,\epsilon_{1},\epsilon_{2},\beta )=\prod_{i,j=1}^{N}N_{i,j}^{\vec{{\bf Y}}}(\Delta) \, , \\
 & Z_{\vec{{\bf Y}}}^{\text{fund-hyper}} (a,\Delta,\epsilon_{1},\epsilon_{2},\beta )=\prod_{i=1}^{N}\prod_{s \in \mathbf{Y}_{i}}
 \left( 1 - e^{ \mathbbm{i} \beta \left( \phi ( a_i , s ) + \Delta + \epsilon_{1} + \epsilon_{2} \right)} \right) \, , \nn \\
 & N_{i,j}^{\vec{{\bf Y}}}( \Delta ) \equiv \prod_{s\in Y_{j}} \left( 1 - e^{\mathbbm{i} \beta \left( E ( a_i - a_j , Y_j , Y_i , s ) + \Delta \right)} \right)
 \prod_{t\in Y_{i}} \left( 1 - e^{\mathbbm{i} \beta \left( \epsilon_{1} + \epsilon_{2} - E ( a_j - a_i , Y_i , Y_j , t ) + \Delta \right)} \right) \nn ,
\end{align}
where for a box $s = ( i , j ) \in \bZ_{\geq 0} \times \bZ_{\geq 0}$ we defined the functions
\be
\begin{aligned} 
 & E ( a , Y_1 , Y_2 , s ) \equiv a - \epsilon_1 L_{Y_2} (s) + \epsilon_2 (A_{Y_1} (s) + 1 ) \, , \\
 & \phi ( a , s ) \equiv a - ( i - 1 ) \epsilon_{1} - ( j - 1 ) \epsilon_{2} \, .
\end{aligned}
\ee
The rest of the symbols above are defined as follows.
\begin{itemize}
\item[--] $\vec{{\bf Y}}$ is a vector of partitions ${\bf Y}_{i}$. A partition
is a non-increasing sequence of non-negative integers which stabilizes
at zero
\be
{\bf Y}_{i}=\left\{ Y_{i\,1}\ge Y_{i\,2}\ge\ldots\ge Y_{i\,n_{i}+1}=0=Y_{i\,n_{i}+2}=Y_{i\,n_{i}+3}=\ldots\right\} .
\ee
We define
\be
 \big|\vec{{\bf Y}} \big|\equiv \sum_{i , j = 1}^{N} Y_{i j} \, .
\ee
\item [--] For a box $s \in Y_{l}$ with coordinates $s = ( i , j )$, we define the \emph{leg length} and the \emph{arm length}
\be
L_{Y_{l}}(s)\equiv Y_{lj}^{\text{T}}-i \, ,\qquad A_{Y_{l}}(s)\equiv Y_{li}-j \, ,
\ee
where $\text{T}$ stands for transpose.
\item [--] $a$ is a complex Cartan subalgebra valued scalar of framing
parameters associated with the gauge group action.
\item [--] $\Delta$ is a flavor symmetry group vector of mass parameters for the hypermultiplets.
\item [--] $q$ is a counting parameter coming from the one instanton
action
\be
q=e^{-\frac{8\pi^{2}\beta}{g_{\text{YM}}^{2}}} \, .
\ee
\end{itemize}

\subsubsection{Effective Seiberg-Witten prepotential}
\label{subsubsec:SW prepotential}

In this section we provide the general form of the perturbative Seiberg-Witten (SW) prepotential for a 5D $\cN=1$ theory with gauge group $G$, coupling constant $g_{\text{YM}}$, and Chern-Simons coupling $k$.
For a theory with hypermultiplets transforming in the representation $\fR_I$ of $G$, the effective prepotential
is related to the Nekrasov's partition function $Z_{\text{Nekrasov}}^{\bC^2 \times S^1}$ as follows \cite{Nekrasov:2002qd,Nekrasov:2003rj,Nakajima:2005fg}:
\be
 \label{5D:prepotential:total}
 2 \pi \mathbbm{i} \cF \equiv - \lim_{\epsilon_1 , \epsilon_2 \to 0} \epsilon_1 \epsilon_2 \log Z_{\text{Nekrasov}}^{\bC^2 \times S^1} (g_{\text{YM}} , k , a , \Delta_{\fR_I} , \epsilon_1, \epsilon_2 , \beta) \, ,
\ee
whose perturbative part can be explicitly written as
\be
 \label{full:pert:prepotential}
 \begin{aligned}
  2 \pi \mathbbm{i} \cF^{\text{pert}} (a , \Delta) & = - \frac{4 \pi^2 \beta}{g_{\text{YM}}^2} \Tr_{\text{F}} (a^2) - \frac{\mathbbm{i} k \beta}{6} \Tr_{\text{F}} (a^3)
  - \frac{1}{\beta^2} \sum_{\alpha \in G} \left[ \Li_3 ( x^\alpha ) + \frac{\mathbbm{i}}{2} g_3 \left( - \alpha( \beta a ) \right) - \zeta(3) \right] \\ &
  + \frac{1}{\beta^2} \sum_{I} \sum_{\rho_I \in \fR_I} \left[ \Li_3 ( x^{\rho_I} y^{\nu_I} ) - \frac{\mathbbm{i}}{2} g_3 \left(  \rho_I( \beta a ) + \nu_I ( \beta \Delta ) \right) - \zeta(3) \right] \, ,
 \end{aligned}
\ee
where the function $g_3(a)$ is defined in \eqref{gfunctions}.
The first term in \eqref{full:pert:prepotential} comes from the classical action while the $\Li_3$ factor is the one-loop contribution of the infinite tower of KK modes on $S^1$ as discussed in \cite{Nekrasov:1996cz}.
The other polynomial terms come explicitly from the limit of the perturbative contribution in \cite{Nakajima:2005fg}.
We interpret them as effective Chern-Simons terms coming from a parity preserving regularization of the path integral, as also discussed in section \ref{subsubsec:1-loop via index theorem}.

\subsection[Partition function on \texorpdfstring{$\cM_4 \times S^1$}{M(4) x S**1}]{Partition function on $\cM_4 \times S^1$}
\label{completePF}

The functional integral under consideration is to be computed in an exact saddle point approximation around the moduli space which comprises the bulk moduli, the fluxes, and the instantons. 
We will assume as in \cite{Bawane:2014uka,Bershtein:2015xfa,Bershtein:2016mxz} that the complete partition function is  given by gluing $d$ copies of $Z_{\text{Nekrasov}}^{\bC^2 \times S^1}$, one for each fixed point of the toric action. At each fixed point, the parameters $a$, $\Delta$, $\epsilon_1$, $\epsilon_2$ in \eqref{K-theoretic:Nekrasov} are replaced with their equivariant version
$a^{(l)}$, $\Delta^{(l)}$, $\epsilon_{1}^{(l)}$, $\epsilon_{2}^{(l)}$, whose explicit form  is explained in appendix \ref{sec:toric geometry} and  given, for simple cases, in examples \ref{ex1}-\ref{ex3}.
In particular, as in \cite{Bershtein:2015xfa}, the gauge magnetic fluxes are incorporated into this expression through
\be
a^{(l)}=a+\epsilon_{1}^{(l)} \mathfrak{p}_{l} + \epsilon_{2}^{(l)} \mathfrak{p}_{l+1} \, .
\ee
The reason for this replacement is discussed in section \ref{subsubsec:fluxes} and it is also easy to see in the index theorem discussed in section \ref{subsubsec:1-loop via index theorem}.
Furthermore, the equivariant chemical potential is given by (see the discussion around \eqref{shift:Delta})
\be
 \Delta^{(l)} = \dot{\Delta} + \epsilon_{1}^{(l)} \mathfrak{t}_{l} + \epsilon_{2}^{(l)} \mathfrak{t}_{l+1} \, , \qquad \dot{\Delta} = \Delta - \big( \epsilon_1^{(l)} + \epsilon_2^{(l)} \big) \, .
\ee
As we will see this gluing  is consistent with the classical and one-loop contributions determined in sections \ref{sebsec:classical} and \ref{subsubsec:1-loop via index theorem}. 

The topologically twisted index of an $\cN = 1$ theory on $\cM_4 \times S^1$ then  reads\footnote{Similar gluing formulae hold for other five-dimensional partition functions \cite{Qiu:2014oqa,Nieri:2013yra,Pasquetti:2016dyl}.}
\be
\label{completePartitionFunction}
Z_{\mathcal{M}_{4}\times S^{1}}=\sum_{\{\fp_{l}\}|\text{semi-stable}}\oint_{\cC} \rd a \prod_{l=1}^{\chi(\mathcal{M}_{4})}
Z_{\text{Nekrasov}}^{\mathbb{C}^{2}\times S^{1}} \big( a^{(l)} , \epsilon_{1}^{(l)} , \epsilon_{2}^{(l)} , \beta , q ; \Delta^{(l)} \big) \, .
\ee

Here, following \cite{Nekrasov:2003vi,Bawane:2014uka,Bershtein:2015xfa,Bershtein:2016mxz}, we restrict the sum in \eqref{completePartitionFunction} to fluxes $\fp_l$ corresponding to semi-stable bundles. It was argued in \cite{Bawane:2014uka,Bershtein:2015xfa,Bershtein:2016mxz} that the sum should be extended to all semi-stable equivariant bundles. These are classified by a set of fluxes $\fp_l$, one for each divisor,\footnote{Remember that there are $d$ divisors but only $d-2$ independent two-cycles in the cohomology of $\cM_4$.}  subject to stability conditions that have been studied by mathematicians \cite{Kool2015}. These conditions are already quite complicated for $N=2$. Summing over all semi-stable equivariant bundles, the authors of \cite{Bawane:2014uka,Bershtein:2015xfa,Bershtein:2016mxz} found perfect agreement with known Donaldson invariant results.
In this paper we have not determined either the correct contour of integration or the correct conditions to be imposed on the fluxes. We expect that the two aspects are related. 

Notice that formula \eqref{completePartitionFunction} is consistent with and generalizes the blowup formula derived in \cite{Nakajima:2005fg},
which just corresponds to the case where $\cM_4$ is the (non-compact) blowup of $\mathbb{C}^2$ at a point.%
\footnote{See \cite[(4.14)]{Nakajima:2005fg}. The blowup of $\mathbb{C}^2$ at a point can be described by a toric fan with ${\vec n}_1=(1,0)$, ${\vec n}_2=(1,1)$ and ${\vec n}_3 =(0,1)$.}

In this paper we will be interested in theories with gauge
group $\U(N)$ or $\USp(N)$. Moreover, we want
the partition function in the large $N$, non-equivariant limit $\epsilon_{1,2} \to 0$.
The appropriate large $N$ limit is defined in the next section. 
We will assume that the instanton contribution to the free energy, defined by 
\be
F_{\mathcal{M}_{4}\times S^{1}}\equiv-\log Z_{\mathcal{M}_{4}\times S^{1}} \, ,
\ee
decays exponentially in any such limit. This is supported by the appearance
of the factor  $q^{|\vec{{\bf Y}}|}$
in \eqref{eq:Nekrasov_partition_function}. In the following sections,
all partition functions are written only for the zero instanton sector.

\subsubsection{The non-equivariant limit}

The Nekrasov's partition function \eqref{K-theoretic:Nekrasov} is singular for $\epsilon_1,\epsilon_2 \to 0$ but the product in \eqref{completePartitionFunction} is perfectly smooth in this limit.
By performing explicitly the limit, we can write the classical and perturbative part of the localized partition function in the non-equivariant limit  as
\bea\label{finalPF}
 Z_{\mathcal{M}_{4}\times S^{1}} =\frac{1}{\left|\mathfrak{W}\right|}\sum_{\{\mathfrak{p}_l\}}
 \oint_{\cC} & \prod_{i=1}^{\text{rk}(G)} \frac{\rd x_{i}}{2\pi\mathbbm{i}x_{i}} \, e^{\frac{4 \pi^{2}\beta}{g_{\text{YM}}^{2}} \Tr_{\text{F}} c(\mathfrak{p}_l) + \frac{\mathbbm{i} k \beta}{2} \Tr_{\text{F}} \left( c(\mathfrak{p}_l ) a \right)}
 \prod_{\alpha\in G} \left(\frac{1-x^{\alpha}}{x^{\alpha/2}}\right)^{d\left( \alpha(\mathfrak{p}_l) \right)} \\
 & \times \prod_{I} \prod_{\rho_I \in \mathfrak{R}_I}\left(\frac{1-x^{\rho_I}y^{\nu_I}}{x^{\rho_I/2}y^{\nu_I/2}}\right)^{d_{\text{hyper}}\left(\rho_I(\mathfrak{p}_l) ,\nu_I(\mathfrak{t}_{l}) \right)} .
\eea
As one can see, the results from  gluing (and taking the non-equivariant limit) precisely match the classical contributions
\eqref{Zcl:localization:computation} and the one-loop determinants evaluated using the index theorem (see \eqref{1-loop:index:theorem:vector} and \eqref{1-loop:index:theorem:hyper}).
 We may also expect some simplification in the sum over fluxes compared with the equivariant formula \eqref{completePartitionFunction}. We expect that, in the non-equivariant limit, the fluxes $\fp_l$ should just correspond to the set of non-equivariant bundles on $\cM_4$.

Formula \eqref{finalPF} has a natural interpretation in terms of the quantum mechanics obtained by reducing the
five-dimensional theory on $\cM_4$. In the reduction on $\cM_4$ in a sector with gauge and background fluxes $\fp_l$ and $\ft_l$,
we will obtain a set of zero modes whose multiplicity is given by the Hirzubruch-Riemann-Roch theorem and coincides with $d$ or $d_{\text{hyper}}$.
They will organize themselves into a set of Fermi or chiral multiplets according to the sign of $d$ and $d_{\text{hyper}}$.
\eqref{finalPF} is then precisely the sum over all sectors of gauge magnetic fluxes of the localization formula for the corresponding
quantum mechanics partition functions, as  derived in \cite{Hori:2014tda}.
This is in complete analogy with the structure of the three- and four-dimensional topologically twisted indices \cite{Benini:2015noa}. 

Some examples of the calculation of the degeneracy are given below. We refer to appendix \ref{sec:toric geometry} for details and notations. 

\begin{example}\label{ex1}
Complex projective space, $\mathbb{P}^2$.
\be
\vec{n}_{1}=(1,0) \, ,\quad\vec{n}_{2}=(0,1) \, ,\quad\vec{n}_{3}=(-1,-1) \, .
\ee
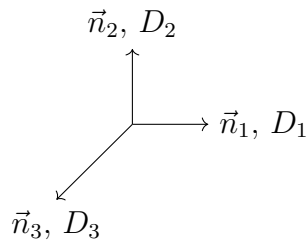
\begin{minipage}[t]{.5\textwidth}
\centering
\vspace{0pt}
\begin{tabular}{|c|c|c|c|}
\hline 
\multicolumn{1}{|c}{$\mathbb{P}^{2}$} & \multicolumn{1}{c}{} & \multicolumn{1}{c}{} & \tabularnewline
\hline 
$l$ & 1 & 2 & 3\tabularnewline
\hline 
$\epsilon_{1}^{(l)}$ & $\epsilon_{1}$ & $\epsilon_{2}-\epsilon_{1}$ & $-\epsilon_{2}$\tabularnewline
\hline 
$\epsilon_{2}^{(l)}$ & $\epsilon_{2}$ & $-\epsilon_{1}$ & $\epsilon_{1}-\epsilon_{2}$\tabularnewline
\hline 
\end{tabular}
\end{minipage}\hfill
\begin{minipage}[t]{.5\textwidth}
\centering
\vspace{0pt}
\begin{tikzpicture}
[scale=0.5 ]
\draw[->] (0,0) -- (2,0) node[right] {${\vec n}_1,\, D_1$};
\draw[->] (0,0) -- (0,2) node[above] {${\vec n}_2, \, D_2$};
\draw[->] (0,0) -- (-2,-2) node[ below ] {${\vec n}_3,\, D_3$};
  \end{tikzpicture}
\end{minipage}
\bea
& c_{\mathbb{P}^{2}} (\fp_l) = \left(\mathfrak{p}_{1}+\mathfrak{p}_{2}+\mathfrak{p}_{3}\right)^2 , \\
& d_{\mathbb{P}^{2}} (\fp_l) = \frac{1}{2}\left(\mathfrak{p}_{1}+\mathfrak{p}_{2}+\mathfrak{p}_{3}+1\right)\left(\mathfrak{p}_{1}+\mathfrak{p}_{2}+\mathfrak{p}_{3}+2\right) , \\
& d_{\mathbb{P}^{2}}^{\text{hyper}} (\fp_l, \ft_l) = - \frac{1}{2}\left(\mathfrak{p}_{1}+\mathfrak{p}_{2}+\mathfrak{p}_{3}+\mathfrak{t}_{1}+\mathfrak{t}_{2}+\mathfrak{t}_{3}-2\right)
\left(\mathfrak{p}_{1}+\mathfrak{p}_{2}+\mathfrak{p}_{3}+\mathfrak{t}_{1}+\mathfrak{t}_{2}+\mathfrak{t}_{3}-1\right) .
\eea
\end{example}
\begin{example}\label{ex2} The product of two spheres, $\mathbb{F}_{0}\simeq\bP^1 \times \bP^1$.
\be
\vec{n}_{1}=(1,0) \, ,\quad\vec{n}_{2}=(0,1) \, ,\quad\vec{n}_{3}=(-1,0) \, ,\quad\vec{n}_{4}=(0,-1) \, .
\ee
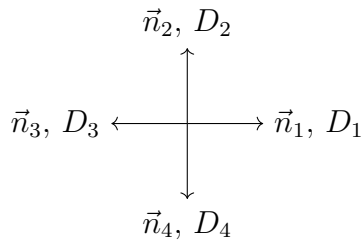
\begin{minipage}[t]{.5\textwidth}
\centering
\vspace{0pt}
\begin{tabular}{|c|c|c|c|c|}
\hline 
\multicolumn{1}{|c}{$\mathbb{F}_{0}$} & \multicolumn{1}{c}{} & \multicolumn{1}{c}{} & \multicolumn{1}{c}{} & \tabularnewline
\hline 
$l$ & 1 & 2 & 3 & 4\tabularnewline
\hline 
$\epsilon_{1}^{(l)}$ & $\epsilon_{1}$ & $\epsilon_{2}$ & $-\epsilon_{1}$ & $-\epsilon_{2}$\tabularnewline
\hline 
$\epsilon_{2}^{(l)}$ & $\epsilon_{2}$ & $-\epsilon_{1}$ & $-\epsilon_{2}$ & $\epsilon_{1}$\tabularnewline
\hline 
\end{tabular}
\end{minipage}\hfill
\begin{minipage}[t]{.5\textwidth}
\centering
\vspace{0pt}
\begin{tikzpicture}
[scale=0.5 ]
\draw[->] (0,0) -- (2,0) node[right] {${\vec n}_1,\, D_1$};
\draw[->] (0,0) -- (0,2) node[above] {${\vec n}_2, \, D_2$};
\draw[->] (0,0) -- (-2,0) node[ left] {${\vec n}_3,\, D_3$};
\draw[->] (0,0) -- (0,-2) node[ below] {${\vec n}_4,\, D_4$};
  \end{tikzpicture}
\end{minipage}
\bea
& c_{\mathbb{F}_{0}} (\fp_l) = 2  \left(\mathfrak{p}_{1}+\mathfrak{p}_{3}\right)\left(\mathfrak{p}_{2}+\mathfrak{p}_{4}\right) , \\
& d_{\mathbb{F}_{0}} (\fp_l) =\left(\mathfrak{p}_{1}+\mathfrak{p}_{3}+1\right)\left(\mathfrak{p}_{2}+\mathfrak{p}_{4}+1\right) , \\
& d_{\mathbb{F}_{0}}^{\text{hyper}} (\fp_l , \ft_l) = - \left(\mathfrak{p}_{1}+\mathfrak{p}_{3}+\mathfrak{t}_{1}+\mathfrak{t}_{3}-1\right)\left(\mathfrak{p}_{2}+\mathfrak{p}_{4}+\mathfrak{t}_{2}+\mathfrak{t}_{4}-1\right) .
\eea
\end{example}
\begin{example}
\label{ex3} $\mathbb{F}_{1}$, the blowup of $\mathbb{P}^2$ at a point.
\be
\vec{n}_{1}=(1,0) \, ,\quad\vec{n}_{2}=(0,1) \, ,\quad\vec{n}_{3}=(-1,1) \, ,\quad\vec{n}_{4}=(0,-1) \, .
\ee
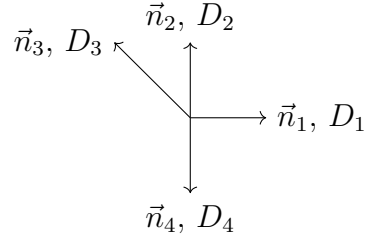
\begin{minipage}[t]{.5\textwidth}
\centering
\vspace{0pt}
\begin{tabular}{|c|c|c|c|c|}
\hline 
\multicolumn{1}{|c}{$\mathbb{F}_{1}$} & \multicolumn{1}{c}{} & \multicolumn{1}{c}{} & \multicolumn{1}{c}{} & \tabularnewline
\hline 
$l$ & 1 & 2 & 3 & 4\tabularnewline
\hline 
$\epsilon_{1}^{(l)}$ & $\epsilon_{1}$ & $\epsilon_{1}+\epsilon_{2}$ & $-\epsilon_{1}$ & $-\epsilon_{2}$\tabularnewline
\hline 
$\epsilon_{2}^{(l)}$ & $\epsilon_{2}$ & $-\epsilon_{1}$ & $-\epsilon_{1}-\epsilon_{2}$ & $\epsilon_{1}$\tabularnewline
\hline 
\end{tabular}
\end{minipage}\hfill
\begin{minipage}[t]{.5\textwidth}
\centering
\vspace{0pt}
\begin{tikzpicture}
[scale=0.5 ]
\draw[->] (0,0) -- (2,0) node[right] {${\vec n}_1,\, D_1$};
\draw[->] (0,0) -- (0,2) node[above] {${\vec n}_2, \, D_2$};
\draw[->] (0,0) -- (-2,2) node[ left] {${\vec n}_3,\, D_3$};
\draw[->] (0,0) -- (0,-2) node[ below] {${\vec n}_4,\, D_4$};
  \end{tikzpicture}
\end{minipage}
\bea
& c_{\mathbb{F}_{1}}(\fp_l ) = \left(\mathfrak{p}_{2}+\mathfrak{p}_{4}\right)\left(2\mathfrak{p}_{1}-\mathfrak{p}_{2}+2\mathfrak{p}_{3}+\mathfrak{p}_{4}\right) , \\
& d_{\mathbb{F}_{1}}(\fp_l) = \frac{1}{2}\left(\mathfrak{p}_{2}+\mathfrak{p}_{4}+1\right)\left(2\mathfrak{p}_{1}-\mathfrak{p}_{2}+2\mathfrak{p}_{3}+\mathfrak{p}_{4}+2\right) , \\
& d_{\mathbb{F}_{1}}^{\text{hyper}}(\fp_l , \ft_l) = - \frac{1}{2}\left(\mathfrak{p}_{2}+\mathfrak{p}_{4}+\mathfrak{t}_{2}+\mathfrak{t}_{4}-1\right)
\left(2\mathfrak{p}_{1}-\mathfrak{p}_{2}+2\mathfrak{p}_{3}+\mathfrak{p}_{4}+2\mathfrak{t}_{1}-\mathfrak{t}_{2}+2\mathfrak{t}_{3}+\mathfrak{t}_{4}-2\right) .
\eea
\end{example}

\subsubsection{A closer look at $\protect\fakebold{\mathbb{\mathbb{P}}} ^1 \times \protect\fakebold{\mathbb{\mathbb{P}}}^1\times S^1$}
\label{subsubsec:naive P1 X P1}

We can compare the previous result with the expectations for the case $\mathbb{P}^1\times \mathbb{P}^1\times S^1$, where the computation, in principle,  can be done by an explicit expansion in modes.  By an obvious generalization of the results  in \cite{Benini:2015noa}, we expect the following partition function
\bea
 \label{SYM:index:S2^2}
 Z_{\mathbb{P}^1\times \mathbb{P}^1\times S^1} = \frac{1}{\left|\mathfrak{W}\right|}\sum_{\fm , \fn}
 \oint_{\cC} & \prod_{i=1}^{\text{rk}(G)} \frac{\rd x_{i}}{2\pi\mathbbm{i}x_{i}} \, e^{\frac{8 \pi^2 \beta}{g_{\text{YM}}^2} \Tr_{\text{F}} ( \fm \fn ) + \mathbbm{i} k \beta \Tr_{\text{F}} ( \fm \fn a )}
 \prod_{\alpha \in G} \bigg( \frac{1 - x^\alpha}{x^{\alpha/2}} \bigg)^{(\alpha(\fm)+ 1) ( \alpha(\fn)+ 1)} \\ &
 \times \prod_{I} \prod_{\rho_I \in \fR_I} \bigg( \frac{x^{\rho_I/2} y^{\nu_I/2}}{1 - x^{\rho_I} y^{\nu_I}} \bigg)^{( \rho_I(\fm) + \nu_I(\fs)  - 1 ) ( \rho_I(\fn)  + \nu_I(\ft)  - 1 )} \, ,
\eea
where $\fm/\fn$ (and $\fs/\ft$) are the gauge (and background) magnetic fluxes on two spheres. The degeneracies come from the Hirzebruch-Riemann-Roch theorem and the classical action comes
from an explicit computation along the lines of section \ref{sebsec:classical}.

We see that the result coincides with \eqref{finalPF} with the replacements
\be
 \fm= \fp_1+\fp_3\, \, \qquad \fn = \fp_2+\fp_4\, , \qquad \fs = \ft_1 +\ft_3\, , \qquad \ft = \ft_2+\ft_4 \, .
\ee
The integral in  \eqref{finalPF} indeed only depends on such combinations. We also expect that, in the non-equivariant limit, the sum over semi-stable equivariant fluxes
reduces to a sum over the two standard fluxes $\fm$ and $\fn$ on  $\mathbb{P}^1\times \mathbb{P}^1$. Such  set of non-equivariant fluxes was indeed used in \cite{Bawane:2014uka},
where the four-dimensional partition function on $\mathbb{P}^1\times \mathbb{P}^1$ has been studied. 

Notice that, in the conventions that we are  using for the background fluxes, what we will call  universal twist \cite{Benini:2015bwz,Bobev:2017uzs} in section
\ref{sec:largeN} corresponds to $\fs=\ft=1$.

\section[Large \texorpdfstring{$N$}{N} limit]{Large $N$ limit}
\label{sec:largeN}

In this section, we analyze the large $N$ limit of some of the topologically twisted indices and other related quantities, finding an interesting structure.

In the large $N$ limit we may expect some simplifications. In particular, instantons are suppressed and the perturbative contribution
to the topologically twisted index of 5D $\cN = 1$ theories discussed in section \ref{sec:localization} becomes exact.
Moreover, we may also expect that the choice of integration contour and the stability conditions on fluxes become simpler at large $N$. In particular, 
we will work under the assumption that, in the large $N$ limit, the fluxes become actually independent.

In the topologically twisted index of 3D $\cN = 2$ theories on $\Sigma_{\fg_1} \times S^1$ \cite{Nekrasov:2014xaa,Benini:2015noa,Benini:2016hjo,Closset:2017zgf} a distinguished role was played by the twisted superpotential  $\wt \cW$ of the two-dimensional theory obtained by compactification on $S^1$ (with infinitely many KK modes). In particular, the partition function can be written as a sum over the set of Bethe vacua 
of the two-dimensional theory, which corresponds to the critical points of  $\wt \cW$ \cite{Nekrasov:2014xaa,Closset:2017zgf}. Moreover, in the large $N$ limit, one particular Bethe vacuum dominates the partition function \cite{Benini:2015eyy}. The natural quantity to consider for the 5D topologically twisted index 
is the SW prepotential $\cF(a)$ of the four-dimensional theory obtained by compactification on $S^1$. There are two reasons to expect that the critical points of $\cF(a)$ play a distinguished role in five dimensions. First,  the partition function of the topologically twisted ${\cal N}=2$ theories in four dimensions  can be split into two contributions, one coming from the integration over the Coulomb plane (usually called the $u$ plane), and the other from the locus where monopoles and dyons become massless, corresponding to the critical points of 
the prepotential $\cF$ \cite{Moore:1997pc}. The integral over the $u$ plane also often reduces to boundary contributions from all the singular points in the moduli space.  Secondly, in the integrable system obtained by placing the four-dimensional theory  on a $\Omega$-background on $\mathbb{R}^4$ with $\epsilon_1 = \hbar$ and $\epsilon_2 = 0$, the Bethe vacua read
\cite{Nekrasov:2009rc}
\be
 \exp \bigg( \mathbbm{i} \frac{\partial \wt\cW_\hbar (a)}{\partial a_j} \bigg) = 1 \, , \qquad j = 1, \ldots , \text{rk}(G) \, ,
\ee
where $\wt\cW_\hbar (a)$ is the twisted superpotential for the two-dimensional effective theory obtained by reducing on the $\Omega$-deformed copy of $\mathbb{R}^2$. We expect that these conditions play a role in the equivariant partition function with $\epsilon_1 = \hbar$ and $\epsilon_2 = 0$.  Since $\wt \cW_\hbar(a)$ has the following expansion as $\hbar \to 0$,
\be
 \wt \cW_\hbar(a) = - \frac{2 \pi}{\hbar} \cF (a) + \ldots \, ,
\ee
we also obtain, in the limit $\hbar \rightarrow 0$, the quantization conditions  (\cf\,\cite[(3.62)]{Nekrasov:2010ka} and \cite[(4.6)]{Luo:2013nxa})
\be
 \label{BAEs:general:F}
 \exp \left( \frac{2 \pi \mathbbm{i}}{\hbar} a_j^{\text{D}} \right) = \exp \bigg( \frac{2 \pi \mathbbm{i}}{\hbar} \frac{\partial \cF(a)}{\partial a_j} \bigg) = 1 \, , \qquad j = 1, \ldots , \text{rk}(G) \, .
\ee
It is possible  that, in taking the non-equivariant limit of the index, the information about the pole configurations of the integrand is partially lost and we need to impose extra conditions following from \eqref{BAEs:general:F}.
For all these reasons, we may think that  the critical points of $\cF(a)$ may play a role in the evaluation of the index. In particular, in analogy with the 3D index, we may expect that, in the large $N$ limit, one particular critical point of $\cF(a)$ dominates the partition function. We will provide some evidence of this picture by evaluating various quantities in the large $N$ limit at the critical point of $\cF(a)$ and showing that they nicely agree with holographic predictions.  Independently from these considerations,  the problem of finding the distribution of critical points of  $\cF(a)$ in the large $N$ limit is interesting in itself and deserves to be studied.

We will then  consider the large $N$ limit of the distribution of critical points of the functional $\cF(a)$ focusing on  two 5D  theories, $\cN = 2$ SYM, which decompactifies to the $\cN = (2,0)$ theory in six dimensions and the $\cN = 1$ $\USp(2N)$ theory with $N_f$ flavors and an antisymmetric matter field, which corresponds to a 5D UV fixed point.
In both cases we find that the value of $\cF(a)$ at its critical points, as a function of flavor fugacities, precisely coincides, in the large $N$ limit, with the partition function of the same theory on $S^5$.
This is in parallel with what was found for the twisted superpotential  of 3D $\cN = 2$ theories in the large $N$ limit \cite{Hosseini:2016tor},  thus re-enforcing the analogy between the two quantities. 

We shall then study the large $N$ limit of the topologically twisted index of 5D $\cN = 1$ theories on $\mathbb{P}^1\times \mathbb{P}^1\times S^1$.
With no effort, we can  replace $\mathbb{P}^1\times \mathbb{P}^1\times S^1$ with the more general manifold $\Sigma_{\fg_2} \times  \Sigma_{\fg_1} \times S^1$ and we will consider this more general case in the following.
The interest of this model is that we can formally dimensionally reduce on $\Sigma_{\fg_2}$  and obtain a three-dimensional theory. In three dimensions  we can guess the form of the partition function in the large $N$ limit and use the results in \cite{Nekrasov:2014xaa,Benini:2015noa,Benini:2016hjo,Closset:2017zgf}. We expect the partition function of the three-dimensional theory to be given as a sum over topological sectors on $\Sigma_{\fg_2}$. We will denote the gauge/flavor magnetic fluxes on $\Sigma_{\fg_1}$ and $\Sigma_{\fg_2}$ by $\fm / \fs$ and $\fn / \ft$, respectively. We also denote by $\Delta$ a complexified chemical potential for the flavor symmetry. In each sector of gauge magnetic flux $\fn$ on $\Sigma_{\fg_2}$ we have a twisted superpotential for the compactified theory satisfying \cite{Nekrasov:2009uh} 
\be
\label{WF}
\frac{\partial \wt \cW(a,\fn, \Delta, \ft)}{\partial \fn_i} = -2 \pi \frac{\partial \cF(a,\Delta)}{\partial a_i} \, , \qquad \frac{\partial \wt \cW(a,\fn, \Delta, \ft)}{\partial \ft} = -2 \pi \frac{\partial \cF(a,\Delta)}{\partial \Delta}    \, .
\ee
One way to determine $\wt \cW$ is to compare the 5D partition function for $\mathbb{P}^1\times \mathbb{P}^1\times S^1$ given in section \ref{sec:localization} with the structure of the topologically twisted index in three dimensions.
This is given by localization as a contour integral of a meromorphic quantity \cite{Benini:2015noa,Benini:2016hjo,Closset:2017zgf}
\be
 \label{twisted3D}
 \sum_{\fm\in \Gamma_\fh} Z^{3\text{D}}_{\text{int}} (a , \fm , \fn) = \sum_{\fm\in \Gamma_\fh} e^{\mathbbm{i} \fm_i \frac{\partial \wt \cW(a , \fn)}{\partial a_i}} Z^{3\text{D}}_{\text{int}} (a, \fm=0, \fn) \, ,
\ee
where $\Gamma_\fh$ is the lattice of gauge magnetic fluxes for the gauge group $G$. By generalizing \eqref{SYM:index:S2^2} from $\mathbb{P}^1\times \mathbb{P}^1$ to $\Sigma_{\fg_1} \times\Sigma_{\fg_2}$ as in \cite{Benini:2016hjo}, and choosing a convenient parameterization for the fluxes, we expect the integrand $Z^{3\text{D}}_{\text{int}} (a , \fm , \fn)$ to be
\bea
 \label{SYM:index:Sigmag2xS1xSigmag10}
 & \bigg( \det\limits_{i j} \frac{\partial^2 \wt \cW (a , \fn)}{\partial a_i \partial a_j} \bigg)^{\fg_1}
 e^{\frac{8 \pi^2 \beta}{g_{\text{YM}}^2} \Tr_{\text{F}} ( \fm \fn ) + \mathbbm{i} k \beta \Tr_{\text{F}} (\fm \fn a) }
 \prod_{\alpha \in G} \bigg( \frac{1 - x^\alpha}{x^{\alpha/2}} \bigg)^{(\alpha(\fm)+ 1 - \fg_1 ) ( \alpha(\fn)+ 1 - \fg_2 )} \\ &
 \times \prod_{I} \prod_{\rho_I \in \fR_I} \bigg( \frac{x^{\rho_I/2} y^{\nu_I/2}}{1 - x^{\rho_I} y^{\nu_I}} \bigg)^{( \rho_I(\fm) + \nu_I(\fs) + \fg_1 - 1 ) ( \rho_I(\fn)  + \nu_I(\ft) + \fg_2 - 1 )} \, .
\eea
We can therefore read off the twisted superpotential
\be
 \label{full:pert:twisted superpotential}
 \begin{aligned}
  \wt \cW^{\text{pert}} (a , \fn , & \Delta , \ft) = - \frac{8 \pi^2 \mathbbm{i} \beta}{g_{\text{YM}}^2} \Tr_{\text{F}} (\fn  a) + \frac{k \beta}{2} \Tr_{\text{F}} (\fn a^2) \\ &
  + \frac{1}{\beta} \sum_{\alpha \in G} \left( \alpha (\fn) + 1 - \fg_2 \right) \left[ \Li_2 ( x^\alpha ) - \frac{1}{2} g_2 \left( - \alpha(\beta a) \right) \right] \\ &
  - \frac{1}{\beta} \sum_I \sum_{\rho_I \in \fR_I} \left( \rho_I(\fn) + \nu_I (\ft) + \fg_2 - 1 \right) \left[ \Li_2 ( x^{\rho_I} y^{\nu_I} ) - \frac{1}{2} g_2 \left( \rho_I(\beta a) + \nu_I (\beta \Delta) \right) \right] \, ,
 \end{aligned}
\ee
that indeed satisfies \eqref{WF}.

The topologically twisted index can then be computed as follows \cite{Nekrasov:2014xaa,Benini:2016hjo}
\bea
 \label{Sigma_g1xM_3:Z}
 & Z^{\text{pert}}_{ \Sigma_{\fg_2} \times ( \Sigma_{\fg_1} \times S^1)} (\fs , \ft , \Delta) =
 \frac{(-1)^{\text{rk}(G)}}{|\mathfrak{W}|} \sum_{\fn \in \Gamma_\fh} \sum_{a = a_{(i)}} Z^{\text{pert}} \big|_{\fm = 0} (a , \fn)
 \bigg( \det\limits_{i j} \frac{\partial^2 \wt \cW^{\text{pert}} (a , \fn)}{\partial a_i \partial a_j} \bigg)^{\fg_1 - 1} \, ,
\eea
where $a_{(i)}$ are the solutions to the Bethe ansatz equations (BAEs)
\be
 \label{BAEs:M3}
 \exp \bigg( \mathbbm{i} \frac{\partial \wt\cW^{\text{pert}} (a , \fn ; \Delta , \ft)}{\partial a_j} \bigg) = 1 \, , \qquad j = 1, \ldots , \text{rk}(G) \, .
\ee
As we will see below, these  BAEs \eqref{BAEs:M3} fix the value of the gauge magnetic fluxes $\fn_i$ in the large $N$ limit, hence, one needs an extra input in order to fix the value of the Coulomb branch parameter $a_i$. Assuming that the right condition to be imposed in the large $N$ limit is \eqref{BAEs:general:F}, we will be able to compute the index for both $\cN = 2$ SYM and the $\USp(2N)$ theory.%
\footnote{An alternative method for evaluating the partition function of the $\USp(2N)$ theory is discussed in appendix \ref{sec:alternative}.}
In particular, in the case of $\cN = 2$ SYM, thought of as compactification of the 6D $\cN = (2,0)$ theory on a circle, the index  computes the elliptic genus of the 2D CFT obtained by compactifying the 6D theory on $\Sigma_{\fg_2} \times \Sigma_{\fg_1}$.  And, indeed, we find the correct Cardy behaviour of the index in the high-temperature limit. This is in parallel with what was found in \cite{Hosseini:2016cyf} for the topologically twisted index of 4D ${\cal N} =1$ SCFTs on $\Sigma_{\fg_1} \times T^2$. 

Finally, we will also consider the large $N$ limit of the twisted superpotential \eqref{full:pert:twisted superpotential} and the distribution of its critical points.
We expect the on-shell value of $\wt \cW^{\text{pert}}$ to compute some physical quantities of the intermediate compactification on  $\Sigma_{\fg_2}$.
We find indeed  that this is the case. For $\cN = 2$ SYM, the critical value of $\wt \cW^{\text{pert}}$ is precisely the trial central charge of the 4D SCFT obtained by compactifying the 6D $(2,0)$ theory on $\Sigma_{\fg_2}$,  computed both in field theory and holographically in \cite{Bah:2011vv,Bah:2012dg}. Quite remarkably, the identification holds for an arbitrary assignment of R-charges for the trial central charge. For the $\USp(2N)$ theory, the critical value of $\wt \cW^{\text{pert}}$, extremized with respect to $\Delta$,  coincides with the $S^3$ free energy of the 3D theory, obtained by compactifying the 5D fixed point on $\Sigma_{\fg_2}$, recently computed holographically in \cite{Bah:2018lyv}.\footnote{The  free energy on $\Sigma_{\fg_2} \times S^3$ as a function of $\Delta$ was explicitly computed in field theory in \cite{Crichigno:2018adf} after the completion of this work and perfectly agrees with the on-shell value of $\wt \cW^{\text{pert}}$ as a function of $\Delta$.}

\subsection[\texorpdfstring{$\cN = 2$}{N=2} super Yang-Mills on \texorpdfstring{$\Sigma_{\fg_2} \times ( \Sigma_{\fg_1} \times S^1)$}{Sigma(g2) x (Sigma(g1) x S**1)}]{$\cN=2$ super Yang-Mills on $\Sigma_{\fg_2} \times ( \Sigma_{\fg_1} \times S^1)$}
\label{subsec:SYM}

A decoupling limit of type IIB string theory on asymptotically locally Euclidean (ALE) spaces predicts the existence of interacting 6D, $\cN = (2,0)$ theories
labelled by an ADE Lie algebra $\fg = (A_{n \geq 1} , D_{n \geq 4} , E_6 , E_7 , E_8)$ \cite{Witten:1995zh}.
The $A_{N-1}$ type can also be realized as the low-energy description of the worldvolume theory of $N$ coincident M5-branes in M-theory \cite{Strominger:1995ac}.
A direct formulation of the $(2,0)$ theory has been a long standing problem. However, it has been argued in \cite{Douglas:2010iu,Lambert:2010iw,Bolognesi:2011nh} that  the $(2, 0)$ theory on a circle $S^1_{(6)}$ of radius $R_6$ is equivalent to the  five-dimensional $\cN = 2$ supersymmetric Yang-Mills (SYM) theory, whose coupling constant is identified with the $S^1_{(6)}$ radius by 
\be
 R_6 = \frac{g_{\text{YM}}^2}{8 \pi^2} \, .
\ee

We are interested in computing the partition function of the $(2,0)$ theory on $\Sigma_{\fg_2} \times ( \Sigma_{\fg_1} \times T^2)$, partially topologically twisted on $\Sigma_{\fg_2} \times \Sigma_{\fg_1}$.
Given the above relation between the $(2,0)$ theory and 5D maximally SYM, and considering the torus $T^2=S^1\times S^1_{(6)}$, this is equivalent to compute the twisted partition function of $\cN=2$ SYM on $\Sigma_{\fg_2} \times ( \Sigma_{\fg_1} \times S^1)$. A further reduction on $S^1$ gives a four-dimensional theory. Recalling that the length of $S^1$ is $\beta$, we see that the (complexified) gauge coupling of the four-dimensional theory can be correctly identified with the modular parameter of the torus $T^2=S^1\times S^1_{(6)}$, 
\be  \tau = \frac{4 \pi \mathbbm{i} \beta}{g_{\text{YM}}^2 } =   \frac{\mathbbm{i} \beta}{2\pi R_6}\, . \ee

The twisted compactification of the $(2,0)$ theory on $\Sigma_{\fg_1} \times \Sigma_{\fg_2}$ gives rise to an $\cN = (0 , 2)$ SCFT in two dimensions \cite{Benini:2013cda}. The holonomy group of $\Sigma_{\fg_1} \times \Sigma_{\fg_2}$ is $\SO(2)_1 \times \SO(2)_2$.
In order to preserve $\cN = (0 , 2)$ supersymmetry in two dimensions we turn on a background Abelian gauge field
coupled to an $\SO(2)^2$ subgroup of $\SO(5)_R$, embedded block-diagonally.
The right-moving trial central charge $c_r (\Delta)$ and the gravitational anomaly $k = c_r - c_l$ for this class of theories were computed in \cite{Benini:2013cda} and, at large $N$, they can be rewritten as (see also appendix \ref{sec:centralcharges(2,0)})
\bea
 \label{cr:cl:6D(2,0):largeN}
 c_r (\fs , \ft, \Delta) \approx c_l (\fs , \ft, \Delta)& \approx \frac{2 N^3 \beta^2}{(2 \pi)^2} \left[ \Delta_1 \Delta_2 ( \ft_1 \fs_2 + \ft_2 \fs_1 ) + ( \Delta_1 \fs_2 + \Delta_2 \fs_1 ) ( \Delta_1 \ft_2 + \Delta_2 \ft_1 ) \right] \\ &
 \approx \frac{N^3 \beta^2}{(2\pi)^2}  \sum_{\varsigma , \varrho = 1}^{2} \ft_\varsigma \fs_\varrho \frac{\partial^2 (\Delta_1\Delta_2)^2}{\partial \Delta_\varsigma\partial \Delta_\varrho} \, , 
\eea
where we introduced the democratic chemical potentials and fluxes for the $\SO(2)^2$ subgroup of $\SO(5)_R$ symmetry
\bea
 \label{democratic:SYM}
 & \Delta_1 = \Delta \, , \qquad \Delta_2 = \frac{2 \pi}{\beta} - \Delta \, , \qquad
 \sum_{\varsigma = 1}^{2} \Delta_\varsigma = \frac{2 \pi}{\beta} \, , \\ &
 \fs_1 = \fs \, , \qquad \fs_2 = 2 ( 1- \fg_1 ) - \fs \, , \qquad
 \sum_{\varsigma = 1}^{2} \fs_\varsigma = 2 ( 1- \fg_1 ) \, , \\ &
 \ft_1 = \ft \, , \qquad \ft_2 = 2 ( 1- \fg_2 ) - \ft \, , \qquad
 \sum_{\varsigma = 1}^{2} \ft_\varsigma = 2 ( 1- \fg_2 ) \, .
\eea

The partition function of the $(2,0)$ theory on $\Sigma_{\fg_2} \times ( \Sigma_{\fg_1} \times T^2)$ is just the elliptic genus of the two-dimensional CFT.
Thus we expect, in the high-temperature limit $\tilde \beta \rightarrow 0$, where $\tilde \beta \equiv - 2 \pi \mathbbm{i} \tau$ is a \emph{fictitious} inverse temperature,%
\footnote{The torus partition function at a given $\tau$ corresponds to a thermal ensemble while the elliptic genus is only counting extremal states. Therefore, the temperature represented by $\im \tau$ is fictitious.}
the partition function to have a Cardy behaviour
\be
\label{expect:N=1*:SYM:Cardy}
 \log Z (\fs , \ft , \Delta) \approx \frac{\mathbbm{i} \pi}{12 \tau} c_r (\fs , \ft , \Delta) \, .
\ee
We will work in the 't Hooft limit
\be
 N \gg 1 \quad \text{ with } \quad \lambda = \frac{g_{\text{YM}}^2 N}{\beta} = \text{fixed} \, ,
\ee
for which the instanton contributions to the partition function are exponentially suppressed. The high-temperature limit of the partition function corresponds to  large $\lambda$. 

\subsubsection{Effective prepotential at large $N$}
\label{subsubsec:SW:SYM}

The effective SW prepotential \eqref{full:pert:prepotential} of $\cN = 2$ SYM can be written as
\bea
 \label{SYM:prepotential}
 \cF (a , \Delta) & = \sum_{i = 1}^{N} \cF^{\text{cl}} (a_i) + \sum_{i \neq j}^{N} \cF^{1\text{-loop}} ( a_i - a_j ) - \sum_{i , j = 1}^{N} \cF^{1\text{-loop}} ( a_i - a_j + \Delta) \\ &
 = \frac{2 \pi \mathbbm{i} \beta}{g_{\text{YM}}^2} \sum_{i = 1}^{N} a_i^2 + \frac{\mathbbm{i}}{2 \pi \beta^2}
 \sum_{i \neq j}^{N} \Li_3 (e^{\mathbbm{i} \beta (a_i - a_j)}) - \frac{\mathbbm{i}}{2 \pi \beta^2} \sum_{i , j = 1}^{N}\Li_3 (e^{\mathbbm{i} \beta (a_i - a_j + \Delta)}) \\ &
 - \frac{1}{4 \pi \beta^2} \sum_{i \neq j}^{N} g_3 \left( \beta ( a_j - a_i ) \right) + \frac{1}{4 \pi \beta^2} \sum_{i , j = 1}^{N} g_3 \left( \beta( a_i - a_j + \Delta ) \right) \, .
\eea
The BAEs \eqref{BAEs:general:F} are then given by
\bea
 \label{BAEs:SYM}
 - \frac{8 \pi^2 N}{\lambda} a_i & = \frac{\mathbbm{i}}{\beta} \sum_{j = 1}^{N} \left[ \Li_2 (e^{\mathbbm{i} \beta (a_i - a_j)}) - \Li_2 (e^{- \mathbbm{i} \beta (a_i - a_j)})
 - \Li_2 (e^{\mathbbm{i} \beta (a_i - a_j + \Delta)}) + \Li_2 (e^{- \mathbbm{i} \beta (a_i - a_j - \Delta)}) \right]  \\ &
 + \frac{\mathbbm{i}}{2 \beta} \sum_{j = 1}^{N} \left[ - g_2 \left( \beta ( a_j - a_i ) \right) + g_2 \left( \beta ( a_i - a_j ) \right) \right] \\ &
 + \frac{\mathbbm{i}}{2 \beta} \sum_{j = 1}^{N} \left[ g_2 \left( \beta( a_i - a_j + \Delta ) \right) - g_2 \left( \beta( a_j - a_i + \Delta ) \right) \right]
 \, .
\eea
In the strong 't Hooft coupling $\lambda \gg 1$ the eigenvalues are pushed apart, \ie\;$| \im (a_i - a_j) | \gg 1$,
and \eqref{BAEs:SYM} can be approximated as 
\bea
 \label{approx:BAEs:SYM}
 \frac{8 \pi^2 N}{\lambda} a_k & \approx \frac{\mathbbm{i}}{2} \Delta ( 2 \pi - \beta \Delta ) \sum_{j = 1}^{N} \sign \left( \im (a_k - a_j) \right) 
 \, .
\eea
The $\sign$ function could be replaced by $\sign( i - j) $ if the eigenvalues $a_i$ are ordered by increasing imaginary part. We thus find the solution
\be
 \label{SYM:sol:BAEs}
 a_k = \frac{\mathbbm{i} \lambda}{16 \pi^2 N} \left[ \Delta ( 2 \pi - \beta \Delta ) (2 k - N - 1) 
 \right] \, .
\ee
It is also interesting to see what the value of the effective SW prepotential \eqref{SYM:prepotential} at the solution \eqref{SYM:sol:BAEs} is.
In the strong 't Hooft coupling $\lambda \gg 1$ we find that
\bea
 \label{SYM:F:approx}
 \cF (a , \Delta) & \approx \frac{2 \pi \mathbbm{i} N}{\lambda} \sum_{i = 1}^{N} a_i^2 - \frac{1}{4 \pi \beta^2}
 \sum_{i , j = 1}^{N} \left[ g_3 \left( \beta ( a_j - a_i ) \right) + g_3 \left( \beta( a_i - a_j + \Delta ) \right) \right] \sign \left( \im (a_i - a_j) \right) \\ &
 = \frac{2 \pi \mathbbm{i} N}{\lambda} \sum_{i = 1}^{N} a_i^2 + \frac{1}{8 \pi} \Delta ( 2 \pi - \beta \Delta ) \sum_{i , j = 1}^{N} (a_i - a_j) \sign \left( \im (a_i - a_j) \right) \\ &
 = - \sum_{j = 1}^{N} \left[ \frac{2 \pi \mathbbm{i} N}{\lambda} \im(a_j) - \frac{\mathbbm{i}}{4 \pi} \Delta ( 2 \pi - \beta \Delta ) ( 2 j - N - 1) \right] \im ( a_j ) \, .
\eea
In order to get the last equality we used the relation
\be
 \sum_{i , j = 1}^{N} | \im ( a_ i - a_j ) | = 2 \sum_{j = 1}^{N} (2 j - 1 - N) \im ( a_j ) \, .
\ee
Plugging the solution \eqref{SYM:sol:BAEs} back into \eqref{SYM:F:approx}, we obtain 
\be
 \label{SYM:F:on-shell}
 \cF ( \Delta ) \approx \frac{\mathbbm{i} \beta g_{\text{YM}}^2 N^3}{384 \pi^3} \left( \Delta_1 \Delta_2 \right)^2 \, .
\ee
In order to obtain \eqref{SYM:F:on-shell} we used
\be
 \label{sum:abs:sign}
 \sum_{k = 1}^{N} (2 k - N - 1)^2 = \frac{1}{3} \left( N^3 - N \right) \, .
\ee

Remarkably, the BAEs \eqref{approx:BAEs:SYM} and the SW prepotential in the large $N$ limit \eqref{SYM:F:approx}, \eqref{SYM:F:on-shell} are \emph{identical}
to matrix model saddle point equations and free energy for the path integral on $S^5$ found in \cite{Minahan:2013jwa} (\cf\,\cite[(4.13), (4.16) and (4.17)]{Minahan:2013jwa}).%
\footnote{One needs to set $\Delta_1 = \pi \left( 1 + \frac{2 \mathbbm{i}}{3} m_{\text{there}} \right)$ and $\Delta_2 = \pi \left( 1 - \frac{2 \mathbbm{i}}{3} m_{\text{there}} \right)$.}

\subsubsection{Effective twisted superpotential at large $N$}
\label{subsubsec:tildeW:SYM}

The effective twisted superpotential of the theory reads 
\bea
 \label{Wt:SYM:pert}
 \wt \cW (a , \fn , \Delta , \ft) & = - \frac{8 \pi^2 \mathbbm{i} \beta}{g_{\text{YM}}^2} \sum_{i = 1}^{N} \fn_i a_i \\ &
 + \frac{1}{\beta} \sum_{i \neq j}^{N} ( \fn_i - \fn_j + 1 - \fg_2 ) \left[ \Li_2 ( e^{\mathbbm{i} \beta (a_i - a_j)} ) - \frac{1}{2} g_2 \left( - \beta (a_i - a_j) \right) \right] \\ &
 - \frac{1}{\beta} \sum_{i , j = 1}^{N} \left( \fn_i - \fn_j + \ft + \fg_2 - 1 \right) \left[ \Li_2 ( e^{\mathbbm{i} \beta (a_i - a_j + \Delta)} ) - \frac{1}{2} g_2 \left( \beta (a_i - a_j + \Delta) \right) \right] \, .
\eea
Let us try to find a solution to the BAEs \eqref{BAEs:M3}. For the twisted superpotential \eqref{Wt:SYM:pert}, they are given by%
\footnote{The BAEs actually read  $\frac{\partial \wt\cW^{\text{pert}}}{\partial a_j} = 2 \pi \fl_j$ where $\fl_j \in \mathbb{Z}$ are angular ambiguities.
Since $\fl_j$'s are generically of order one they are negligible in the final solution and we set them to zero in the following.}
\be
 \begin{aligned}
  \label{SYM:BAEs:n:a}
  & \frac{8 \pi^2 \mathbbm{i} \beta}{g_{\text{YM}}^2} \frac{\fn_i}{1 - \fg_2} = \mathbbm{i} \sum_{j (\neq i)}^{N} \left[ \left( \frac{\fn_i - \fn_j}{1 - \fg_2} + 1 \right) \Li_1 (e^{\mathbbm{i} \beta ( a_i - a_j )})
  - \left( \frac{\fn_j - \fn_i}{1 - \fg_2} + 1 \right) \Li_1 (e^{- \mathbbm{i} \beta ( a_i - a_j )}) \right] \\ &
  + \frac12 \sum_{j (\neq i)}^{N} \left[ \left( \frac{\fn_i - \fn_j}{1 - \fg_2} + 1 \right) g_1 \left( - \beta ( a_i - a_j ) \right)
  - \left( \frac{\fn_j - \fn_i}{1 - \fg_2} + 1 \right) g_1 \left( \beta ( a_i - a_j ) \right) \right] \\ &
  - \mathbbm{i} \sum_{j = 1}^{N} \left[ \left( \frac{\fn_i - \fn_j + \ft}{1 - \fg_2} - 1 \right) \Li_1 ( e^{\mathbbm{i} \beta (a_i - a_j + \Delta)} )
  - \left( \frac{\fn_j - \fn_i + \ft}{1 - \fg_2} - 1 \right) \Li_1 ( e^{- \mathbbm{i} \beta (a_i - a_j - \Delta)} ) \right] \\ &
  + \frac{1}{2} \sum_{j = 1}^{N} \left[ \left( \frac{\fn_i - \fn_j + \ft}{1 - \fg_2} - 1 \right) g_1 \left( \beta (a_i - a_j + \Delta) \right)
  - \left( \frac{\fn_j - \fn_i + \ft}{1 - \fg_2} - 1 \right) g_1 \left( - \beta (a_i - a_j - \Delta) \right) \right] .
 \end{aligned}
\ee
In the strong 't Hooft coupling limit, \ie\;$\lambda \gg 1$, \eqref{SYM:BAEs:n:a} can be approximated as
\be
 \label{SYM:sol:BAEs:tildeW}
 \fn_i \approx \frac{\mathbbm{i} \lambda \beta}{16 \pi^2 N} (\Delta_1 \ft_2 + \Delta_2 \ft_1) \sum_{j (\neq i)}^{N} \sign \left( \im ( a_i - a_j ) \right) \, .
\ee
The $\sign$ function in \eqref{SYM:sol:BAEs:tildeW} could be replaced by $\sign( i - j) $ if the eigenvalues $a_i$ are ordered by increasing imaginary part.
We thus find that
\be
 \label{SYM:ni}
 \fn_k \approx \frac{\mathbbm{i} \lambda \beta}{16 \pi^2 N} ( \Delta_1 \ft_2 + \Delta_2 \ft_1 ) (2 k - N - 1) \, .
\ee
Note that the above result is neither \emph{real} nor \emph{integer}.
This is peculiar to the limit $\lambda \gg 1$. Since the $\fn_k$'s are large we treat them effectively as a continuous variable.
We also consider the above result as a complex saddle point contribution to the partition function.

The twisted superpotential \eqref{Wt:SYM:pert} at strong 't Hooft coupling limit can be approximated as
\be
 \begin{aligned}
  \frac{\wt \cW (a , \fn , \Delta , \ft)}{1 - \fg_2} & \approx - \frac{8 \pi^2 \mathbbm{i} N}{\lambda (1 - \fg_2)} \sum_{i = 1}^{N} \fn_i a_i \\ &
  - \frac{1}{2 \beta} \sum_{i , j = 1}^{N} \left( \frac{\fn_i - \fn_j}{1 - \fg_2} + 1 \right) g_2 \left( - \beta (a_i - a_j) \right) \sign \left( \im (a_i - a_j ) \right) \\ &
  + \frac{1}{2 \beta} \sum_{i , j = 1}^{N} \left( \frac{\fn_i - \fn_j + \ft}{1 - \fg_2} - 1 \right) g_2 \left( \beta (a_i - a_j + \Delta) \right) \sign \left( \im (a_i - a_j ) \right) \, .
 \end{aligned}
\ee
This can be further simplified to
\be
 \label{SYM:tildeW:simplified}
 \begin{aligned}
  \wt \cW (a , \fn , \Delta , \ft) & \approx \frac{8 \pi^2 N}{\lambda} \sum_{j = 1}^{N} \im ( a_j ) \left[ \fn_j - \frac{\mathbbm{i} \lambda \beta}{16 \pi^2 N} ( \Delta_1 \ft_2 + \Delta_2 \ft_1 ) ( 2 j - N - 1 ) \right] \\ &
 - \frac{\beta}{4} \Delta_1 \Delta_2 \sum_{i , j = 1}^{N} ( \fn_i - \fn_j ) \sign \left( \im (a_i - a_j ) \right) \, .
 \end{aligned}
\ee
Plugging \eqref{SYM:ni} back into \eqref{SYM:tildeW:simplified}, all the dependence on $a_i$ goes away and we are left with
\be
 \label{SYM:tildeW:on-shell:n}
 \begin{aligned}
  \wt \cW (\Delta , \ft) & \approx - \frac{\beta}{4} \Delta_1 \Delta_2 \sum_{i , j = 1}^{N} ( \fn_i - \fn_j ) \sign \left( \im (a_i - a_j ) \right) \\ &
  = - \frac{\mathbbm{i} \beta g_{\text{YM}}^2 N^3}{96 \pi^2} \Delta_1 \Delta_2 ( \Delta_1 \ft_2 + \Delta_2 \ft_1 ) \, .
 \end{aligned}
\ee
This can be more elegantly rewritten as
\be
 \label{index theorem:tildeW:F}
 \wt \cW (\Delta , \ft) \approx - 2 \pi \sum_{\varsigma = 1}^{2} \ft_\varsigma \frac{\partial \cF(\Delta)}{\partial \Delta_\varsigma} \, ,
\ee
where $\cF(\Delta)$ is given in \eqref{SYM:F:on-shell}.

\paragraph*{$\wt \cW(\Delta , \ft)$ and the 4D central charge.}

Remarkably, we find the following relation between the twisted superpotential \eqref{SYM:tildeW:on-shell:n} and the conformal anomaly coefficient $a (\Delta , \ft)$
of the four-dimensional $\cN=1$ theory that is obtained by compactifying the 6D $\cN = (2,0)$ theory on $\Sigma_{\fg_2}$  \cite{Bah:2011vv,Bah:2012dg},
\be
 \label{tildeW:a central charge}
 \wt \cW (\Delta , \ft) \approx - \frac{ 8 \pi^2}{27 \beta \tau} a (\Delta , \ft) \, ,
\ee
where $\tau$ is the four-dimensional coupling constant.

Indeed, the central charge of the 4D theory, at large $N$, can be written as
\be
 \label{a4D}
 a (\Delta , \ft) \approx -\frac{9 N^3}{128} \sum_{\varsigma=1}^2 \ft_\varsigma \frac{\partial (\hat \Delta_1\hat\Delta_2)^2}{\partial \hat\Delta_\varsigma} \, ,
\ee
where we used $\hat\Delta_\varsigma = \beta \Delta_\varsigma / \pi$, satisfying $\hat \Delta_1+\hat \Delta_2=2$, to parameterize a trial R-symmetry of the 4D $\cN=1$ theory.
\eqref{a4D} can be easily derived from the results in \cite{Bah:2011vv,Bah:2012dg} (\cf\,for example \cite[(2.22)]{Bah:2012dg}). It can be also more straightforwardly 
derived as in appendix \ref{sec:centralcharges(2,0)}.

Curiously, the same relation between the twisted superpotential and the central charge in \eqref{tildeW:a central charge} was found for a class of $\cN = 1$ gauge theories on $S^2 \times T^2$, with a partial topological twist along $S^2$ \cite{Hosseini:2016cyf}.

\subsubsection{Partition function at large $N$}
\label{subsubsec:logZ:SYM}

The topologically twisted index of 5D $\cN = 2$ SYM on $\Sigma_{\fg_2} \times ( \Sigma_{\fg_1} \times S^1 )$ reads
\bea
 \label{SYM:index:Sigmag2xS1xSigmag1}
 Z (y , \fs , \ft) = \frac{1}{N!} \sum_{\{\fm , \fn\} \in \bZ^N} & \oint_{\cC} \prod_{i = 1}^{N} \frac{\rd x_i}{2 \pi \mathbbm{i} x_i}
 e^{\frac{8 \pi^2 \beta}{g_{\text{YM}}^2} ( \fm_i - \fm_j ) ( \fn_i - \fn_j )}
 \bigg( \det\limits_{i j} \frac{\partial^2 \wt \cW (a , \fn)}{\partial a_i \partial a_j} \bigg)^{\fg_1} \\ &
 \times \prod_{i \neq j}^{N} \bigg( \frac{1 - x_i / x_j}{\sqrt{x_i / x_j}} \bigg)^{( \fm_i - \fm_j + 1 - \fg_1 ) ( \fn_i - \fn_j + 1 - \fg_2 )} \\ &
 \times \prod_{i , j = 1}^{N} \bigg( \frac{\sqrt{x_i y / x_j}}{1 - x_i y / x_j} \bigg)^{( \fm_i - \fm_j + \fs + \fg_1 - 1 ) ( \fn_i - \fn_j + \ft + \fg_2 - 1 )} \, .
\eea
This can be evaluated using \eqref{Sigma_g1xM_3:Z}. We can write
\bea
 Z (y , \fs , \ft) & = \frac{(-1)^{N}}{N!}
 \bigg( \frac{\sqrt{y}}{1 - y} \bigg)^{N ( \fg_1 + \fs - 1 ) ( \fg_2 + \ft - 1 )}
 \sum_{\fn \in \bZ^N} \sum_{a = a_{(i)}} \bigg( \det\limits_{i j} \frac{\partial^2 \wt \cW (a , \fn)}{\partial a_i \partial a_j} \bigg)^{\fg_1 - 1} \\ &
 \times \prod_{i \neq j}^{N} \bigg( \frac{1 - x_i / x_j}{\sqrt{x_i / x_j}} \bigg)^{( 1 - \fg_1 ) ( \fn_i - \fn_j + 1 - \fg_2 )}
 \bigg( \frac{\sqrt{x_i y / x_j}}{1 - x_i y / x_j} \bigg)^{( \fs + \fg_1 - 1 ) ( \fn_i - \fn_j + \ft + \fg_2 - 1 )} \, .
\eea
We are interested in the logarithm of the partition function in the strong 't Hooft coupling limit.
The only piece which survives in this limit is given by
\bea
 \log Z^{(1)} & \equiv \log \prod_{i \neq j}^{N} \bigg( \frac{1 - x_i / x_j}{\sqrt{x_i / x_j}} \bigg)^{( 1 - \fg_1 ) ( \fn_i - \fn_j + 1 - \fg_2 )}
 \bigg( \frac{\sqrt{x_i y / x_j}}{1 - x_i y / x_j} \bigg)^{( \fs + \fg_1 - 1 ) ( \fn_i - \fn_j + \ft + \fg_2 - 1 )} \\ &
 = - \sum_{i \neq j}^{N} ( 1 - \fg_1 ) ( \fn_i - \fn_j + 1 - \fg_2 ) \left[ \Li_1 ( e^{\mathbbm{i} \beta ( a_i - a_j )} ) - \frac{\mathbbm{i}}{2} g_1 \left( - \beta ( a_i - a_j ) \right) \right] \\ &
 + \sum_{i \neq j}^{N} ( \fs + \fg_1 - 1 ) ( \fn_i - \fn_j + \ft + \fg_2 - 1 ) \left[ \Li_1 ( e^{\mathbbm{i} \beta ( a_i - a_j + \Delta )} ) + \frac{\mathbbm{i}}{2} g_1 \left( \beta ( a_i - a_j + \Delta ) \right) \right] \, .
\eea
In the strong 't Hooft coupling limit it can be approximated as
\bea
 \label{SYM:logZ:total}
 \log Z^{(1)} & \approx - \frac{\mathbbm{i} \beta}{4} \sum_{i \neq j}^{N} \left[ ( a_i - a_j ) ( \ft_1 \fs_2 + \ft_2 \fs_1 ) + ( \fn_i - \fn_j ) ( \Delta_1 \fs_2 + \Delta_2 \fs_1 ) \right] \sign \left( \im ( a_i - a_j ) \right) \\ &
 = \frac{\beta}{2} \sum_{j = 1}^{N} (2 j - 1 - N) \left[ ( \ft_1 \fs_2 + \ft_2 \fs_1 ) \im ( a_j ) - \mathbbm{i} ( \Delta_1 \fs_2 + \Delta_2 \fs_1 ) \fn_j \right] \, ,
\eea
where we assumed that the eigenvalues $a_i$ are ordered by increasing imaginary part, and used the relation
\be
 \sum_{i , j = 1}^{N} ( \fn_i - \fn_j ) \sign ( i - j ) = 2 \sum_{j = 1}^{N} ( 2 j - N - 1 ) \fn_i \, .
\ee
Plugging the solutions \eqref{SYM:sol:BAEs} and \eqref{SYM:ni} back into \eqref{SYM:logZ:total}, we find that
\bea
 \label{SYM:Final:logZ:index theorem}
 \log Z & \approx \frac{\beta g_{\text{YM}}^2 N^3}{96 \pi^2}  \left[ \Delta_1 \Delta_2 ( \ft_1 \fs_2 + \ft_2 \fs_1 ) + ( \Delta_1 \fs_2 + \Delta_2 \fs_1 ) ( \Delta_1 \ft_2 + \Delta_2 \ft_1 ) \right] \\ &
 = \mathbbm{i} \sum_{\varsigma = 1}^2 \fs_\varsigma \frac{\partial \wt \cW (\ft , \Delta)}{\partial \Delta_\varsigma}
 = - 2 \mathbbm{i} \pi \sum_{\varsigma , \varrho = 1}^2 \fs_\varsigma \ft_\varrho \frac{\partial^2 \cF (\Delta)}{\partial \Delta_\varsigma \partial \Delta_\varrho} \\ &
 = \frac{g_{\text{YM}}^2N^3}{192 \pi^2   \beta} \sum_{\varsigma , \varrho = 1}^2 \fs_\varsigma \ft_\varrho \frac{\partial^2 (\beta \Delta_1 \Delta_2)^2}{\partial \Delta_\varsigma \partial \Delta_\varrho} \, ,
\eea
where $\wt \cW (\ft , \Delta)$ is given in \eqref{SYM:tildeW:on-shell:n} and we used \eqref{index theorem:tildeW:F} in writing the last equality.
\eqref{SYM:Final:logZ:index theorem} can be also expressed in terms of the trial right-moving central charge of the 2D $\cN = (0,2)$ SCFT, see \eqref{cr:cl:6D(2,0):largeN}, as
\be
 \label{N=1*:SYM:Cardy}
 \log Z (\fs , \ft , \Delta) \approx \frac{\mathbbm{i} \pi}{12 \tau} c_r (\fs , \ft , \Delta) \approx - \frac{8 \mathbbm{i} \pi^2}{27 \beta \tau} \sum_{\varsigma = 1}^2 \fs_\varsigma \frac{\partial a ( \ft , \Delta)}{\partial \Delta_\varsigma} \, .
\ee
In writing the second equality we used the relation \eqref{tildeW:a central charge}.
In \cite{Hosseini:2016cyf},  the very same Cardy behaviour \eqref{N=1*:SYM:Cardy} of the partition function in the high-temperature limit has been proved for a class of $\cN = 1$ gauge theories on $S^2 \times T^2$, with a partial topological twist along $S^2$.

\subsubsection{Counting states and the $\cI$-extremization principle}
\label{I-extremization principle}

The topologically twisted index of the 6D $(2,0)$ theory on $ \Sigma_{\fg_2} \times \Sigma_{\fg_1} \times T^2$ can be interpreted as a trace over a Hilbert space of states on $\Sigma_{\fg_2} \times \Sigma_{\fg_1} \times S^1$:
\bea
 \label{EG}
 Z (\fs, \ft, \Delta) = \Tr_{\Sigma_{\fg_2} \times \Sigma_{\fg_1} \times S^1}  (-1)^F q^{H_L}  y^J \, ,
\eea
where $q=e^{2 \pi \mathbbm{i} \tau}$, $y=e^{\mathbbm{i} \beta \Delta}$ and the Hamiltonian $H_L$ on $\Sigma_{\fg_2} \times \Sigma_{\fg_1} \times S^1$ explicitly depends on the magnetic fluxes $\fs_\varsigma , \ft_\varsigma$ $(\varsigma = 1, 2)$.
The number of supersymmetric ground states $d_{\text{micro}}$ with momentum $n$ and electric charge $\fq$ under the Cartan subgroup of the flavor symmetry commuting with the Hamiltonian -- in the microcanonical ensemble -- is then given by
the Fourier transform of \eqref{EG} with respect to $(\tau , \Delta)$:
\be
 \label{ch:1:micro:density}
 d_{\text{micro}} (\fs , \ft ,n, \fq) = - \frac{\mathbbm{i} \beta}{(2 \pi)^2} \int_{\mathbbm{i} \mathbb{R}} \rd \tilde \beta \int_{0}^{\frac{2\pi}{\beta}} \rd \Delta \, Z (\fs, \ft, \Delta) \, e^{ \tilde \beta n - \mathbbm{i} \beta  \Delta \fq} \, ,
\ee
where $\tilde \beta=-2 \pi \mathbbm{i} \tau$ and the corresponding integration is over the imaginary axis.

In the  limit of large charges, we may use the saddle point approximation. Consider for simplicity $\fq=0$.
The number of supersymmetric ground states $d_{\text{micro}}$ with charges $(\fs ,\ft , n)$ can be obtained by extremizing
\be
 \label{SBH}
 \cI_{\text{SCFT}} (\tilde \beta , \Delta) \equiv \log Z (\fs, \ft, \Delta) + \tilde \beta n
 = \frac{N^3}{24 \tilde \beta} \sum_{\varsigma , \varrho = 1}^2 \fs_\varsigma \ft_\varrho \frac{\partial^2 (\beta \Delta_1 \Delta_2)^2}{\partial \Delta_\varsigma \partial \Delta_\varrho} + n \tilde \beta \, ,
\ee
with respect to $\Delta$ and $\tilde \beta$, \ie 
\be
 \frac{\partial \cI (\tilde \beta , \Delta)}{\partial \Delta} = 0 \, , \qquad \frac{\partial \cI (\tilde \beta , \Delta)}{\partial \tilde \beta} = 0 \, ,
\ee
and evaluating it at its extremum
\be
 \log d_{\text{micro}} (\fs , \ft ,n, 0) =  \cI \big|_{\text{crit}} (\fs , \ft, n) \, .
\ee
Given \eqref{N=1*:SYM:Cardy}, we see that the extremization with respect to $\Delta$  is the $c$-extremization principle \cite{Benini:2012cz,Benini:2013cda}
and sets the trial right-moving central charge $c_r (\fs , \ft , \Delta)$ to its \emph{exact} value $c_{\text{CFT}} \approx c_r \approx c_l$ in the IR.
For the case at our disposal, \eqref{cr:cl:6D(2,0):largeN} has a critical point at
\be
 \label{crit:Delta}
 \bar \Delta = - \frac{2 \pi}{\beta} \frac{\ft_1 \fs_2 + \ft_2 \fs_1 - \ft_1 \fs_1}{\fs_1 ( \ft_1 - 2 \ft_2 ) + \fs_2 ( \ft_2 - 2 \ft_1 )} \, ,
\ee
and its value at $\bar \Delta$ reads
\be
 \label{exact:cr}
 c_{\text{CFT}} (\fs , \ft) \approx - 2 N^3 \frac{\ft_1^2 \fs_2^2 + \ft_1 \ft_2 \fs_1 \fs_2 + \ft_2^2 \fs_1^2}{\fs_1 ( \ft_1 - 2 \ft_2) + \fs_2 ( \ft_2 - 2 \ft_1 )} \, .
\ee
Extremizing $\cI (\tilde \beta , \Delta)$ with respect to $\tilde\beta$ yields
\be
 \label{crit:tbeta}
 \bar {\tilde \beta} (\fs , \ft , n)
 = \pi \sqrt{\frac{c_{\text{CFT}} (\fs , \ft)}{6 n} } \, .
\ee
Plugging back \eqref{crit:Delta} and \eqref{crit:tbeta} into $\cI (\tilde \beta , \Delta)$, we find that
\bea\label{Cardys3}
 \cI_{\text{SCFT}}\big|_{\text{crit}} (\fs , \ft , n)  = 2 \pi \sqrt{ \frac{ n\, c_{\text{CFT}}(\fs , \ft)}{6}} \, .
\eea
This is obviously nothing else than Cardy formula \cite{Cardy:1986ie}.

This procedure corresponds to the $\cI$\emph{-extremization} principle used for BPS black holes (strings) \cite{Benini:2015eyy,Benini:2016rke,Hosseini:2016cyf,Hosseini:2018qsx} in the context of the AdS$_{4 (5)}$ /CFT$_{3 (4)}$ correspondence 
and contains two basic pieces of information:
\begin{enumerate}
 \item \label{step1} extremizing the index unambiguously determines the exact R-symmetry of the SCFT in the IR;
 \item the value of the index at its extremum is the (possibly regularized) number of ground states.
\end{enumerate}
We will apply this counting to  supergravity black strings and black holes in section \ref{sec:4D domain-walls}.

\subsection[\texorpdfstring{$\USp(2N)$}{USp(2N)} theory with matter on \texorpdfstring{$\Sigma_{\fg_2} \times ( \Sigma_{\fg_1} \times S^1)$}{Sigma(g2) x (Sigma(g1) x S**1)}]{$\USp(2N)$ theory with matter on $\Sigma_{\fg_2} \times ( \Sigma_{\fg_1} \times S^1)$}
\label{subsec:USp(2N)}

In this section we focus on gauge theories with a conjectured massive type IIA dual \cite{Intriligator:1997pq} (see also \cite{Brandhuber:1999np,Bergman:2012kr,Morrison:1996xf,Seiberg:1996bd}) and compute their partition function at large $N$.
The dual supergravity backgrounds have a warped AdS$_6 \times S^4$ geometry.
The stringy root of this theories is in type I' string theory as strongly coupled microscopic theories on the intersection of $N$ D4-branes and $N_f$ D8-branes and orientifold planes.
The worldvolume theory on the $N$ D4-branes plus their images is a $\USp(2N)$ gauge theory with $N_f$ hypermultiplets in the fundamental representation and one hypermultiplet $\mathcal{A}$ in the antisymmetric representation of $\USp(2N)$.
In addition to the $\SU(2)$ R-symmetry the theory has an $\SU(2)_M \times \SO(2 N_f) \times \U(1)_I$ global symmetry:
$\SU(2)_M$ acts on $\cA$ as a doublet,
$\SO(2 N_f)$ is the flavor symmetry associated to the fundamental hypermultiplets, and $\U(1)_I$ is the topological symmetry associated to the conserved instanton number current $j = \ast \Tr (F \wedge F)$.
At the fixed point, the $\SO(2N_f) \times \U(1)_I$ part of the symmetry is enhanced non-perturbatively to an exceptional group $\E_{N_f + 1}$,
due to the instanton which becomes massless at the origin of the Coulomb branch of the fixed point theory \cite{Seiberg:1996bd}.
Finally, in the large $N$ limit the free energy of the $\USp(2 N)$ theory on $S^5$ reads \cite{Jafferis:2012iv}
\be
 F_{S^5} \approx - \frac{9 \sqrt{2} \pi N^{5/2}}{5 \sqrt{8 - N_f}} \, .
\ee

The Cartan of $\USp(2N)$ has $N$ elements which we denote by $u_i$, $i = 1, \ldots, N$.
We normalize the weights of the fundamental representation of $\USp(2N)$ to be $\pm e_i$ (so they form a basis of unit vectors for $\bR^N$).
The antisymmetric representation thus has weights $\pm e_i \pm e_j$ with $i > j$ and $N-1$ zero weights,
and the roots are $\pm e_i \pm e_j$ with $i > j$ and $\pm 2 e_i$.
As we shall see below, at large $N$, $u_{i} = \cO(N^{1/2})$ (see \eqref{largeN:ansatz} with $\alpha = 1/2$).
Hence, the contributions with nontrivial instanton numbers are exponentially suppressed in the large $N$ limit.

We set the length of $S^1$ to one, \ie\;$\beta = 1$, throughout this section.

\subsubsection{Effective prepotential at large $N$}
\label{subsubsec:SW:USp(2N)}

The effective prepotential of the theory is then given by \eqref{full:pert:prepotential}:
\bea
 \label{prepotential:USp(2N)}
 \cF (a_i, \Delta_K) & = \sum_{i = 1}^{N} \left[ \cF^{\text{pert}} (\pm 2 a_i) - \sum_{f = 1}^{N_f} \cF^{\text{pert}} (\pm a_i + \Delta_f) \right] \\ &
 + \sum_{i > j}^{N} \left[ \cF^{\text{pert}} (\pm a_i \pm a_j) - \cF^{\text{pert}} (\pm a_i \pm a_j + \Delta_m) \right]
 + (N - 1) \cF^{\text{pert}} (\Delta_m) \, ,
\eea
where the index $K$ labels all the matter fields in the theory.
Here, we introduced the notation $\cF (\pm a) \equiv \cF (a) + \cF (- a)$.
Notice that the contribution of the last term to the large $N$ prepotential is of $\cO(N^2)$ and thus subleading.
Let us analyze the effective prepotential \eqref{prepotential:USp(2N)} assuming that the eigenvalues grow in the large $N$ limit.
We restrict to $\im a_i > 0$ due to the Weyl reflections of the $\USp(2N)$ group.
We consider the following large $N$ saddle point eigenvalue distribution ansatz:
\be
 \label{largeN:ansatz}
 a_j = \mathbbm{i} N^{\alpha} t_j \, ,
\ee
for some number $0 < \alpha < 1$ to be determined later. At large $N$, we define the continuous function $t_j = t (j / N)$ and we introduce the density of eigenvalues
\be
 \rho(t) = \frac{1}{N} \frac{\rd j}{\rd t} \, ,
\ee
normalized so that $\int \rd t \rho(t) = 1$. In the large $N$ limit the sums over $N$ become Riemann integrals, for example,
\be
 \sum_{j = 1}^{N} \to N \int \rd t \rho (t) \, .
\ee

Consider the first line in \eqref{prepotential:USp(2N)}:
\be
 \begin{aligned}
  & \frac{\mathbbm{i}}{2 \pi} \sum_{i = 1}^{N} \bigg[ \Li_3 (e^{2 \mathbbm{i} a_i}) + \Li_3 (e^{- 2 \mathbbm{i} a_i}) - \sum_{f = 1}^{N_f} \Li_3 (e^{\mathbbm{i} (a_i + \Delta_f)})
  - \sum_{f = 1}^{N_f} \Li_3 (e^{- \mathbbm{i} (a_i - \Delta_f)}) \bigg] \\ &
  - \frac{1}{4 \pi} \sum_{i = 1}^{N} \bigg[ g_3 ( - 2 a_i ) + g_3 ( 2 a_i ) + \sum_{f = 1}^{N_f} g_3 ( a_i + \Delta_f ) + \sum_{f = 1}^{N_f} g_3 ( - a_i + \Delta_f ) \bigg] \, .
 \end{aligned}
\ee
The second line is of $\cO (N^{2 \alpha +1})$ and thus subleading in the large $N$ limit (as we see below).
Using the ansatz \eqref{largeN:ansatz} we may write
\be
 \begin{aligned}
  \label{cont:gauge:fund}
  \cF^{(0)} \approx & \frac{\mathbbm{i} N}{2 \pi} \int \rd t \rho(t) \bigg[ \Li_3 (e^{- 2 N^\alpha t}) + \Li_3 (e^{2 N^\alpha t})
  - \sum_{f = 1}^{N_f} \Li_3 (e^{- N^{\alpha} t + \mathbbm{i} \Delta_f}) - \sum_{f = 1}^{N_f} \Li_3 (e^{N^{\alpha}  t + \mathbbm{i} \Delta_f}) \bigg] \\ &
  \approx - \frac{\mathbbm{i} (8 - N_f)}{12 \pi} N^{1 + 3 \alpha} \int \rd t \rho(t) t^3 \left[ \Theta (t) - \Theta (-t) \right] + \cO (N^{2 \alpha +1}) \\ &
  = - \frac{\mathbbm{i} (8 - N_f)}{12 \pi} N^{1 + 3 \alpha} \int \rd t \rho(t) | t |^3 + \cO (N^{2 \alpha +1})\, ,
 \end{aligned}
\ee
where $\Theta (t)$ is the Heaviside theta function.
Now, let us focus on the second line of \eqref{prepotential:USp(2N)}. Consider the following terms
\bea
 \cF^{(1)}_{\text{hyper}} & = - \frac{\mathbbm{i}}{2 \pi} \sum_{i > j}^{N} \left[ \Li_3 (e^{\mathbbm{i} (a_i - a_j + \Delta_m)}) + \Li_3 (e^{- \mathbbm{i} (a_i - a_j - \Delta_m)}) \right] \\ &
 - \frac{1}{4 \pi} \sum_{i > j}^{N} \left[ g_3 (a_i - a_j + \Delta_m) + g_3 (a_j - a_i + \Delta_m) \right] \, .
\eea
At large $N$, we obtain
\be
 \begin{aligned}
  \label{F:cont:as:I}
  \cF^{(1)}_{\text{hyper}} & \approx - \frac{\mathbbm{i} N^2}{4 \pi} \int \rd t \rho(t) \int \rd t' \rho(t') \left[ \Li_3 (e^{- N^{\alpha} (t - t') + \mathbbm{i} \Delta_m}) + \Li_3 (e^{ N^{\alpha} (t - t') + \mathbbm{i} \Delta_m}) \right] \\ &
  - \frac{N^2}{8 \pi} \int \rd t \rho(t) \int \rd t' \rho(t') \left[ (\pi - \Delta_m ) N^{2 \alpha} (t - t')^2 + 2 g_3 ( \Delta_m ) \right] \\ &
  \approx \frac{N^{2}}{4 \pi} \int \rd t \rho(t) \int \rd t' \rho(t') g_3 \left( - \mathbbm{i} N^{\alpha} (t' - t) + \Delta_m \right) \Theta ( t' - t) \\ &
  + \frac{N^{2}}{4 \pi} \int \rd t \rho(t) \int \rd t' \rho(t') g_3 \left( - \mathbbm{i} N^{\alpha} (t - t') + \Delta_m \right) \Theta ( t - t') \\ &
  - \frac{N^2}{8 \pi} \int \rd t \rho(t) \int \rd t' \rho(t') \left[ (\pi - \Delta_m ) N^{2 \alpha} (t - t')^2 + 2 g_3 ( \Delta_m ) \right] \\ &
  = \frac{\mathbbm{i} N^{2}}{4 \pi} \int \rd t \rho(t) \int \rd t' \rho(t') \left( \frac{1}{6} N^{3 \alpha} \left| t - t' \right|^3 - g_2 ( \Delta_m ) N^{\alpha} \left| t - t' \right| \right) \, .
 \end{aligned}
\ee
Next, consider
\be
 \label{gauge:USp(2N):short:roots}
 \cF^{(1)}_{\text{vector}} = \frac{\mathbbm{i}}{2 \pi} \sum_{i > j}^{N} \left[ \Li_3 (e^{\mathbbm{i} (a_i - a_j)}) + \Li_3 (e^{- \mathbbm{i} (a_i - a_j)}) + \frac{\mathbbm{i}}{2} g_3 (a_j - a_i) + \frac{\mathbbm{i}}{2} g_3 (a_i - a_j) \right] \, .
\ee
Its contribution can be simply obtained by, see \eqref{Z2:gs},
\be
 \cF^{(1)}_{\text{vector}} = - \cF^{(1)}_{\text{hyper}} \big|_{\Delta_m = 2 \pi} \, .
\ee
We thus find
\be
 \label{cont:gauge:as:I}
 \cF^{(1)}_{\text{vector}} + \cF^{(1)}_{\text{hyper}} \approx \frac{\mathbbm{i}}{4 \pi} \left[ \frac{\pi^2}{3} - g_2 ( \Delta_m ) \right] N^{2 + \alpha} \int \rd t \rho(t) \int \rd t' \rho(t') \left| t - t' \right| \, .
\ee
The next term that we shall consider is the following
\bea
 \begin{aligned}
  \cF^{(2)}_{\text{hyper}} & = - \frac{\mathbbm{i}}{2 \pi} \sum_{i > j}^{N} \left[ \Li_3 (e^{\mathbbm{i} (a_i + a_j + \Delta_m)}) + \Li_3 (e^{- \mathbbm{i} (a_i + a_j - \Delta_m)}) \right] \\ &
  - \frac{1}{4 \pi} \sum_{i > j}^{N} \left[ g_3 (a_i + a_j + \Delta_m) + g_3 ( - a_i - a_j + \Delta_m) \right] \, .
  \end{aligned}
\eea
The first term in the first line is exponentially suppressed in the large $N$ limit (since $\im a_i > 0 \, , \forall i $).
We then get
\be
 \begin{aligned}
  \cF^{(2)}_{\text{hyper}} & \approx - \frac{\mathbbm{i} N^2}{4 \pi} \int \rd t \rho(t) \int \rd t' \rho(t') \Li_3 (e^{N^{\alpha} (t + t')+ \mathbbm{i} \Delta_m}) \\ &
  - \frac{N^2}{8 \pi} \int \rd t \rho(t) \int \rd t' \rho(t') \left[ (\pi - \Delta_m ) N^{2 \alpha} (t + t')^2 + 2 g_3 ( \Delta_m ) \right] \\ &
  \approx \frac{N^2}{4 \pi} \int \rd t \rho(t) \int \rd t' \rho(t') g_3 (- \mathbbm{i} N^{\alpha} (t + t') + \Delta_m) \\ &
  - \frac{N^2}{8 \pi} \int \rd t \rho(t) \int \rd t' \rho(t') \left[ (\pi - \Delta_m ) N^{2 \alpha} (t + t')^2 + 2 g_3 ( \Delta_m ) \right] \\ &
  = \frac{\mathbbm{i} N^{2}}{4 \pi} \int \rd t \rho(t) \int \rd t' \rho(t') \left[ \frac{1}{6} N^{3 \alpha} (t + t')^3 - g_2 ( \Delta_m ) N^{\alpha} (t+ t')^2 \right] \, .
 \end{aligned}
\ee
The last term that we shall consider reads
\be
 \cF^{(2)}_{\text{vector}} = \frac{\mathbbm{i}}{2 \pi} \sum_{i > j}^{N} \left[ \Li_3 (e^{\mathbbm{i} (a_i + a_j)}) + \Li_3 (e^{- \mathbbm{i} (a_i + a_j)}) + \frac{\mathbbm{i}}{2} g_3 (- a_i - a_j) + \frac{\mathbbm{i}}{2} g_3 (a_i + a_j) \right] \, .
\ee
Its contribution can be simply obtained by, see \eqref{Z2:gs},
\be
 \cF^{(2)}_{\text{vector}} = - \cF^{(2)}_{\text{hyper}} \big|_{\Delta_m = 2 \pi} \, .
\ee
We thus find that
\be
 \label{cont:gauge:as:II}
 \cF^{(2)}_{\text{vector}} + \cF^{(2)}_{\text{hyper}} \approx \frac{\mathbbm{i}}{4 \pi} \left[ \frac{\pi^2}{3} - g_2 ( \Delta_m ) \right] N^{2 + \alpha} \int \rd t \rho(t) \int \rd t' \rho(t') (t+ t') \, .
\ee
Putting \eqref{cont:gauge:fund}, \eqref{cont:gauge:as:I}, and \eqref{cont:gauge:as:II} together we obtain the final expression for the effective prepotential at large $N$:
\bea
 \label{LargeN:Usp(2N):F}
 \cF \left[ \rho(t) , \Delta_m \right] & = \cF^{(0)} + \sum_{\vartheta = 1}^{2} \left( \cF^{(\vartheta)}_{\text{vector}} + \cF^{(\vartheta)}_{\text{hyper}} \right) \\ &
 \approx - \frac{\mathbbm{i} (8 - N_f)}{12 \pi} N^{1 + 3 \alpha} \int_{0}^{t_*} \rd t \rho(t) | t |^3
 - \mu \left( \int_{0}^{t_*} \rd t \rho (t)- 1 \right) \\ &
 + \frac{\mathbbm{i}}{4 \pi} \left[ \frac{\pi^2}{3} - g_2 ( \Delta_m ) \right] N^{2 + \alpha} \int_{0}^{t_*} \rd t \rho(t) \int_{0}^{t_*} \rd t' \rho(t') \left[ | t - t' | + ( t + t' ) \right] \, ,
\eea
where we added the Lagrange multiplier $\mu$ for the normalization of $\rho(t)$.
$\alpha$ will be determined to be $1/2$ by the competition between the first and the last term in \eqref{LargeN:Usp(2N):F}, and therefore $\cF \propto N^{5/2}$.
Remarkably, the effective prepotential \eqref{LargeN:Usp(2N):F} equals the large $N$ expression of the $S^5$ free energy computed in \cite[(3.4)]{Jafferis:2012iv} (see also \cite[(3.14)]{Chang:2017mxc}).
This is in complete analogy with the observation made in \cite{Hosseini:2016tor}.
There, the effective twisted superpotential of a three-dimensional $\cN=2$ theory on $A$-twisted $\Sigma_{\fg_1} \times S^1$ was shown to be equal to the $S^3$ free energy of the same $\cN=2$ theory, both evaluated at large $N$.

Extremizing \eqref{LargeN:Usp(2N):F} with respect to the continuous function $\rho(t)$ we find the following saddle point equation
\bea
 \frac{2 \mathbbm{i} \pi \mu}{N^{5/2}} = \frac{(8 - N_f)}{6} | t' |^3 - \left[ \frac{\pi^2}{3}-g_2(\Delta_m) \right] \int_{0}^{t_*} \rho (t) \left[ | t - t' | + ( t + t' ) \right] \, .
\eea
On the support of $\rho (t)$ the solution reads
\bea
 \label{LargeN:USp(2N):sol}
 & \rho (t) = \frac{2 | t | }{t_{*}^2} \, , \qquad
 t_* = \frac{2}{\sqrt{8 - N_f}} \left[ \frac{\pi^2}{3} -g_2 ( \Delta_m ) \right]^{1/2} \, , \\ &
 \mu = \frac{4 \mathbbm{i}}{3 \pi} \frac{N^{5/2}}{\sqrt{8 - N_f}} \left[ \frac{\pi^2}{3} - g_2 ( \Delta_m ) \right]^{3/2} \, .
\eea
Evaluating \eqref{LargeN:Usp(2N):F} at \eqref{LargeN:USp(2N):sol} yields%
\footnote{$\cF (\Delta_m) = \frac25 \mu (\Delta_m)$ due to a virial theorem for the large $N$ prepotential \eqref{LargeN:Usp(2N):F}.}
\bea
 \label{LargeN:USp(2N):prepotential}
 \cF (\Delta_m) \approx \frac{2^{3/2} \mathbbm{i}}{15 \pi} \frac{N^{5/2}}{\sqrt{8 - N_f}} \left[ \Delta_m (2 \pi - \Delta_m ) \right]^{3/2} \, .
\eea
Finally, let us rewrite \eqref{LargeN:USp(2N):prepotential} as
\be
 \label{USp(2N):F:on-shell}
 \cF (\Delta) \approx \frac{2^{3/2} \mathbbm{i}}{15 \pi} \frac{N^{5/2}}{\sqrt{8 - N_f}} \left( \Delta_1 \Delta_2 \right)^{3/2} \, ,
\ee
for later use. Here we introduced the democratic chemical potentials
\be
 \Delta_1 = \Delta_m \, , \qquad \Delta_2 = 2 \pi - \Delta_m \, .
\ee

\subsubsection{Effective twisted superpotential at large $N$}
\label{subsubsec:tildeW:USp(2N)}

We are interested in the large $N$ limit of the effective twisted superpotential
\bea
 \label{Wt:USp(2N):pert}
 \wt \cW^{\text{pert}} (a , \fn , \Delta , \ft) & = \sum_{i = 1}^{N} \bigg[ \wt \cW^{\text{pert}} (\pm 2 a_i) - \sum_{f = 1}^{N_f} \wt \cW^{\text{pert}} (\pm a_i + \Delta_f) \bigg] \\ &
 + \sum_{i > j}^{N} \left[ \wt \cW^{\text{pert}} (\pm a_i \pm a_j) - \wt \cW^{\text{pert}} (\pm a_i \pm a_j + \Delta_m) \right]
 + (N - 1) \wt \cW^{\text{pert}} (\Delta_m) \, .
\eea
We consider the following ansatz for the large $N$ saddle point eigenvalue distribution
\be
 a_i = \mathbbm{i} N^{\alpha} t_i
 \, , \qquad \fn_i = \mathbbm{i} N^{\alpha} \fN_i
 \, .
\ee
Consider the first line in \eqref{Wt:USp(2N):pert}:
\be
 \begin{aligned}
  \wt \cW^{(0)} & = \sum_{i = 1}^{N} ( \pm 2 \fn_i + 1 - \fg_2 ) \left[ \Li_2 (e^{\pm 2 \mathbbm{i} a_i}) - \frac{1}{2} g_2 (\mp 2 a_i) \right] \\ &
  - \sum_{f = 1}^{N_f} \sum_{i = 1}^{N} ( \fn_i + \ft_f + \fg_2 - 1 ) \left[ \Li_2 (e^{\mathbbm{i} (a_i + \Delta_f)}) - \frac{1}{2} g_2 ( a_i + \Delta_f ) \right] \\ &
  - \sum_{f = 1}^{N_f} \sum_{i = 1}^{N} ( - \fn_i + \tilde \ft_f + \fg_2 - 1 ) \left[ \Li_2 (e^{- \mathbbm{i} (a_i - \tilde \Delta_f)}) - \frac{1}{2} g_2 ( - a_i + \tilde \Delta_f ) \right] .
 \end{aligned}
\ee
The $g_2$ terms are of $\cO (N^{2})$ and thus subleading in the large $N$ limit. Hence,
\be
 \begin{aligned}
  \label{Wt:cont:gauge:fund}
  & \wt \cW^{(0)} \approx \mathbbm{i} \frac{(8 - N_f)}{2} N^{1 + 3 \alpha} \int \rd t \rho(t) \fN (t) t^2 \sign (t) \, .
 \end{aligned}
\ee
Now, let us focus on the second line of \eqref{Wt:USp(2N):pert}. Consider the following terms
\bea
 \wt\cW^{(1)}_{\text{hyper}} & = - \sum_{i > j}^{N} ( \fn_i - \fn_j + \ft_m + \fg_2 - 1 ) \left[ \Li_2 (e^{\mathbbm{i} (a_i - a_j + \Delta_m)}) - \frac12 g_2 (a_i - a_j + \Delta_m)  \right] \\ &
 - \sum_{i > j}^{N} ( \fn_j - \fn_i + \ft_m + \fg_2 - 1 ) \left[ \Li_2 (e^{- \mathbbm{i} (a_i - a_j - \Delta_m)}) - \frac12 g_2 (a_j - a_i + \Delta_m)  \right] .
\eea
At large $N$, we obtain
\be
 \begin{aligned}
  \label{Wt:cont:as:I}
  \wt\cW^{(1)}_{\text{hyper}} & \approx - \frac{\mathbbm{i}}{4} N^{2 + 3 \alpha} \int \rd t \rho(t) \int \rd t' \rho(t') ( \fN (t) - \fN(t') ) ( t - t' )^2 \sign ( t - t' ) \\ &
  + \frac{\mathbbm{i}}{4} g_2 ( \Delta_1 ) N^{2 + \alpha} \int \rd t \rho(t) \int \rd t' \rho(t') ( \fN (t) - \fN(t') ) \sign (t - t') \\ &
  + \frac{\mathbbm{i}}{8} ( \Delta_1 - \Delta_2 ) ( \ft_1 - \ft_2 ) N^{2 + \alpha} \int \rd t \rho(t) \int \rd t' \rho(t') | t - t' | \, ,
 \end{aligned}
\ee
where we used the democratic chemical potentials and fluxes
\be
 \Delta_1 = \Delta_m \, , \qquad \Delta_2 = 2 \pi - \Delta_m \, , \qquad
 \ft_1 = \ft_m \, , \qquad \ft_2 = 2 ( 1 - \fg_2 ) - \ft_m \, .
\ee
The similar contribution coming from the vector multiplet can be obtained by using
\be
 \label{tW:hyper:map:gauge}
 \wt \cW^{\text{pert}}_{\text{vector}} = - \wt \cW^{\text{pert}}_{\text{hyper}} |_{\ft_m = 2 ( 1 - \fg_2 ) , \, \Delta_m = 2 \pi} \, .
\ee
It reads
\bea
 \wt\cW^{(1)}_{\text{vector}} & \approx \frac{\mathbbm{i}}{4} N^{2 + 3 \alpha} \int \rd t \rho(t) \int \rd t' \rho(t') ( \fN (t) - \fN(t') ) ( t - t' )^2 \sign ( t - t' ) \\ &
  - \frac{\mathbbm{i} \pi^2}{6} N^{2 + \alpha} \int \rd t \rho(t) \int \rd t' \rho(t') ( \fN (t) - \fN(t') ) \sign (t - t') \\ &
  - \frac{\mathbbm{i} \pi}{2} ( 1- \fg_2 ) N^{2 + \alpha} \int \rd t \rho(t) \int \rd t' \rho(t') | t - t' | \, .
\eea
We thus find that
\bea
 \label{Wt:cont:gauge:as:I}
 \wt\cW^{(1)}_{\text{vector}} + \wt\cW^{(1)}_{\text{hyper}} & \approx - \frac{\mathbbm{i}}{4} \Delta_1 \Delta_2 N^{2 + \alpha} \int \rd t \rho(t) \int \rd t' \rho(t') ( \fN (t) - \fN(t') ) \sign (t - t') \\ &
 - \frac{\mathbbm{i}}{4} ( \Delta_1 \ft_2 + \Delta_2 \ft_1 ) N^{2 + \alpha} \int \rd t \rho(t) \int \rd t' \rho(t') | t - t' | \, .
\eea
The next term we shall consider is given by
\bea
 \begin{aligned}
  \wt\cW^{(2)}_{\text{vector}} + \wt\cW^{(2)}_{\text{hyper}} & =  \sum_{i > j}^{N} ( \fn_i + \fn_j + 1 - \fg_2 ) \left[ \Li_2 (e^{\mathbbm{i} (a_i + a_j)}) - \frac12 g_2 ( - a_i -  a_j )  \right] \\ &
  + \sum_{i > j}^{N} ( - \fn_i - \fn_j + 1 - \fg_2 ) \left[ \Li_2 (e^{- \mathbbm{i} (a_i + a_j)}) - \frac12 g_2 ( a_i + a_j )  \right] \\ &
  - \sum_{i > j}^{N} ( \fn_i + \fn_j + \ft_m + \fg_2 - 1 ) \left[ \Li_2 (e^{\mathbbm{i} (a_i + a_j + \Delta_m)}) - \frac12 g_2 (a_i + a_j + \Delta_m)  \right] \\ &
  - \sum_{i > j}^{N} ( - \fn_i - \fn_j + \ft_m + \fg_2 - 1 ) \left[ \Li_2 (e^{- \mathbbm{i} (a_i + a_j - \Delta_m)}) - \frac12 g_2 ( - a_i - a_j + \Delta_m)  \right] .
  \end{aligned}
\eea
In the large $N$ limit it can be approximated as
\bea
 \label{Wt:cont:gauge:as:II}
 \wt\cW^{(2)}_{\text{vector}} + \wt\cW^{(2)}_{\text{hyper}} & \approx - \frac{\mathbbm{i}}{4} \Delta_1 \Delta_2 N^{2 + \alpha} \int \rd t \rho(t) \int \rd t' \rho(t') ( \fN (t) + \fN(t') ) \\ &
 - \frac{\mathbbm{i}}{4} ( \Delta_1 \ft_2 + \Delta_2 \ft_1 ) N^{2 + \alpha} \int \rd t \rho(t) \int \rd t' \rho(t') ( t + t' ) \, .
\eea
Putting \eqref{Wt:cont:gauge:fund}, \eqref{Wt:cont:gauge:as:I}, and \eqref{Wt:cont:gauge:as:II} together we obtain the final expression for the effective twisted superpotential at large $N$:
\bea
 \label{LargeN:Usp(2N):Wt}
 \wt \cW \left[ \rho(t) , \fN(t) , \Delta_m , \ft_m \right] & = \wt \cW^{(0)} + \sum_{\vartheta = 1}^{2} \left( \wt \cW^{(\vartheta)}_{\text{vector}} + \wt \cW^{(\vartheta)}_{\text{hyper}} \right) \\ &
 \approx \mathbbm{i} \frac{(8 - N_f)}{2} N^{1 + 3 \alpha} \int_{0}^{t_*} \rd t \rho(t) \fN (t) t^2
 - \mathbbm{i} \gamma \left( \int_{0}^{t_*} \rd t \rho (t)- 1 \right) \\ &
 - \frac{\mathbbm{i}}{4} ( \Delta_1 \Delta_2 ) N^{2 + \alpha} \int_{0}^{t_*} \rd t \rho(t) \int_{0}^{t_*} \rd t' \rho(t') ( \fN (t) - \fN(t') ) \sign (t - t') \\ &
 - \frac{\mathbbm{i}}{4} ( \Delta_1 \Delta_2 ) N^{2 + \alpha} \int_{0}^{t_*} \rd t \rho(t) \int_{0}^{t_*} \rd t' \rho(t') ( \fN (t) + \fN(t') ) \\ &
 - \frac{\mathbbm{i}}{4} ( \Delta_1 \ft_2 + \Delta_2 \ft_1 ) N^{2 + \alpha} \int_{0}^{t_*} \rd t \rho(t) \int_{0}^{t_*} \rd t' \rho(t') \left[ | t - t' | + ( t + t' ) \right] ,
\eea
where we added the Lagrange multiplier $\gamma$ for the normalization of $\rho(t)$.
In order to have a non-trivial saddle point we need to set $1 + 3 \alpha = 2 + \alpha$, implying that $\alpha = 1 / 2$.
The twisted superpotential thus scales as $N^{5 / 2}$.
Setting to zero the variation with respect to $\rho(t)$, we get the equation
\bea
 \frac{\gamma}{N^{5/2}} & = \frac{(8 - N_f)}{2} \fN (t') t'^2
 - \frac12 ( \Delta_1 \ft_2 + \Delta_2 \ft_1 ) \int_{0}^{t_*} \rho(t) \left[ | t + t' | + ( t + t' ) \right] \\ &
 - \frac12 ( \Delta_1 \Delta_2 ) \int_{0}^{t_*} \left[ ( \fN (t) - \fN(t') ) \sign ( t - t' ) + ( \fN (t) + \fN(t') ) \right] \, .
\eea
On the support of $\rho (t)$ the solution is given by
\bea
 \label{LargeN:USp(2N):sol:Wt}
 & \rho(t) = \frac{2 | t |}{t_*^2} \, , \qquad t_* = \frac{(2 \Delta_1 \Delta_2)^{1/2}}{\sqrt{8 - N_{f}}} \, , \qquad
 \fN (t) = \frac12 \left( \frac{\ft_1}{\Delta_1} + \frac{\ft_2}{\Delta_2} \right) t \, , \\ &
 \gamma = - \frac{N^{5/2}}{\sqrt{8 - N_{f}}} ( 2 \Delta_1 \Delta_2 )^{1/2} ( \Delta_1 \ft_2 + \Delta_2 \ft_1 ) \, .
\eea
Evaluating \eqref{LargeN:Usp(2N):Wt} at \eqref{LargeN:USp(2N):sol:Wt} yields%
\footnote{$\wt\cW (\Delta_m , \ft_m) = \frac{2 i}{5} \gamma (\Delta_m , \ft_m)$ due to a virial theorem for the large $N$ twisted superpotential \eqref{LargeN:Usp(2N):Wt}.}
\be
 \label{LargeN:USp(2N):superpotential}
 \wt \cW (\Delta_m , \ft_m) \approx - \frac{2^{3/2} \mathbbm{i} N^{5/2}}{5 \sqrt{8 - N_{f}}} ( \Delta_1 \Delta_2 )^{1/2} ( \Delta_1 \ft_2 + \Delta_2 \ft_1 ) \, .
\ee
This can be more elegantly rewritten as
\be
 \label{USp(2N):index theorem:tildeW:F}
 \wt \cW (\Delta_m , \ft_m) \approx - 2 \pi \sum_{\varsigma = 1}^{2} \ft_\varsigma \frac{\partial \cF(\Delta)}{\partial \Delta_\varsigma} \, ,
\ee
where $\cF(\Delta)$ is given in \eqref{USp(2N):F:on-shell}.

\paragraph*{$\wt \cW$ and the free energy on $\Sigma_{\fg_2} \times S^3$.} 

Remarkably, we find the following relation between the twisted superpotential \eqref{USp(2N):index theorem:tildeW:F} and the $S^3$ free energy 
of the 3D $\cN=2$ theory that is obtained by compactifying the 5D theory on $\Sigma_{\fg_2}$ with fluxes $\ft_\varsigma$:
\be\label{relationUSP}
 \wt \cW (\bar\Delta_m , \ft) \approx \frac{\mathbbm{i} \pi}{2} \, F_{\Sigma_{\fg_2} \times S^3} (\ft) \, ,
 \ee
 where $\wt \cW (\bar\Delta_m , \ft)$ is evaluated at its extremum
\be
 \frac{\bar\Delta_m}{\pi} = \frac{5 \ft_1 - 3 \ft_2 \pm \sqrt{9 \ft_1^2 - 14 \ft_1 \ft_2 + 9 \ft_2^2}}{4 (\ft_1 - \ft_2)} \, .
\ee

Indeed, the $S^3$ free energy was very recently computed holographically in \cite{Bah:2018lyv} and it reads
\be\label{FS3}
 F_{\Sigma_{\fg_2} \times S^3} (\ft) \approx \frac{16 \pi}{5} \frac{ (1-\fg_2) N^{5/2}}{\kappa \sqrt{8 - N_f}} \frac{(z^2-\kappa^2)^{3/2} \left ( \sqrt{\kappa^2 + 8 z^2} -\kappa\right)}{\left ( 4 z^2 -\kappa^2
+\kappa \sqrt{\kappa^2 + 8 z^2}\right )^{3/2}}\, ,
\ee
where $\kappa =1$ for $\fg_2=0$ and $\kappa = -1$ for $\fg_2>1$ and the variable $z$ parameterizes the fluxes $\ft_\varsigma$.
In the special case of the torus, $\fg_2=1$, the above expression should be replaced by
\be
 \label{FS32}
 F_{\Sigma_{\fg_2} \times S^3}(\ft) \approx \frac{4 \sqrt{2} \pi}{5} \frac{N^{5/2}}{ \sqrt{8 - N_f}}  | z | \, .
\ee

It is now a simple exercise to check that \eqref{relationUSP} holds identically for all $\fg_2$ upon identifying
\be
 \ft_1 = ( 1 - \fg_2 ) \left( 1 + \frac{z}{\kappa} \right) \, , \qquad   \ft_2 = ( 1 - \fg_2 ) \left( 1- \frac{z}{\kappa} \right) \, .
\ee

We  expect that \eqref{relationUSP} holds also off-shell
 \be\label{relationUSP2}
 \wt \cW (\Delta , \ft) \approx \frac{\mathbbm{i} \pi}{2}\, F_{\Sigma_{\fg_2} \times S^3} (\Delta, \ft) \, ,
 \ee
 where $F_{\Sigma_{\fg_2} \times S^3} (\Delta, \ft)$ is the free energy as a function of a trial R-symmetry.
 \eqref{relationUSP} would correspond then to the statement that the $S^3$ free energy of the 3D theory is obtained by extremizing  $F_{\Sigma_{\fg_2} \times S^3} (\Delta, \ft)$ with respect to $\Delta$ \cite{Jafferis:2010un}.\footnote{The  free energy on $\Sigma_{\fg_2} \times S^3$ as a function of $\Delta$ was explicitly computed in field theory in \cite{Crichigno:2018adf} after the completion of this work. The result in \cite{Crichigno:2018adf} perfectly agrees with \eqref{relationUSP2}.}

\subsubsection{Partition function at large $N$}
\label{subsubsec:logZ:USp(2N)}

The topologically twisted index of the $\USp(2N)$ theory with matter on $\Sigma_{\fg_2} \times ( \Sigma_{\fg_1} \times S^1 )$ reads
\bea
 \label{USp(2N):index:Sigmag2xS1xSigmag1}
 Z^{\text{pert}} (y , \fs , \ft) & = \frac{(-1)^{N}}{2^N N!}
 \sum_{\fn \in \Gamma_\fh} \sum_{a = a_{(i)}} \bigg( \det\limits_{i j} \frac{\partial^2 \wt \cW (a , \fn)}{\partial a_i \partial a_j} \bigg)^{\fg_1 - 1} \\ &
 \times \bigg( \frac{y_m^{1/2}}{1 - y_m} \bigg)^{(N-1) ( \fs_m + \fg_1 - 1 ) ( \ft_m + \fg_2 - 1 )}
 \prod_{g = \pm 1} \prod_{i = 1}^{N} \bigg( \frac{1 - x_i^{2 g}}{x_i^{g}} \bigg)^{( 1 - \fg_1 ) ( 2 g \fn_i + 1 - \fg_2 )} \\ &
 \times \prod_{i = 1}^{N} \prod_{f = 1}^{N_f} \bigg( \frac{x_i^{1 / 2} y_f^{1 / 2}}{1 - x_i y_f} \bigg)^{( \fs_f + \fg_1 - 1 ) ( \fn_i + \ft_f + \fg_2 - 1)}
 \bigg( \frac{x_i^{- 1 / 2} \tilde y_f^{1 / 2}}{1 - x_i^{-1} \tilde y_f} \bigg)^{( \tilde \fs_f + \fg_1 - 1 ) ( - \fn_i + \tilde \ft_f + \fg_2 - 1)} \\ &
 \times \prod_{g = \pm 1} \prod_{i > j}^{N} \bigg( \frac{1 - (x_i x_j )^{g}}{( x_i x_j )^{g / 2}} \bigg)^{( 1 - \fg_1 ) ( g \fn_i + g \fn_j + 1 - \fg_2 )}
 \bigg( \frac{1 - (x_i / x_j )^{g}}{( x_i / x_j )^{g / 2}} \bigg)^{( 1 - \fg_1 ) ( g \fn_i - g \fn_j + 1 - \fg_2 )} \\ &
 \times \prod_{g = \pm 1} \prod_{i > j}^{N} \bigg( \frac{( x_i x_j )^{g / 2} y_m^{1/2}}{1 - (x_i x_j )^{g} y_m} \bigg)^{( \fs_m + \fg_1 - 1 ) ( g \fn_i + g \fn_j + \ft_m + \fg_2 - 1 )} \\ &
 \times \prod_{g = \pm 1} \prod_{i > j }^{N} \bigg( \frac{( x_i / x_j )^{g / 2} y_m^{1/2}}{1 - (x_i / x_j )^{g} y_m} \bigg)^{( \fs_m + \fg_1 - 1 ) ( g \fn_i - g \fn_j + \ft_m + \fg_2 - 1 )}
 \, .
\eea
The products $\prod_{i = 1}^{N}$ are of $\cO (N^{2})$ and thus subleading in the large $N$ limit. Then we consider the last line in \eqref{USp(2N):index:Sigmag2xS1xSigmag1}:
\bea
 \log Z^{(1)}_{\text{hyper}} & = ( \fs_m + \fg_1 - 1 ) \sum_{i >j }^{N} ( \fn_i - \fn_j + \ft_m + \fg_2 - 1 ) \left[ \Li_1 (e^{\mathbbm{i} (a_i - a_j + \Delta_m)}) + \frac{\mathbbm{i}}{2} g_1 (a_i - a_j + \Delta_m)  \right] \\ &
 + ( \fs_m + \fg_1 - 1 ) \sum_{i > j}^{N} ( \fn_j - \fn_i + \ft_m + \fg_2 - 1 ) \left[ \Li_1 (e^{- \mathbbm{i} (a_i - a_j - \Delta_m)}) + \frac{\mathbbm{i}}{2} g_1 (a_j - a_i + \Delta_m)  \right] .
\eea
At large $N$, it can be approximated as
\be
 \begin{aligned}
  \label{logZ:cont:as:I}
  \log Z^{(1)}_{\text{hyper}} & \approx - \frac{1}{8} ( \Delta_1 - \Delta_2 ) ( \fs_1 - \fs_2 ) N^{2 + \alpha} \int \rd t \rho(t) \int \rd t' \rho(t') ( \fN (t) - \fN(t') ) \sign (t - t') \\ &
  - \frac{1}{8} ( \ft_1 - \ft_2 ) ( \fs_1 - \fs_2 ) N^{2 + \alpha} \int \rd t \rho(t) \int \rd t' \rho(t') | t - t' | \, ,
 \end{aligned}
\ee
where we introduced the democratic fluxes
\be
 \fs_1 = \fs_m \, , \qquad \fs_2 = 2 ( 1 - \fg_1 ) - \fs_m \, .
\ee
The similar contribution coming from the vector multiplet can be obtained by using
\be
 \label{logZ:hyper:map:gauge}
 \log Z^{\text{pert}}_{\text{vector}} = - \log Z^{\text{pert}}_{\text{hyper}} |_{\fs_m = 2 ( 1 - \fg_1 ) , \, \ft_m = 2 ( 1 - \fg_2 ) , \, \Delta_m = 2 \pi} \, .
\ee
It reads
\bea
 \log Z^{(1)}_{\text{vector}} & \approx \frac{\pi}{2} ( 1- \fg_1 ) N^{2 + \alpha} \int \rd t \rho(t) \int \rd t' \rho(t') ( \fN (t) - \fN(t') ) \sign (t - t') \\ &
  + \frac12( 1- \fg_2 ) ( 1- \fg_1 ) N^{2 + \alpha} \int \rd t \rho(t) \int \rd t' \rho(t') | t - t' | \, .
\eea
We thus find that
\bea
 \label{logZ:cont:gauge:as:I}
 \log Z^{(1)}_{\text{vector}} + \log Z^{(1)}_{\text{hyper}} & \approx \frac14 ( \Delta_1 \fs_2 + \Delta_2 \fs_1 ) N^{2 + \alpha} \int \rd t \rho(t) \int \rd t' \rho(t') ( \fN (t) - \fN(t') ) \sign (t - t') \\ &
 + \frac14 ( \ft_1 \fs_2 + \ft_2 \fs_1 ) N^{2 + \alpha} \int \rd t \rho(t) \int \rd t' \rho(t') | t - t' | \, .
\eea
The terms coming from the roots $e_1 + e_2$ and $- e_1 - e_2$ are treated as in \eqref{logZ:cont:as:I}. They can be approximated at large $N$ as
\bea
 \label{logZ:cont:gauge:as:II}
 \log Z^{(2)}_{\text{vector}} + \log Z^{(2)}_{\text{hyper}} & \approx \frac14 ( \Delta_1 \fs_2 + \Delta_2 \fs_1 ) N^{2 + \alpha} \int \rd t \rho(t) \int \rd t' \rho(t') ( \fN (t) + \fN(t') ) \\ &
 + \frac14 ( \ft_1 \fs_2 + \ft_2 \fs_1 ) N^{2 + \alpha} \int \rd t \rho(t) \int \rd t' \rho(t') ( t + t' ) \, .
\eea
Putting everything together we obtain the following functional for the logarithm of the partition function at large $N$:
\bea
 \label{LargeN:Usp(2N):logZ}
 \frac{\log Z}{N^{5/2}} & \approx \frac14 ( \Delta_1 \fs_2 + \Delta_2 \fs_1 ) \int_{0}^{t_*} \rd t \rho(t) \int_{0}^{t_*} \rd t' \rho(t') \left[ ( \fN (t) - \fN(t') ) \sign (t - t') + ( \fN (t) + \fN(t') ) \right] \\ &
 + \frac14 ( \ft_1 \fs_2 + \ft_2 \fs_1 ) \int_{0}^{t_*} \rd t \rho(t) \int_{0}^{t_*} \rd t' \rho(t') \left[ | t - t' | + ( t + t' ) \right] .
\eea
Finally we take the solution to the BAEs \eqref{LargeN:USp(2N):sol:Wt}, plug it back into \eqref{LargeN:Usp(2N):logZ} and compute the integral. We obtain
\bea
 \log Z ( \Delta_m , \ft_m , \fs_m ) & \approx \frac{\sqrt{2} N^{5/2}}{5 \sqrt{8 - N_f}}
 \left[ \frac{( \Delta_1 \fs_2 + \Delta_2 \fs_1 ) ( \Delta_1 \ft_2 + \Delta_2 \ft_1 )}{( \Delta_1 \Delta_2)^{1/2}}
 + 2 ( \Delta_1 \Delta_2 )^{1/2} ( \ft_1 \fs_2 + \ft_2 \fs_1 ) \right] \\ &
 = \frac{\sqrt{2} N^{5/2}}{5 \sqrt{8 - N_f}} \frac{\Delta_1 \fs_2 ( \Delta_1 \ft_2 + 3 \Delta_2 \ft_1 ) + \Delta_2 \fs_1 ( 3 \Delta_1 \ft_2 + \Delta_2 \ft_1 )}{( \Delta_1 \Delta_2 )^{1/2}} .
\eea
Remarkably, this can be rewritten as
\be
 \label{USp(2N):Final:logZ:index theorem}
 \log Z ( \Delta_m , \ft_m , \fs_m ) \approx \mathbbm{i} \sum_{\varsigma = 1}^2 \fs_\varsigma \frac{\partial \wt \cW (\ft , \Delta)}{\partial \Delta_\varsigma}
 = - 2 \mathbbm{i} \pi \sum_{\varsigma , \varrho = 1}^2 \fs_\varsigma \ft_\varrho \frac{\partial^2 \cF (\Delta)}{\partial \Delta_\varsigma \partial \Delta_\varrho} \, ,
\ee
where $ \wt \cW (\ft , \Delta)$ and $\cF (\Delta)$ are given in \eqref{LargeN:USp(2N):superpotential} and \eqref{USp(2N):F:on-shell}, respectively.

In analogy with \cite{Benini:2015eyy,Benini:2016rke}, we expect that  the extremization of the topologically twisted index \eqref{USp(2N):Final:logZ:index theorem} reproduces the entropy of asymptotically AdS$_6$ black holes
in massive type IIA supergravity with magnetic fluxes $\ft_m$ and $\fs_m$, and horizon topology AdS$_2 \times \Sigma_{\fg_1} \times \Sigma_{\fg_2}$.
Unfortunately, such black holes are still to be found. The only known example is the black hole dual to the universal twist, \ie \,$\ft_m= 1- \fg_2$ and $\fs_m=1 -\fg_1$,
whose entropy was computed in \cite{Bobev:2017uzs}, using  gauged supergravity in six dimensions and elaborating on the results in \cite{Naka:2002jz}.  The result for the entropy \cite[(4.36)]{Bobev:2017uzs} has been recently corrected by a factor of two in \cite{Suh:2018tul}.  Our index \eqref{USp(2N):Final:logZ:index theorem} for the appropriate values of the chemical potentials for the universal twist, \ie\;$\Delta_1=\Delta_2=\pi$, predicts 
\be S_{\text{BH}} \approx \frac{8 \sqrt{2} \pi}{5}(1-\fg_1)(1-\fg_2) \frac{N^{5/2}}{\sqrt{8-N_f}} \approx -\frac{8}{9} (1-\fg_1)(1-\fg_2) F_{S^5} \, ,\ee
and  correctly matches  the result in \cite{Suh:2018tul}.%
\footnote{We thank P. Marcos Crichigno, Dharmesh Jain and Brian Willett for pointing out a numerical mistake in our computation in the first version of this paper.}

\section[4D black holes from AdS\texorpdfstring{$_7$}{(7)} black strings]{4D black holes from AdS$_7$ black strings}
\label{sec:4D domain-walls}

Our primary interest in this section is to understand the $\cI$-extremization principle \eqref{SBH} for the topologically twisted index of the 6D $\cN = (2,0)$ theory in terms of holography and, in particular, in terms of the attractor mechanism \cite{Ferrara:1995ih,Ferrara:1996dd} in ${\cal N}=2$ supergravity.

We shall consider the supergravity dual of two-dimensional $\cN = (0 , 2)$ SCFTs obtained by compactifying a stack of M5-branes on $\Sigma_{\fg_1} \times \Sigma_{\fg_2}$.
These solutions were constructed in \cite{Benini:2013cda} and can be viewed as black strings in seven dimensions, interpolating between the maximally supersymmetric AdS$_7$ vacuum at infinity
and the near-horizon AdS$_3 \times \Sigma_{\fg_1} \times \Sigma_{\fg_2}$ geometry. This leads to a natural holographic interpretation of these black strings as RG flows across dimensions
--- we have a flow from the 6D $\cN = (2,0)$ theory in the UV to a 2D $\cN = (0,2)$ SCFT in the IR. The BPS black strings in AdS$_7$, {\it \'a la} Maldacena-Nu\~nez \cite{Maldacena:2000mw},
preserve supersymmetry due to the topological twist on the internal space $\Sigma_{\fg_1} \times \Sigma_{\fg_2}$. 

Let us  briefly review these solutions. We work with a $\U(1)^2$ consistent truncation \cite{Liu:1999ai}
of the $\SO(5)$ maximal $(\cN = 4)$ gauged supergravity in seven dimensions \cite{Pernici:1984xx}, obtained by reducing the eleven-dimensional supergravity on $S^4$ \cite{Nastase:1999cb,Nastase:1999kf}.
It contains the metric, a three-form gauge potential $S_5$, two Abelian gauge fields $A^I$ in the Cartan of $\SO(5)$ and two real scalars $\lambda_I$ $(I = 1 , 2)$.
The solution can then be written as%
\footnote{For notational convenience we use $\Sigma_{\sigma} \equiv \Sigma_{\fg_\sigma}$. The relations between the magnetic fluxes $\fs_I$, $\ft_I$ in \eqref{BB7d}
and $a_\sigma$, $b_\sigma$ in \cite[(5.26)]{Benini:2013cda} are the following: $a_1 = - \fs_1 / \eta_1$, $b_1 = - \fs_2 / \eta_1$, $a_2 = - \ft_1 / \eta_2 $, $b_2 = - \ft_2 / \eta_2 $ .}
\bea
 & \rd s_7^2 = e^{2 f(r)} (- \rd t^2 + \rd z^2 + \rd r^2) + \sum_{\sigma=1}^2 e^{2 g_\sigma (r)} \rd s^2_{\Sigma_\sigma} \, , \\
 & S_5 = - \frac{1}{32 \sqrt{3} \eta_1 \eta_2} (\ft_1 \fs_2 + \ft_2 \fs_1) e^{- 4 (\lambda_1 + \lambda_2) - 2 ( g_1 + g_2 )} \rd t \wedge \rd z \wedge \rd r \\
 & F^I = \frac{\fs^I }{4 \eta_1} \text{vol} \left( \Sigma_1 \right) + \frac{\ft^I }{4 \eta_2} \text{vol} \left( \Sigma_2 \right) \, , \qquad
 \lambda_I = \lambda_I (r) \, ,
 \label{BB7d}
\eea
where $\rd s^2_{\Sigma} = \rd \theta^2 + f_{\kappa}^2 (\theta) \rd \varphi^2$ defines the metric on a surface $\Sigma$ of constant scalar curvature $2\kappa$, with $\kappa = \pm 1$, and
\be
 f_\kappa(\theta) = \frac{1}{\sqrt{\kappa}} \sin(\sqrt{\kappa}\theta) = 
 \left\{\begin{array}{l@{\quad}l} \sin\theta\, & \kappa=+1\,, \\                                             
 \sinh\theta\, & \kappa=-1\,. \end{array}\right. 
\ee
The volume of $\Sigma_\sigma$ is given by
\be
 \text{vol}\left (\Sigma_\sigma \right )= \int f_\kappa(\theta) \rd \theta \wedge \rd \phi = 2 \pi \eta_\sigma \, , \qquad
 \eta_\sigma =
 \left\{\begin{array}{l@{\quad}l} 2 | \fg_\sigma - 1 | \, & \fg \neq 0 \, , \\                                             
 1 \, & \fg = 0 \, . \end{array} \right. 
\ee 
Moreover, $F^I = \rd A^I$ and $f(r)$, $g_\sigma (r)$, $\lambda_I(r)$ are functions of the radial coordinate only.\
In the AdS$_3$ region the scalars are fixed in terms of the magnetic charges:
\bea
 & e^{10\lambda_1} = \frac{(\fs_1 \ft_2^2 + \ft_1^2 \fs_2)(\fs_1^2 \ft_2 +\ft_1 \fs_2^2)(\fs_1 \ft_2 + \ft_1 \fs_2 -\fs_2 \ft_2)^2}{(\fs_1^2 \ft_2^2 +\ft_1^2 \fs_2^2 +\fs_1\fs_2 \ft_1 \ft_2)(\fs_1 \ft_2 + \ft_1 \fs_2 -\fs_1 \ft_1)^3} \, , \\
 & e^{10\lambda_2} = \frac{(\fs_1 \ft_2^2 + \ft_1^2 \fs_2)(\fs_1^2 \ft_2 +\ft_1 \fs_2^2)(\fs_1 \ft_2 + \ft_1 \fs_2 -\fs_1 \ft_1)^2}{(\fs_1^2 \ft_2^2 +\ft_1^2 \fs_2^2 +\fs_1\fs_2 \ft_1 \ft_2)(\fs_1 \ft_2 + \ft_1 \fs_2 -\fs_2 \ft_2)^3}  \, ,
\eea
and the warp factors read
\bea
& e^{2 g_1} = - \frac{1}{4 \eta_1}e^{-4\lambda_1-4\lambda_2}(\fs_1 e^{2\lambda_1} +\fs_2 e^{2\lambda_2}) \equiv R^2_{\Sigma_1}\, , \\ &
e^{2 g_2} = - \frac{1}{4 \eta_2} e^{-4\lambda_1-4\lambda_2}(\ft_1 e^{2\lambda_1} +\ft_2 e^{2\lambda_2}) \equiv R^2_{\Sigma_2}\, ,  \\
& e^f= \frac{ \fs_2 \ft_2 -\fs_1 \ft_2 -\ft_1 \fs_2}{\fs_1 \ft_1 +\fs_2 \ft_2 -2 \fs_1 \ft_2 -2 \ft_1 \fs_2} \frac{e^{2\lambda_2}}{r} \equiv \frac{R_{\text{AdS}_3}}{r} \, .
\eea
As shown in \cite{Benini:2013cda} the holographic central charge $c_{\text{sugra}} (\fs , \ft)$ can be computed, {\it \'a la} Brown-Henneaux \cite{Brown1986}, as
\be
 c_{\text{sugra}} (\fs , \ft) = \frac{8 N^3}{\pi^2} \text{vol} \left(\Sigma_1 \times \Sigma_2 \right ) R_{\text{AdS}_3}R^2_{\Sigma_1}R^2_{\Sigma_2} \, ,
\ee
and it matches the CFT result \eqref{exact:cr}. 

If we now add a momentum $n$ along the circle inside AdS$_3$ and do a compactification along this circle, we obtain a static black hole in six-dimensional gauged supergravity.
By a standard argument, the entropy of  such black hole is given by the number of states of the CFT with momentum $n$, and is therefore given by the Cardy formula \eqref{Cardys3}
\bea\label{Cardys32}
S_{\text{BH}} (\fs , \ft , n) = \cI_{\text{SCFT}}\big|_{\text{crit}} (\fs , \ft , n)  = 2 \pi \sqrt{ \frac{ n\, c_{\text{CFT}}(\fs , \ft)}{6}} \, .
\eea
As discussed in section \ref{I-extremization principle}, the entropy $S_{\text{BH}} (\fs , \ft , n)$ is the result of extremizing the functional $\cI_{\text{SCFT}} (\tilde \beta , \Delta)$. 
Both for dyonic BPS black holes in AdS$_4$ \cite{Benini:2015eyy,Benini:2016rke,Hosseini:2017fjo,Benini:2017oxt} and BPS black strings in AdS$_5$ \cite{Hosseini:2016cyf,Hosseini:2018qsx,Hristov:2018lod} the $\cI$-extremization principle has been identified with the attractor mechanism in 4D $\cN = 2$ gauged supergravity \cite{Cacciatori:2009iz,DallAgata:2010ejj}.
We thus expect that the $\cI$-extremization principle \eqref{SBH} corresponds to the attractor mechanism in six-dimensional gauged supergravity. 
Unfortunately, not much is known about such  mechanism in six  dimensions. Therefore, our strategy is to first reduce the seven-dimensional gauged supergravity on $\Sigma_{\fg_2}$ down to five dimensions
and then do a further reduction on the circle inside AdS$_3$ to four-dimensional $\cN = 2$ gauged supergravity, where the attractor mechanism for static BPS black holes is well-understood. An analogous argument has been used in \cite{Amariti:2016mnz} to explain the extremization of the R-symmetry in the two-dimensional CFT.

\subsection{Attractor mechanism}
\label{subsec:attractor}

In 4D ${\cal N}=2$ gauged supergravity%
\footnote{We refer to \cite{Hristov:2014eza} and the appendices of \cite{Hosseini:2017mds} for notations and more details about gauged supergravity in five and four dimensions.}
with $n_{\text{V}}$ vector multiplets $(\Lambda = 0 , 1 , \ldots , n_{\text{V}})$ the Bekenstein-Hawking entropy of a static BPS black hole with horizon topology $\Sigma_\sigma$, with a charge vector $\cQ = (p^\Lambda,q_\Lambda)$, can be be obtained by extremizing \cite{DallAgata:2010ejj}
\bea
 \label{attractor:mechanism:4D}
 \cI_{\text{sugra}} (X^\Lambda) = \frac{2 \pi \mathbbm{i} \eta_\sigma}{4 G_{\text{N}}^{(4)}} \frac{ q_\Lambda X^\Lambda - p^\Lambda F_\Lambda}{g_\Lambda X^\Lambda - g^\Lambda F_\Lambda} \, ,
\eea
with respect to the symplectic sections $X^\Lambda$ such that its value at the critical point is \emph{real}. Here,
\be
 F_\Lambda \equiv \frac{\partial \cF_{\text{sugra}} (X^\Lambda)}{\partial X^\Lambda} \, ,
\ee
with ${\cal F}_{\text{sugra}} (X^\Lambda)$ being the prepotential, $\cG = (g^\Lambda , g_\Lambda)$ is 
the vector of magnetic $g^\Lambda$ and electric $g_\Lambda$ Fayet-Iliopoulos (FI) parameters, and $G_{\text{N}}^{(4)}$ is the four-dimensional Newton's constant.%
\footnote{The magnetic and electric charges  are defined as $\int_{\Sigma_\sigma} F^\Lambda = \text{vol}(\Sigma_\sigma) p^\Lambda$ and $\int_{\Sigma_\sigma} G_\Lambda = \text{vol}(\Sigma_\sigma) q_\Lambda$
where $G_\Lambda = 8 \pi G_\text{N}^{(4)} \delta (\mathscr{L} \rd \text{vol}_4)/\delta F^\Lambda$. In a frame with purely electric gauging $g_\Lambda$,
the charges are quantized as $\eta_\sigma g_\Lambda p^\Lambda \in \bZ$ and $\eta_\sigma q_\Lambda / ( 4G_\text{N}^{(4)} g_\Lambda) \in \bZ$, not summed over $\Lambda$.}
In general, in gauged supergravity, $\cF_{\text{sugra}} (X^\Lambda)$ is a homogeneous function of degree two, so we can equivalently define $\hat X^\Lambda \equiv X^\Lambda/(g_\Lambda X^\Lambda - g^\Lambda F_\Lambda)$ and extremize
\bea
 \label{attractor}
 \cI_{\text{sugra}} (\hat X^\Lambda) = - \frac{2 \pi \mathbbm{i} \eta_\sigma}{4 G_{\text{N}}^{(4)}}  \left ( p^\Lambda F_\Lambda (\hat X^\Lambda) - q_\Lambda \hat X^\Lambda \right ) \, .
\eea
The critical value of \eqref{attractor} $\bar X^\Lambda$ determines the value of the physical scalars $z^i$ at the horizon,
and the entropy of the black hole is then given by evaluating the functional \eqref{attractor} at its extremum
\be
 S_{\text{BH}} (p^\Lambda , q_\Lambda) = \cI_{\text{sugra}} (\bar X^\Lambda) \, .
\ee
This is the so-called \emph{attractor mechanism} \cite{Ferrara:1995ih,Ferrara:1996dd}, stating that the area of the black hole horizon is given in terms of the conserved charges and is independent of the asymptotic moduli.

In a general gauged supergravity with $n_{\text{H}}$ hypermultiplets, in addition to the gravitino and gaugino BPS equations, one needs to impose the BPS equations for the hyperino.
Altogether these equations become algebraic in the near-horizon limit and fix the horizon value of the scalars in the vector multiplets $z^I=X^I / X^0$ and the hyperscalars $q^u$, $u=1,\cdots , 4n_{\text{H}}$.
The full set of equations can be found in \cite{Klemm:2016wng}. In general, the hyperino equations at the horizon just yield a set of linear constraints on the sections $X^\Lambda$.
In simple models, this can be used to integrate out all massive fields at the horizon and write an effective theory with only massless vectors to which we can directly apply \eqref{attractor}.
This approach has been used in \cite{Hosseini:2017fjo,Benini:2017oxt} to reproduce the entropy of AdS$_4 \times S^6$ black holes in massive type IIA supergravity and, as we will see below, also works here. 

\subsection{Localization meets holography}
\label{sec:localization meets holography}

In order to use the attractor mechanism \eqref{attractor}, we need to determine the matter content and the prepotential of the four-dimensional $\cN = 2$ gauged supergravity.
This can be done in two steps, by first reducing on $\Sigma_{\fg_2}$, and then on a circle. Fortunately, both reductions have been worked out in the literature, respectively in  \cite{Szepietowski:2012tb} and  \cite{Hristov:2014eza},
and we can use the results reported there.

The consistent truncation of seven-dimensional $\cN = 4$ $\SO(5)$ gauged supergravity reduced on a Riemann surface has been discussed in \cite{Szepietowski:2012tb}.
It contains two vector multiplets and a charged hypermultiplet.
The vector multiplet scalars in 5D $\cN = 2$ supergravity are parameterized by a set of constrained \emph{real} scalars $L^I$ satisfying
\bea
 \frac{1}{6} c_{IJK} L^I L^J L^K = 1 \, ,
\eea
where the symmetric coefficients $c_{IJK}$ can be identified with the 't Hooft cubic anomaly coefficients \cite{Tachikawa:2005tq}, and for the model at hand $c_{123}=1$ is the only nonzero component.
The prepotentials and Killing vectors associated to the hypermultiplet can be found in \cite{Szepietowski:2012tb} but we will not need the explicit form of them here. 

We can proceed further by adding a momentum $n$ along $S^1 \subset \text{AdS}_3$.
The near-horizon geometry of the 5D black string is then $\text{BTZ} \times \Sigma_{\fg_1}$, where the metric for the extremal BTZ reads \cite{Banados:1992wn}
\bea
 \label{BTZ}
 \rd s_3^2= \frac14 \frac{ - \rd t^2+ \rd r^2}{r^2} + \rho \left[ \rd z +\left( - \frac14 + \frac{1}{2 \rho r} \right) \rd t \right]^2 \, .
\eea
Here, the parameter $\rho$ is related to the electric charge $n$.
This is \emph{locally} equivalent to AdS$_3$, since there exists locally only one constant curvature metric in three dimensions, and solves the same BPS equations; however, they are inequivalent globally.
Compactifying the 5D black string on the circle \cite{Hristov:2014eza} we obtain a static BPS black hole in four dimensions,
with magnetic charges $(\fs_1,\fs_2)$ and  electric charge $n$.
It can be thought as a domain wall which interpolates between an AdS$_2 \times \Sigma_{\fg_1}$ near-horizon region and an asymptotic \emph{non-AdS$_4$} vacuum.

The 4D $\cN = 2$ gauged supergravity is the STU model $(n_{\text{V}} = 3)$ coupled to a charged hypermultiplet \cite{Hristov:2014eza}. It has the prepotential
\bea 
 {\cal F}_{\text{sugra}} (X^{\Lambda}) = - \frac{1}{6} \frac{c_{IJK} X^I X^J X^K}{X^0} = - \frac{X^1 X^2 X^3}{X^0} \, .
\eea
The new vector multiplet corresponds to the isometry of the compactification circle and is associated with $\Lambda=0$. The physical scalars $z^I= X^I/X^0$ are now \emph{complex}.
Their imaginary part is proportional to the five-dimensional real scalars $L^I$ and the real part is given by the component  of the gauge fields along the compactified direction.
In the near-horizon region, the hyperscalars have a nonzero expectation value and, consequently, one of the gauge fields\footnote{This vector is called $c_\mu$ in  \cite{Szepietowski:2012tb}.} becomes massive.  The only physical role of the hypermultiplet is indeed to Higgs one of the gauge fields leaving  an effective theory  with only two massless vector multiplets. We can write down the effective theory as follows.
The hyperino BPS equation can be obtained by reducing to four dimensions equation  \cite[(A.22)]{Szepietowski:2012tb}. After setting the massive gauge field to zero and considering constant scalars at the horizon, we find that%
\footnote{We are  using \cite[(23) and (36)]{Szepietowski:2012tb}. The relations between parameters are the following: $- 4 \eta_2 p_I = \ft_I$ and $L^I_{\text{here}} = X^{I}_{\text{there}}$.
Note that, in the black string of \cite{Benini:2013cda}, the 7D gauge coupling has been fixed as $m = 2$.}
\be
 4 \eta_2 X^3 + \ft_1 X^2 + \ft_2 X^1 =0 \, .
\ee
We see that the hyperino BPS equation only imposes a linear constraint among the sections, which can be used to write an effective prepotential
\bea
 \label{effective:prepotential}
 \cF_{\text{sugra}} (X^{\Lambda}) = \frac{1}{4 \eta_2} \frac{X^1 X^2 (\ft_1 X^2 + \ft_2 X^1)}{X^0}
 = \frac{1}{8 \eta_2 X^0} \sum_{I = 1}^2 \ft_I \frac{\partial ( X^1 X^2)^2}{\partial X^I} \, ,
\eea
for the sections $X^0$, $X^1$, $X^2$ corresponding to the massless vectors at the horizon.

We can now use \eqref{attractor}. In our case, the vector of FI parameters reads \cite{Hristov:2014eza}
\be
 \cG = ( 0 , 0 , 0 , 0 , g , g ) \, .
\ee
The charge vector of the 4D black hole is also given by
\be
 \cQ = - \frac{1}{g \eta_1} \left( 0 , \fs^1 , \fs^2 , 4 g^2 G_{\text{N}}^{(4)} n , 0 , 0 \right) \, ,
\ee
so that \eqref{attractor} can be written as
\be
 \label{attractor:effective:theory}
  \cI_{\text{sugra}} (\hat X^\Lambda) = \frac{\mathbbm{i} \pi}{16 g G_{\text{N}}^{(4)} \eta_2 \hat X^0}
  \sum_{I , J = 1}^2 \fs_I \ft_J \frac{\partial ( \hat X^1 \hat X^2)^2}{\partial \hat X^I \partial \hat X^J} - 2 \mathbbm{i} \pi g n \hat X^0 \, .
\ee
We may fix the constant $g = 4$ in the following. Upon identifying
\be
 \hat X^I \equiv \frac{\beta \Delta_I}{2 \pi g} \, , \qquad \hat X^0 \equiv \frac{\mathbbm{i} \tilde \beta}{2 \pi g} \, ,
\ee
and using the relations between field theory and gravitational parameters\footnote{The reduction in \cite{Szepietowski:2012tb} and  \cite{Hristov:2014eza} is done on a Riemann surface and a circle of volume $2\pi \eta_2$ and $2\pi$, respectively.}
 at large $N$
\be
 N^3 = \frac{3 \pi^2}{16 G_{\text{N}}^{(7)}} \, , \qquad
 G_{\text{N}}^{(4)} = \frac{G_{\text{N}}^{(7)}}{\text{vol}(\Sigma_2 \times S^1)} = \frac{G_{\text{N}}^{(7)}}{(2 \pi)^2 \eta_2} \, ,
\ee
we find that the attractor mechanism \eqref{attractor:effective:theory} elegantly matches the field theory result \eqref{SBH}. Explicitly, we can write
\be
 \label{localization:meets:holography}
 \cF_{\text{sugra}} (\hat X^\Lambda) \propto \wt \cW (\Delta , \tilde \beta) \, , \qquad \cI_{\text{sugra}} (\hat X^\Lambda) = \cI_{\text{SCFT}} (\Delta , \tilde \beta) \, ,
\ee
where $\wt \cW$ is given in \eqref{SYM:tildeW:on-shell:n}. 

We also see that the $\cI$-extremization principle  for the topologically twisted index of the 6D $\cN = (2,0)$ theory
correctly leads to the microscopic counting for the entropy of the six-dimensional black holes obtained by compactifying on a circle the black string solutions of \cite{Benini:2013cda}.

\section{Discussion and future directions}
\label{sec:discussion and outlook}

In this paper we took the first few steps towards the derivation and evaluation of the topologically twisted index of five-dimensional $\cN=1$ gauge theories.
There are countless aspects that we have just briefly addressed in this paper.

In particular, the structure of the index at finite $N$ needs to be studied in more details. It was argued in \cite{Bawane:2014uka,Bershtein:2015xfa,Bershtein:2016mxz} that the equivariant partition function is summed over a set of magnetic fluxes that satisfy complicated semi-stability conditions.
These conditions have been studied by mathematicians \cite{Kool2015} but they become increasingly complicated with $N$ and are almost intractable already for $N=3$ or $N=4$. A related problem is the choice of
integration contour. This is selected by supersymmetry, but it is not determined so simply in our approach. We expect, in analogy with three and four dimensions, that some sort of Jeffrey-Kirwan prescription   \cite{JeffreyKirwan}
is at work. It would be very interesting if the correct determination of the contour allows to simplify the final expression for the matrix model and also the semi-stability constraints on the fluxes.

In the paper we conjectured that the Seiberg-Witten prepotential $\cF(a)$ in five dimensions should play the role of the twisted superpotential  in three and four dimensions and, in particular, its critical points should be relevant 
for the evaluation of the topologically twisted index. In three and four dimensions, a large set of partition functions, not only the twisted indices,
can be written as a sum over Bethe vacua \cite{Closset:2016arn,Closset:2017zgf,Closset:2017bse,Closset:2018ghr}.
It would be interesting to see if $\cF(a)$ plays, at least partially, a similar role in five dimensions.
For this reason, it would be interesting to evaluate explicitly the index for simple theories in five dimensions and compare the results for dual pairs \cite{Bergman:2013aca,Bergman:2015dpa}.

Also our computations at large $N$ is based on the assumption of the importance of the critical points of $\cF(a)$.
The large $N$ results can be explicitly tested against holographic predictions.
It would be particularly interesting, from this point of view, to find a class of AdS$_6$ black holes depending non-trivially on a set of magnetic fluxes \cite{Hosseini:2018mIIA}.  This would allow to test the results in section \ref{sec:largeN} and compare with possible alternatives (like, for example, the one discussed in appendix \ref{sec:alternative}).

In the large $N$ analysis we considered for simplicity two particular theories, $\cN = 2$ SYM  and the $\USp(2N)$ UV fixed point.
We expect that our formalism and our general results \eqref{index:theorem:F:5D}, \eqref{index:theorem:W:5D} and \eqref{index:theorem:logZ:5D} extend to other theories,
with no particular complications. We also considered just the case of $\mathbb{P}^1\times \mathbb{P}^1\times S^1$ (which can be trivially generalized to $\Sigma_{\fg_1}\times \Sigma_{\fg_2}\times S^1$).
The main reason is that, for a factorized manifold, we can perform a dimensional reduction to three dimensions and use the results for three-dimensional topologically twisted indices, which are well developed.
One main ingredient of the analysis was the twisted superpotential of such compactification, defined for each sector of the gauge magnetic flux on $\Sigma_{\fg_2}$.
However, let us notice that, even for a compactification on $\cM_4\times S^1$, we can formally define a twisted superpotential. This can be  defined through \eqref{WF},
by considering the theory on the equivariant background with $\epsilon_1=\hbar$ and $\epsilon_2=0$ and by  gluing the corresponding Nekrasov's partition functions,
or just by analogy with \eqref{full:pert:twisted superpotential}. Once we have this twisted superpotential $\wt \cW(a,\fn)$, that depends on the Coulomb variables $a_i$ and gauge fluxes $\fn_i$, we can extremize both $\cF(a)$ and $\wt \cW(a,\fn)$ with respect to $a_i$ and $\fn_i$.  A generalization of \eqref{Sigma_g1xM_3:Z} would then give an expression for the topologically twisted index.
We do not know if this approach leads to the correct result, but we have reasons to expect that, in the large $N$ limit, the result for different $\cM_4$ should be very similar.
In particular, one can easily see from appendix \ref{sec:centralcharges(2,0)} that the two-dimensional trial central charge of the compactification of the $\cN = (2,0)$ theory on $\cM_4$
depends on $\cM_4$ only through the topological factor $p_1(\cM_4) + 2 \chi(\cM_4)$.
It would be very interesting to investigate further this issue.

We plan to came back to all these points in the near future.

\section*{Acknowledgements}

We would like to thank Francesco Benini, Giulio Bonelli, Kentaro Hori, Francesco Sala, Alessandro Tanzini, Masahito Yamazaki, Yutaka Yoshida and Gabi Zafrir for useful discussions and comments. We thank, in particular,  P. Marcos Crichigno, Dharmesh Jain and Brian Willett for pointing out a numerical mistake in section \ref{subsec:USp(2N)}  in the first version of this paper and Minwoo Suh for sharing his results  \cite{Suh:2018tul} before publication. Special thanks go to Yuji Tachikawa for several illuminating conversations and bringing important references to our attention.
We also like to acknowledge the collaboration with Kiril Hristov and Achilleas Passias on related topics.
The work of SMH and IY was supported by World Premier International Research Center Initiative (WPI Initiative), MEXT, Japan.
AZ is partially supported by the INFN and ERC-STG grant 637844-HBQFTNCER.
IY gratefully acknowledges support from the Simons Center for Geometry and Physics, Stony Brook University at which some of the research for this paper was performed.

\begin{appendix}

\section{Toric geometry}
\label{sec:toric geometry}

A manifold $\cM_4$ of complex dimension two is toric if it admits a $(\mathbb{C}^*)^2$ action and $(\mathbb{C}^*)^2$ itself is dense in $\cM_4$.
Smooth toric surfaces can be obtained by gluing together copies of $\mathbb{C}^2$ \cite{fulton1993introduction}.
In this appendix we briefly review this construction focusing on the aspects used in the paper.   

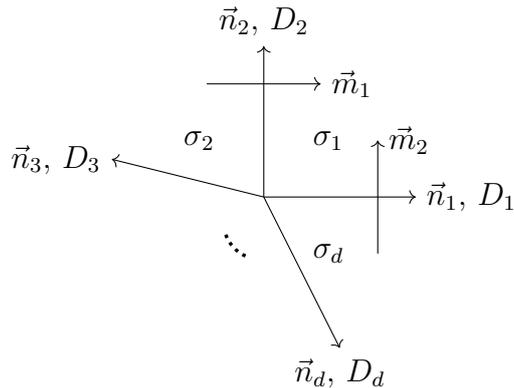
\begin{figure}
\centering
\begin{tikzpicture}
[scale=0.5 ]
\draw[->] (0,0) -- (4,0) node[right] {${\vec n}_1,\, D_1$};
\draw[->] (0,0) -- (0,4) node[above] {${\vec n}_2, \, D_2$};
\draw[->] (0,0) -- (-4,1) node[ left] {${\vec n}_3,\, D_3$};
\draw [line width=1.3, dotted] (-1,-1) arc [radius=1, start angle=200, end angle= 250];
\draw[->] (0,0) -- (2,-4) node[below ] {${\vec n}_d,\, D_d$};

\draw[->] (-1.5,3) -- (1.5,3) node[right] {${\vec m}_1$};
\draw[->] (3,-1.5) -- (3,1.5) node[right] {${\vec m}_2$};

\draw (1.7,1.5)  node {$\sigma_1$};
\draw (-1.7,1.5)  node {$\sigma_2$};
\draw (1.7,-1.5)  node {$\sigma_d$};
  \end{tikzpicture}
 \caption{A toric fan for a two-dimensional complex manifold. }
 \label{fig:toricfan}
  \end{figure}

A compact toric variety $\cM_4$ of complex dimension two is described a set of $d$ integer vectors ${\vec n}_l$ in
the lattice $N=\mathbb{Z}^2$ such that the angle between any pair of adjacent vectors is less than $\pi$,
as in figure \ref{fig:toricfan}. We order the vectors such that ${\vec n}_l$ and ${\vec n}_{l+1}$ are adjacent and we also identify ${\vec n}_{d+1}={\vec n}_1$.
The variety is smooth if ${\vec n}_l$ and ${\vec n}_{l+1}$ are a basis for the lattice $N=\mathbb{Z}^2$. We will assume from now on that our varieties are smooth. 

Consider also the dual lattice $M=N^*$, equipped with the natural pairing $\langle {\vec m}, {\vec n}\rangle = \sum_{i=1}^2 m_i n_i \in \mathbb{Z}$ for ${\vec m}\in M$, ${\vec n}\in N$.
Points in $N$ are associated with one-parameter subgroups of $(\mathbb{C}^*)^2$ and points in $M$ with holomorphic functions on $(\mathbb{C}^*)^2$.
In particular, we associate points ${\vec m}\in M$ with monomial functions $z_1^{m_1} z_2^{m_2}$ and define a natural $(\mathbb{C}^*)^2$ action  on the variables $z_i$.  

Each pair of adjacent vectors $({\vec n}_l,{\vec n}_{l+1})$ defines a two-dimensional cone $\sigma_l$  in the real vector space $N_{\mathbb{R}}= N \otimes_{\mathbb{Z}} \mathbb{R}$,
\bea
 \sigma_l = \{ \lambda_1 {\vec n}_l + \lambda_2 {\vec n}_{l+1} \, |\,  \lambda_i\ge 0\} \, .
\eea 
We can associate an affine variety $V_{\sigma_l}$ (which in the smooth case is a copy of $\mathbb{C}^2$) to each $\sigma_l$ as follows.
Consider the dual cone 
\bea
 \check \sigma_l= \{ {\vec m} \in M_{\mathbb{R}} \, | \, \langle {\vec m} , {\vec u}\rangle \ge 0\, \text{ for } \, {\vec u} \in \sigma_l \} \, ,
\eea
in $M_{\mathbb{R}}= M \otimes_{\mathbb{Z}} \mathbb{R}$. The lattice of integer points in $\check\sigma_l$ are generated by the primitive integer vectors ${\vec m}_l$ normal to the faces of $\sigma_l$ and pointing inwards.%
\footnote{A vector is primitive if its components are relatively prime.} We define $V_{\sigma_l}$ as the affine variety whose set of holomorphic functions is $\{ z_1^{\mu_1} z_2^{\mu_2}\}$ for all integer vectors $(\mu_1,\mu_2)\in \check\sigma_l$.
$V_{\sigma_l}$ is just isomorphic to $\mathbb{C}^2$. For example, consider the case ${\vec n}_1=(1,0)$ and ${\vec n}_2=(0,1)$.
The dual cone is then generated by ${\vec m}_1=(1,0)$ and ${\vec m}_2=(0,1)$ and contains all the integer points in the first quadrant.
The corresponding set of functions $ z_1^{\mu_1} z_2^{\mu_2}$ with $\mu_1\ge 0, \, \mu_2\ge 0$ are precisely the holomorphic functions on $\mathbb{C}^2$.
We also associate the vector ${\vec n}_1$ with the open dense subset where $z_2\ne 0$ and the vector ${\vec n}_2$ with the open dense subset where $z_1\ne 0$.
Since ${\vec n}_l$ and ${\vec n}_{l+1}$  are a basis for $N=\mathbb{Z}^2$, the generic case can be always reduced to the previous one by a change of lattice basis.
Explicitly, we can just replace in the previous example $z_1$ and $z_2$ with local coordinates $z_1^{(l)}= z_1^{m_{l,1}} z_2^{m_{l,2}}$ and $z_2^{(l)}= z_1^{m_{l+1,1}} z_2^{m_{l+1,2}}$
that parameterize a copy of $\mathbb{C}^2$, where ${\vec m}_{l}$ and ${\vec m}_{l+1}$ are the primitive integer vectors orthogonal to ${\vec n}_{l+1}$ and ${\vec n}_l$ and pointing inwards in $\sigma_l$, respectively. 

The smooth toric variety $\cM_4$ is then constructed by gluing together the $d$ affine varieties $V_{\sigma_l}$, isomorphic to $\mathbb{C}^2$, by identifying the dense open subset associated with ${\vec n}_l$ in $V_{\sigma_{l-1}}$ and $V_{\sigma_l}$.
This is completely analogous to the construction of $\mathbb{P}_1$ as a gluing of two copies of $\mathbb{C}$.
The action of the torus $(\mathbb{C}^*)^2$ on the variables $z_1$ and $z_2$ extends naturally to a global action on $V$.
Each chart $V_{\sigma_l}$ contains a special point, the origin $(0,0)\in \mathbb{C}^2$, which is invariant under the torus action.
These are the only invariant points and each of these belongs precisely to one chart.
We then see that there exactly are $d$ fixed points under the torus action $(\mathbb{C}^*)^2$, one for each of the two-dimensional cones $\sigma_l$.

Each vector ${\vec n}_l$ determines a divisor $D_l$ in $\cM_4$ and a corresponding line bundle.%
\footnote{In a local chart corresponding to ${\vec n}_1=(1,0)$ and ${\vec n}_2=(0,1)$, $D_1$ restricts to $z_1=0$ and $D_2$ to $z_2=0$.}
There are precisely two relations among them given by
\bea
 \label{relation}
 \sum_{l=1}^d \langle {\vec m}_k , {\vec n}_l \rangle D_l =0 \, , \qquad k=1,2 \, ,
\eea
where ${\vec m}_1=(1,0)$ and ${\vec m}_2=(0,1)$ is a basis for $M$.
This means that there are $d-2$ independent two-cycles in $\cM_4$.
The intersection number of $D_l$ and $D_{l^\prime}$, with $l\ne l^\prime$, is one if ${\vec n}_l$ and ${\vec n}_{l^\prime}$ are adjacent and otherwise is zero. 

As we saw, the partition function on $\cM_4\times S^1$ localizes at the $d$ fixed points of $(\mathbb{C}^*)^2$.
Each contributes a copy of the Nekrasov's partition function of the corresponding chart $V_{\sigma_l}$, twisted by the magnetic flux.
The total magnetic flux is given by a divisor $\sum_{l=1}^d \fp_l D_l$ in $\cM_4$, but the fixed point in the chart $V_{\sigma_l}$ will only feel the contribution of the flux coming from the divisors $D_l$ and $D_{l+1}$.
In the case of the chart $\mathbb{C}^2$, specified by the vectors ${\vec n}_1=(1,0)$ and ${\vec n}_2=(0,1)$, the local variables $z_1$ and $z_2$ parameterize the tangent space around the fixed point $z_1=z_2=0$.
In this case we write a copy of the Nekrasov's partition function with equivariant parameters $\epsilon_1$ and $\epsilon_2$ and Coulomb variable $a$ given by
\bea
 a+ \epsilon_1  \fp_1 +\epsilon_2 \fp_2 \, .
\eea
The replacement for a generic chart $V_{\sigma_l}$ is then easily obtained.
The equivariant parameters are replaced by the action of $(\mathbb{C}^*)^2$ on the local parameters $z_1^{(l)}$ and $z_2^{(l)}$:
\bea
 \epsilon_1^{(l)}= {\vec m}_l \cdot {\vec \epsilon} \, , \qquad \epsilon_2^{(l)}= {\vec m}_{l+1} \cdot {\vec \epsilon} \, ,
\eea
where ${\vec \epsilon}=(\epsilon_1,\epsilon_2)$ and ${\vec m}_{l}$ and ${\vec m}_{l+1}$ are the primitive integer vectors orthogonal to ${\vec n}_{l+1}$ and ${\vec n}_l$, respectively, and the Coulomb parameter by
\bea
 a^{(l)} = a +  \epsilon_1^{(l)} \fp_l + \epsilon_2^{(l)} \fp_{l+1}\, .
\eea

We can also use toric geometry to evaluate the contribution of the magnetic flux to the Hirzebruch-Riemann-Roch index.
This can be done be evaluating  \eqref{indexfixedpoint} or using \eqref{HRR}. The two results obviously coincide, \eqref{indexfixedpoint} being the localization formula for \eqref{HRR}.
For completeness, let us also show how to evaluate \eqref{HRR}  using  toric geometry techniques.  
In the case where $E$ is a line bundle, the index \eqref{HRR} reads
\bea
 \int_{\cM_4} \text{ch} (E) \text{td} (\cM_4)  = \chi(\cM_4) +  \int_{\cM_4} \left ( \text{td}_1(\bP^2) c_1(E) + \frac{1}{2} c_1(E)^2 \right ) \, .
\eea
We need the following general information about toric variety \cite{fulton1993introduction}.
The holomorphic Euler characteristic of every smooth toric four-manifold is one and $c_1(\cM_4)$ is associated with the divisor $\sum_{l=1}^d D_l$.
We can then evaluate the integral of products of Chern classes with the intersection of the corresponding divisors.
Since $\text{td}_1(\cM_4) = c_1(\cM_4)/2$ and $c_1(E) = \sum_{l=1}^d \fp_l D_l$ the index is given by
\bea\label{indexHRR}
 1+ \frac12 \bigg( \sum_{l=1}^d \fp_l D_l \bigg) \cdot \bigg( \sum_{l=1}^d (\fp_l +1) D_l \bigg) \, .
\eea
The intersection of divisors can be computed with the rules given above.
For example, $\bF_1$ is the toric manifold specified by the vectors  ${\vec n}_1=(1,0)$,  ${\vec n}_2=(0,1)$, ${\vec n}_3=(-1,1)$ and ${\vec n}_4=(0,-1)$.
The relations \eqref{relation} gives $D_3=D_1$ and $D_4=D_1+D_2$.
By combining these relations with the fact that $D_i \cdot D_j=1$ if ${\vec n}_i$ and ${\vec n}_j$ are adjacent and $D_i \cdot D_j=0$ otherwise (for $i\ne j$),
we can determine the nonzero intersections among the independent divisors $D_1$ and $D_2$: $D_1^2=0$, $ D_1\cdot D_2=1$ and $D_2^2=-1$.
We then easily obtain the index
\bea
 \frac12 ( \fp_2+\fp_4 + 1 ) ( 2 \fp_1-\fp_2 +2 \fp_3 +\fp_4 + 2 ) \, ,
\eea
which correctly reproduces the results in example \ref{ex3}. 

Notice that, by comparing \eqref{indexHRR} with \eqref{indexfixedpoint}, we can derive a general formula
for the self intersections of divisors 
\be
D_{l} \cdot D_{l} =
\begin{cases}
- ( n_{l+1,1} + n_{l-1,1} ) / n_{l,1} \, , & \text{if } \, n_{l,1} \ne 0 \\
- ( n_{l+1,2} + n_{l-1,2} ) / n_{l,2} \, , & \text{if } \, n_{l,2} \ne 0
\end{cases} \, .
\ee
The two expressions agree when both conditions are met.

\section{Metric and spinor conventions}
\label{sec: spinor conventions}

We work in Euclidean signature in five dimensions. We use $m,n,p,\ldots$
for spacetime and $a,b,c,\ldots$ for tangent space indices. Spacetime
indices are lowered and raised by $g_{mn}$ and its inverse $g^{mn}$.
Tangent space indices are lowered and raised by $\delta_{ab}$ and
$\delta^{ab}$. Repeated indices are summed. The two sets of indices
are related using a vielbein $e_{m}^{\phantom{m}a}$, such that
\be
g_{mn}=e_{m}^{\phantom{m}a}e_{n}^{\phantom{n}a} \, .
\ee
We define the Levi-Civita connection
\be
\Gamma^{m}_{\phantom{m}np}=\frac{1}{2}g^{mt}\left(\partial_{n}g_{tp}+\partial_{p}g_{nt}-\partial_{t}g_{np}\right) \, ,
\ee
and the spin connection
\be
\omega_{m}^{\phantom{m}ab}=e_{n}^{\phantom{n}a}\nabla_{m}e^{n b} \,  .
\label{eq:Spin_Connection}
\ee
The Riemann tensor is defined as
\be
R_{mn}^{\phantom{mn}ab}(e)=\partial_{m}\omega_{n}^{\phantom{n}ab}-\partial_{n}\omega_{m}^{\phantom{m}ab}+\omega_{m}^{\phantom{m}ac}\omega_{nc}^{\phantom{nc}b}-\omega_{m}^{\phantom{m}ac}\omega_{nc}^{\phantom{nc}b} \, .
\ee
The Ricci scalar is then given by
\be
R(e)=e_{a}^{n}e_{b}^{m}R_{mn}^{\phantom{mn}ab}(e) \, .
\ee

The spin group is $\text{Spin}(5)\simeq \USp(4)$.
A spinor is a section $\zeta^{\alpha}$ of the pseudo-real $\mathbf{4}$
representation of $\USp(4)$ with $\alpha \in \{1, \ldots, 4\}$.
The Clifford algebra is generated by $\Gamma^{a}$ with $a \in \{1, \ldots, 5\}$ and 
\be
 \left\{ \Gamma^{a},\Gamma^{b}\right\} =2\delta^{ab} \, .
\ee
The $\Gamma$ matrices have index structure $\left(\Gamma^{a}\right)^{\alpha}_{\phantom{\alpha}\beta}$.
We define the Pauli matrices 
\be
\tau^{1}=\begin{pmatrix}0 & 1\\
1 & 0
\end{pmatrix} \, ,\qquad\tau^{2}=\begin{pmatrix}0 & -\mathbbm i\\
\mathbbm i & 0
\end{pmatrix} \, ,\qquad\tau^{3}=\begin{pmatrix}1 & 0\\
0 & -1
\end{pmatrix} \, ,
\ee
and
\be
\sigma^{i}=-\bar{\sigma}^{i}=-\mathbbm{i}\tau^{i} \, , \qquad \sigma^{4}=\bar{\sigma}^{4}=\mathbbm{1}_{2} \, .
\ee
A possible choice of the $\Gamma^{a}$ is given by
\bea
\Gamma^{i}=\begin{pmatrix}0 & \sigma^{i}\\
\bar{\sigma}^{i} & 0
\end{pmatrix} \, ,\qquad i\in\{ 1,\ldots,4\} \, ,\\
\Gamma^{5}=\Gamma^{1}\Gamma^{2}\Gamma^{3}\Gamma^{4}=\begin{pmatrix}\mathbbm{1}_{2} & 0\\
0 & -\mathbbm{1}_{2}
\end{pmatrix} \, .
\eea
We define the $\Gamma$ matrices with multiple indices using permutations
as
\be
\Gamma^{a_{1}a_{2}\ldots a_{p}}\equiv\frac{1}{p!}\sum_{\sigma\in\text{perm}(p)} \sign (\sigma)\prod_{i=1}^{p}\Gamma^{a_{\sigma(i)}} \, ,
\ee
such that 
\be
\Gamma^{ab}=\frac{1}{2}\left(\Gamma^{a}\Gamma^{b}-\Gamma^{b}\Gamma^{a}\right).
\ee
We also define
\be
\sigma^{ij}\equiv\frac{1}{2}(\sigma^{i}\bar{\sigma}^{j}-\sigma^{j}\bar{\sigma}^{i}) \, , \qquad \bar{\sigma}^{ij}\equiv\frac{1}{2}(\bar{\sigma}^{i}\sigma^{j}-\bar{\sigma}^{j}\sigma^{i}) \, .
\ee
The covariant derivatives for spinors are defined by 
\be
\nabla_{m}\xi=\partial_{m}\xi+\frac{1}{4}\omega_{m}^{\phantom{m}ab}\Gamma_{ab}\xi \, .
\ee
We also define a spinor Lie derivative along a Killing vector field $v$ as 
\be
\mathcal{L}_{v}\xi\equiv v^{m}\nabla_{m}\xi+\frac{1}{2}\nabla_{m}v_{n}\Gamma^{mn}\xi \, .
\ee

The $\Gamma^{a}$ satisfy
\be
\left(\Gamma^{a}\right)^{\dagger}=\Gamma^{a} \, ,
\ee
where $\dagger$ denotes the conjugate transpose. We define a charge conjugation matrix 
\be
C_{\alpha\beta}\equiv\begin{pmatrix} \sigma^{2} & 0 \\
0 & \sigma^{2}
\end{pmatrix} \, ,
\ee
satisfying
\be
\left(\Gamma^{a}\right)^{\text{T}}=C\Gamma^{a}C^{-1} \, , \qquad
C^{*}=C \, , \qquad
C^{\dagger}=-C \, , \qquad
C^{*}C=-\mathbbm{1}_{4} \, ,
\ee
where $\text{T}$ denotes transposition. Spinor bilinears are defined as
\be
\xi\Gamma^{a_{1}a_{2}\ldots a_{p}}\eta\equiv\xi^{\alpha}C_{\alpha\beta}\left(\Gamma^{a_{1}a_{2}\ldots a_{p}}\right)^{\beta}_{\phantom{\beta}\gamma}\eta^{\gamma} \, .
\ee

The R-symmetry group of the five-dimensional $\mathcal{N}=1$ super-algebra is $\SU(2)_{\text{R}}$
with invariant antisymmetric tensor $\varepsilon^{IJ}$ such that
\be
\varepsilon^{12}=-\varepsilon^{21}=1 \, .
\ee
An $\SU(2)$ Majorana spinor is a doublet $\zeta^{\alpha}_{\phantom{\alpha}I}$
with $I\in\{ 1,2\} $ in the fundamental representation
of $\SU(2)_{\text{R}}$ satisfying the condition
\be
{\zeta_{\alpha}^{*}}^{I}=C_{\alpha\beta}\varepsilon^{IJ}\zeta^{\beta}_{\phantom{\beta}J} \, .
\ee

If the manifold $\mathcal{M}_4$ is not spin, all spinors are valued instead in an appropriate bundle associated with the choice of a $\text{spin}^c$ structure.
This choice can be made canonically for an almost complex manifold, but doing so may require a redefinition of certain background fluxes. See \cite{Festuccia:2016gul} 
for examples in the context of localization and \cite{Salamon96spingeometry} for a complete reference.
\section{Large $N$ 't Hooft anomalies of 4D/2D SCFTs from M5-branes}
\label{sec:centralcharges(2,0)}

In this appendix we provide a simple formula, at large $N$, in order to extract the 't Hooft anomaly coefficients of four-dimensional $\cN = 1$ (or two-dimensional $\cN = (0,2)$ field theories) 
that arise from M5-branes wrapped on a Riemann surface $\Sigma_{\fg_2}$ (or a four-manifold $\cM_4$).
The trial 't Hooft anomaly coefficients of this class of theories can be extracted by integrating the eight-form anomaly polynomial $I_8$ of
the 6D $\cN = (2,0)$ theory over $\Sigma_{\fg_2}$ or $\cM_4$  \cite{Alday:2009qq,Bah:2011vv,Bah:2012dg,Benini:2013cda}.
The anomaly eight-form of the 6D theory of type $\fg = (A_{n \geq 1} , D_{n \geq 4} , E_6 , E_7 , E_8)$ reads \cite{Harvey:1998bx,Intriligator:2000eq,Yi:2001bz}
\be
 \label{total:8form}
 I_8[\fg] = r_\fg I_8 [1] + d_\fg h_\fg \frac{p_2(NW)}{24} \, ,
\ee
where $I_8[1]$ is the anomaly eight-form of one M5-brane \cite{Witten:1996hc}, $NW$ is the $\SO(5)$ R-symmetry bundle, and $p_2(NW)$ is its second Pontryagin class.
Here we have denoted the rank, the dimension, and the Coxeter number of the Lie algebra $\fg$ by $r_\fg$, $d_\fg$ and $h_\fg$, respectively.
For the $A_{N-1}$ theory, in the large $N$ limit, the anomaly eight-form is simply given by
\be
 \label{8form2}
 I_8[A_{N-1}] \approx \frac{N^3}{24} e_1^2 e_2^2 \, ,
\ee
where $e_\varsigma$ $(\varsigma = 1, 2)$ are the Chern roots of $NW$.

Let us first consider the compactification of 6D theories on a Riemann surface $\Sigma_{\fg_2}$.
The prescription in \cite{Bah:2011vv,Bah:2012dg} for computing the anomaly coefficient $a(\hat\Delta)$ of the 4D SCFT amounts to first replace the Chern roots $e_\varsigma$ in \eqref{8form2} with
\be
 \label{replacement1}
 e_\varsigma \to - \frac{\ft_\varsigma}{2 ( 1 - \fg_2 )} x + \hat \Delta_\varsigma  c_1(F) \, ,
\ee
implementing the topological twist along $\Sigma_{\fg_2}$, and then integrate the $I_8 [A_{N-1}]$ on $\Sigma_{\fg_2}$:
\be
 \label{I8:integrated:Sigma}
 I_6 = \int_{\Sigma_{\fg_2}} I_8 [A_{N-1}] \, .
\ee
Here $x$ is the Chern root of the tangent bundle to $\Sigma_{\fg_2}$, $c_1(F)$ is a flux coupled to the R-symmetry, $\ft_\varsigma$ are the fluxes parameterizing the twist and $\hat\Delta_\varsigma$ parameterize the trial R-symmetry.
They fulfill the following constraints
\be
 \ft_1 + \ft_2 = 2 ( 1 - \fg_2 ) \, , \qquad \hat \Delta_1+\hat \Delta_2 = 2 \, .
\ee
On the other hand, the anomaly six-form of a 4D SCFT, at large $N$, reads
\be
 \label{I6:a}
 I_6 \approx \frac{16}{27} a(\hat\Delta) c_1(F)^3 \, .
\ee
The 't Hooft anomaly coefficient $a(\hat\Delta)$, at large $N$, where $c=a$, can then be read off 
by expanding $I_8 [A_{N-1}] $ at first order in $x$, using \eqref{I8:integrated:Sigma} --- $\int_{\Sigma_{\fg_2}} x = 2 ( 1 - \fg_2 )$ --- and comparing the result with \eqref{I6:a}.
It is simply given by
\be
 a(\hat\Delta) \approx - \frac{9 N^3}{128} \sum_{\varsigma=1}^2 \ft_\varsigma \frac{\partial (\hat\Delta_1\hat \Delta_2)^2}{\partial \hat \Delta_\varsigma} \, .
\ee
This is \eqref{a4D} in the main text.

Consider now the compactification of 6D field theories on $\Sigma_{\fg_1}\times \Sigma_{\fg_2}$ \cite{Benini:2013cda}.
We first need to replace 
\be
 \label{replacemen21}
 e_\varsigma \to - \frac{\fs_\varsigma}{2(1-\fg_1)} x_1 - \frac{\ft_\varsigma}{2(1-\fg_2)} x_2+ \hat \Delta_\varsigma  c_1(F) \, ,
\ee
where $x_\varsigma$ are now the Chern roots of the tangent bundles to $\Sigma_{\fg_\varsigma}$, and $\ft_\varsigma$/$\fs_\varsigma$ the fluxes on $\Sigma_{\fg_2}$/$\Sigma_{\fg_1}$, with
\be
 \ft_1 + \ft_2 = 2 ( 1 - \fg_2 ) \, , \qquad \fs_1 + \fs_2 = 2 ( 1 - \fg_1) \, .
\ee
Then we integrate $I_8 [A_{N-1}]$ over $\Sigma_{\fg_1} \times \Sigma_{\fg_2}$ using $\int_{\Sigma_{\fg_\varsigma}} x_\varrho=2(1-\fg_\varsigma) \delta_{\varsigma \varrho}$.
The result should be compared with the four-form anomaly polynomial of the two-dimensional SCFT that, in the large $N$ limit, where $c_l=c_r$, reads
\be
 I_4 \approx \frac{c_l(\hat\Delta)}{6}  c_1(F)^2 \, .
\ee 
This time only the term proportional to $x_1 x_2$ in \eqref{8form2} contributes and  we obtain 
\be
 c_l(\hat\Delta) \approx \frac{N^3}{4} \sum_{\varsigma,\varrho=1}^2 \ft_\varsigma \fs_\varrho \frac{\partial^2 (\hat\Delta_1\hat \Delta_2)^2}{\partial \hat \Delta_\varsigma\partial \hat\Delta_\varrho} \, .
\ee
This is \eqref{cr:cl:6D(2,0):largeN} with the identification $\beta\Delta_\varsigma/\pi=\hat\Delta_\varsigma$.

For completeness, we also study the compactification of 6D field theories on a four-manifold $\cM_4$ with a single flux on $\cM_4$.
Denoting with $x_1$ and $x_2$ the Chern roots of the tangent bundle to $\cM_4$, the topological twist can be implemented by \cite{Benini:2013cda}
\be
 \label{replacement3}
 e_\varsigma \rightarrow - \fr_\varsigma (x_1+x_2) + \hat \Delta_\varsigma  c_1(F) \, ,
\ee
where $\fr_1+\fr_2=1$. The integration of $I_8[A_{N-1}]$ over $\cM_4$ can be done by noticing that the integrals
\be
 p_1(\cM_4) = 3 \, \sigma(\cM_4) = \int_{\cM_4} (x_1^2+x_2^2) \, ,\qquad \chi(\cM_4) =\int_{\cM_4} x_1 x_2 \, 
\ee
give the first Pontryagin number and the Euler number of $\cM_4$. The result is then simply
\be
 c_l (\hat\Delta) \approx \frac{N^3}{8} \left( p_1(\cM_4) + 2 \chi(\cM_4) \right) \sum_{\varsigma , \varrho = 1}^2 \fr_\varsigma \fr_\varrho  \frac{\partial^2 (\hat\Delta_1\hat \Delta_2)^2}{\partial \hat \Delta_\varsigma \partial \hat \Delta_\varrho} \, .
\ee

We see that the 't Hooft anomaly coefficients of 4D (or 2D) field theories, which are obtained by wrapping M5-branes on $\Sigma_{\fg_2}$ (or $\cM_4$), can always be written as (multiple) applications of operators of the form
$\sum_{\varsigma=1}^2 \fr_\varsigma \partial_{\hat\Delta_\varsigma}$ to the function $(\hat\Delta_1\hat \Delta_2)^2$, appearing in the eight-form anomaly polynomial through the term $p_2(NW)$. 

\section{An alternative large $N$ saddle point for the $\USp(2N)$ theory}
\label{sec:alternative}

In this appendix, for completeness,  we discuss a possible alternative method for evaluating \eqref{Sigma_g1xM_3:Z} in the saddle point approximation.
Solving \eqref{BAEs:M3} in the large $N$ limit gives a relation between $a_{(i)}$ and $\fn_i$.
Eliminating $a_{(i)}$, we can write \eqref{Sigma_g1xM_3:Z} as a sum over the fluxes $\fn_i$:
\bea
 \label{Sigma_g1xM_3:Z2}
 & Z^{\text{pert}}_{ \Sigma_{\fg_2} \times ( \Sigma_{\fg_1} \times S^1)} (\fs , \ft , \Delta) =
 \frac{(-1)^{\text{rk}(G)}}{|\mathfrak{W}|} \sum_{\fn \in \Gamma_\fh}  Z^{\text{pert}} \big|_{\fm = 0} (a(\fn) , \fn)
 \bigg( \det\limits_{i j} \frac{\partial^2 \wt \cW^{\text{pert}} (a(\fn) , \fn)}{\partial a_i \partial a_j} \bigg)^{\fg_1 - 1} \, ,
\eea
where $a_{(i)}(\fn)$ is the large $N$ solution to \eqref{BAEs:M3}.
Each term in the sum is an exponentially large function of $N$, and we can use again the saddle point approximation to find the dominant contribution to the partition function.
While for $\cN=2$ SYM this method fails since \eqref{BAEs:M3} completely fixes the values of $\fn_i$, it works for the $\USp(2N)$ theory.
For completeness, we quote the result for the partition function evaluated with this approach
\bea
 \log Z ( \Delta_m , \ft_m , \fs_m ) 
 = \frac{4 N^{5/2}}{5 \sqrt{8 - N_f}} \sqrt{ ( \Delta_1 \ft_2 +  \Delta_2 \ft_1 ) ( \Delta_1 \fs_2 +  \Delta_2 \fs_1 ) ( \fs_1 \ft_2 +  \fs_2 \ft_1 )}\,  .
\eea
Notice that this is different from \eqref{USp(2N):Final:logZ:index theorem}.
It coincides with \eqref{USp(2N):Final:logZ:index theorem} only in the case of the universal twist, $\ft_m= 1- \fg_2$ and $\fs_m=1 -\fg_1$.
It would be very interesting to compare the two alternative results with the entropy of asymptotically AdS$_6$ black holes in massive type IIA supergravity, with magnetic fluxes $\ft_m$ and $\fs_m$ \cite{Hosseini:2018mIIA}.

\section{Polylogarithms}
\label{sec:PolyLog}

Polylogarithms $\Li_s (z)$ are defined by
\be
 \Li_s (z) = \sum_{n = 1}^{\infty} \frac{z^n}{n^s} \, ,
\ee
for $|z| < 1$ and by analytic continuation outside the disk. They satisfy the following relations
\be
\partial_a \Li_s (e^{\mathbbm{i} a}) = \mathbbm{i} \Li_{s-1}(e^{\mathbbm{i} a}) \, , \qquad \qquad \Li_s(e^{\mathbbm{i} a}) = \mathbbm{i} \int_{+ \mathbbm{i} \infty}^a \Li_{s-1}(e^{\mathbbm{i} a '}) \, \rd a' \, .
\ee
For $s \geq 1$, the functions have a branch point at $z=1$ and we shall take the principal determination with a cut $[1 , + \infty)$ along the real axis.
The functions $\Li_s(e^{\mathbbm{i} a})$ are periodic under $a \to a + 2 \pi$ and have branch cut discontinuities along the vertical line $[0, - \mathbbm{i} \infty)$ and its images.
For $0< \re a < 2\pi$, polylogarithms fulfill the following inversion formula%
\footnote{The inversion formul\ae{} in the domain $- 2 \pi < \re a < 0$ are obtained by sending $a \to - a$.}
\bea
\label{PolyLog:inversion formulae}
\Li_s (e^{\mathbbm{i} a}) + (-1)^s \Li_s (e^{- \mathbbm{i} a}) = - \frac{(2 \mathbbm{i} \pi)^s}{s!} B_s \left( \frac{a}{2 \pi} \right) \equiv \mathbbm{i}^{s - 2} g_s (a) \, ,
\eea
where $B_s (a)$ are the Bernoulli polynomials. In this paper we need, in particular,
\be\label{gfunctions}
 g_2 ( a ) = \frac{a^2}{2} - \pi a + \frac{\pi^2}{3} \, , \qquad g_3 ( a ) = \frac{a^3}{6} - \frac{\pi}{2} a^2 + \frac{\pi^2}{3} a \, .
\ee
One can find the formul\ae{} in the other regions by periodicity. Let us also mention that
\be
 \label{Z2:gs}
 g_s (2 \pi - a) = (-1)^s g_s(a) \, .
\ee
Finally, assuming $0 < \Delta < 2 \pi$, we find that
\be
 \label{PolyLog:asymptotic}
 \Li_s (e^{t + \mathbbm{i} \Delta}) \sim \mathbbm{i}^{s-2} g_s( - \mathbbm{i} t + \Delta ) \, , \qquad \text{ as } t \to \infty \, .
\ee

\end{appendix}

\bibliographystyle{ytphys}

\bibliography{5D_indices}

\end{document}